\documentclass[sn-basic]{sn-jnl}

\usepackage{graphicx}%
\usepackage{multirow}%
\usepackage{ltablex}
\usepackage{amsmath,amssymb,amsfonts}%
\usepackage{amsthm}%
\usepackage{mathrsfs}%
\usepackage[title]{appendix}%
\usepackage{xcolor}%
\usepackage{textcomp}%
\usepackage{manyfoot}%
\usepackage{booktabs}%
\usepackage{algorithm}%
\usepackage{algorithmicx}%
\usepackage{algpseudocode}%
\usepackage{listings}%
\usepackage{ulem}
\usepackage{wasysym}
\usepackage{mhchem}

\begin{document}

\title[Volatiles loss from terrestrial-type (exo)planets]{Upper atmosphere dynamics and drivers of volatiles loss from terrestrial-type (exo)planets}


\author*[1,2]{\fnm{Daria} \sur{Kubyshkina}}\email{daria.kubyshkina@oeaw.ac.at}

\author[3,4]{\fnm{M.J.} \sur{Way}}\email{michael.way@nasa.gov}

\author[5,6]{\fnm{Iannis} \sur{Dandouras}}
\equalcont{These authors contributed equally to this work.}

\author[1]{\fnm{Helmut} \sur{Lammer}}
\equalcont{These authors contributed equally to this work.}

\author[7]{\fnm{Antonino~Francesco} \sur{Lanza}}
\equalcont{These authors contributed equally to this work.}

\author[11]{\fnm{Manasvi} \sur{Lingam}}
\equalcont{These authors contributed equally to this work.}

\author[1,8]{\fnm{Rumi} \sur{Nakamura}}
\equalcont{These authors contributed equally to this work.}

\author[9]{\fnm{Moa} \sur{Persson}}
\equalcont{These authors contributed equally to this work.}

\author[1]{\fnm{Manuel} \sur{Scherf}}
\equalcont{These authors contributed equally to this work.}

\author[10]{\fnm{Kanako} \sur{Seki}}\email{seki@g.ecc.u-tokyo.ac.jp}
\equalcont{These authors contributed equally to this work.}

\affil[1]{\orgdiv{Space Research Institute}, \orgname{Austrian Academy of Sciences}, \orgaddress{\street{Schmiedlstrasse 6}, \city{Graz}, \postcode{8042}, \country{Austria}}}

\affil[2]{\orgdiv{Space Research and Planetology}, \orgname{Physics Institute, University of Bern}, \orgaddress{\street{Geselschaftsstrasse 6}, \city{Bern}, \postcode{CH-3012}, \country{Switzerland}}}

\affil[3]{
\orgname{NASA Goddard Institute for Space Studies}, \orgaddress{\street{2880 Broadway}, \city{New York}, \postcode{10025}, \state{New York}, \country{USA}}}

\affil[4]{
\orgname{Theoretical Astrophysics, Department of Physics and Astronomy, Uppsala University}, \orgaddress{\city{Uppsala}, \postcode{75310}}, \country{Sweden}}

\affil[5]{
\orgname{Institut de Recherche en Astrophysique et Planétologie, Université de Toulouse/CNRS/CNES}, \orgaddress{\city{Toulouse}, \country{France}}}

\affil[6]{
\orgname{Center for Space Research and Technology, Academy of Athens}, \orgaddress{\city{Athens}, \country{Greece}}}

\affil[7]{
\orgname{INAF-Osservatorio Astrofisico di Catania}, \orgaddress{\street{via S.~Sofia, 78}, \city{Catania}, \postcode{95123}, \country{Italy}}}

\affil[8]{
\orgname{International Space Science Institute}, \orgaddress{\street{Hallerstrasse 6}, \city{Bern}, \country{Switzerland}}}

\affil[9]{
\orgname{Swedish Institute of Space Physics}, \orgaddress{ \city{Uppsala}, \country{Sweden}}}

\affil[10]{
\orgname{Research Center for Advanced Science and Technology (RCAST), The University of Tokyo}, \orgaddress{\street{Komaba 4-6-1}, \city{Meguro-ku}, \postcode{153-8904}, \state{Tokyo}, \country{Japan}}}

\affil[11]{
\orgname{Florida Institute of Technology}, \orgaddress{\city{Melbourne}, \country{USA}}}


\abstract{Volatile loss from exoplanetary atmospheres and its possible implications for the longevity of habitable surface conditions is a topic of vigorous debate currently. {The vast majority of the habitable zone terrestrial-like exoplanets known to date orbit low-mass M- and K-dwarf stars and are subject to the conditions drastically different to those of terrestrial planets in the Solar System. In particular, they orbit far closer to their host stars than similar planets around G-dwarfs similar to the Sun. }
Therefore they receive higher X-ray and UV fluxes, even though luminosities of M-{ and K-}dwarfs are lower than those of heavier stars. Furthermore, due to their slower evolution, M-dwarfs retain high activity on the gigayear timescales.
The combination of these two effects has led to claims that most terrestrial planets orbiting M-dwarfs may have their atmospheres stripped from the higher X-ray and UV fluxes of their host stars. Opposing this are researchers who point out that volatile inventories for terrestrial exoplanets are ill-constrained, and hence, they may be able to ``weather the storm" of these higher X-ray and UV fluxes. In this chapter, we focus on exploring volatile loss in the upper atmospheres of terrestrial planets in our solar system and applications to those in exoplanetary systems {around stars of different types}.}



\maketitle
\section{Introduction}\label{sec:actual_introduction}
The escape of volatile elements is one of the key processes in the atmospheric evolution of terrestrial planets orbiting in the classical habitable zone \cite{Kasting1993,Kopparapu2013} or closer to their host stars. In the early stages of planetary evolution ($\sim$100-500\,Myr),  escape can occur through catastrophic impactor events \citep[e.g.][]{Schlichting2015Icar..247...81S,Kegerreis2020ApJ...897..161K} or extreme hydrodynamic outflow \citep[e.g.][]{owen_wu2016boil-off,Lammer2008SSRv..139..399L,Lammer2020SSRv..216...74L}. Depending on the properties of the host star, planetary mass, and orbit, these can fully erode or significantly deplete the primordial (hydrogen-dominated) atmosphere and delay the formation of {steam atmospheres degassed from the solidifying magma ocean and volcanically degassed} secondary atmospheres. 
Following the early phase of extreme escape, atmospheres of low-mass planets, even if otherwise stable, can remain vulnerable to the loss of volatiles over gigayear timescales. This is accomplished via various processes ranging from hydrodynamic outflow powered by stellar high-energy irradiation \citep[X-ray and extreme ultraviolet, EUV, together with XUV; e.g.][]{king2021MNRAS.501L..28K} to the erosion by stellar winds \citep[e.g. ][]{Kislyakova2014A&A...562A.116K,Modi2023MNRAS.525.5168M}. Besides further depleting the bulk of atmospheric species, it can alter the atmospheric composition through the fractionation of lighter and heavier species. This in turn may affect a wide range of processes including global atmospheric dynamics.

The majority of atmospheric escape processes are initiated in upper atmospheres, which are commonly defined as the atmospheric region above the mesopause {(the boundary between the mesosphere and thermosphere, where atmospheric temperature reaches its minimum)}. Here the atmospheric species are no longer affected by turbulent mixing, and their distribution becomes stratified. As this region is dominated by molecular diffusion, it follows that each species, $i$, has its own scale height, $H_i$. In the lower atmosphere this can be described by a single scale height, $H$, as eddy diffusion controls the atmospheric mixing. Lower and upper atmosphere can thus be defined as the homo- and hetero-sphere, respectively. The upper atmosphere includes the thermosphere, where most of the high-energy stellar radiation is absorbed, and most photoionization occurs. It also includes the exosphere, where the atmosphere becomes rarefied and transitions into the collisionless regime. For different planets, however, this general definition can imply rather different local conditions, and qualitatively (but not quantitatively) similar upper atmosphere features can lead to drastically different implications for atmospheric dynamics.
\begin{figure}
    \centering
    \includegraphics[width=0.7\linewidth]{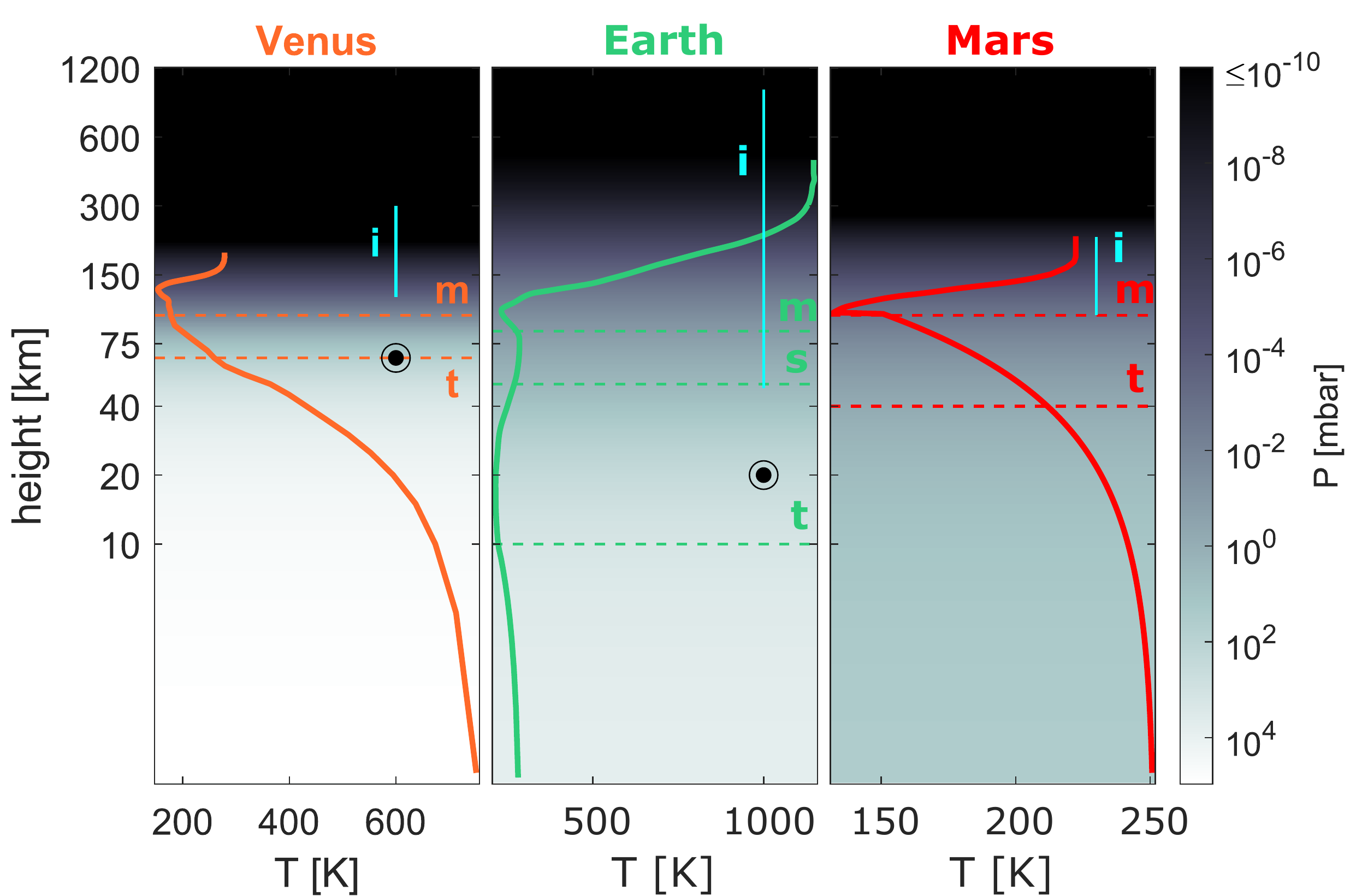}
    \caption{Height scales of atmospheres of Solar System's terrestrial planets. The solid lines show the temperature profiles against altitude, while the background color reflects the atmospheric pressure. Horizontal dashed lines denote the average positions of atmospheric boundaries: tropopause (``t''), stratopause (``s''), and mesopause (``m''). Vertical cyan lines depict the extension of ionospheres (``i'') under typical conditions. The ``$\odot$'' symbol denotes the 100 mbar level.}
    \label{fig:height_scales}
\end{figure}

In Fig.\,\ref{fig:height_scales}, we illustrate the height and pressure scales of Venus, Earth and Mars\footnote{{Fig.\,\ref{fig:height_scales} is illustrative and combines the data from \citet{Zasova2006CosRe..44..364Z} (for Venus), \citet{Minzner1977RvGSP..15..375M} (the US Standard Atmosphere, for Earth), \citet{Haberle1999JGR...104.8957H} (for Mars) with the upper atmosphere models from \citet{johnstone2018} and \citet{Amerstorfer2017}.}}.
Due to stratification, the lighter atmospheric species are ionised at higher altitudes compared to their heavier counterparts; thus, the heating in the thermosphere is most commonly attributed to  photoionisation of atmospheric hydrogen. The peak temperatures in this region depend strongly upon stellar input, planetary mass, and composition and size of the atmosphere defining the atmospheric scale height. Thus, the nitrogen-oxygen dominated atmosphere of Earth is more susceptible to heating compared to the CO$_2$ dominated atmospheres of Mars and Venus. This results in a much more extended exosphere and ionosphere. At the same time, the sizes and thermal structures of the atmospheres of Mars and Venus are drastically different despite their similar compositions due to the differences in mass and stellar irradiation.

Even though thermal escape processes are generally associated with photoionisation of H and photodissociation of H$_2$, which occur at high altitudes and typically within a narrow altitude range (around $\sim0.01-10$\,mbar pressures), photochemistry is effective over a wider range of altitudes. It becomes particularly relevant if one considers not just the reactions directly involving stellar photons (i.e. photoionisation, -dissociation, and -excitation of various atoms and molecules) but also the cascade of chemical reactions involving their products (e.g. secondary photoionisation, charge exchange, or recombination). This leads to the further redistribution of the absorbed energy, which can drastically change the predictions of atmospheric escape models \citep[e.g.][]{Shematovich2014A&A...571A..94S,GMunoz2023A&A...672A..77G,Kawamura2024ApJ...967...95K}. Therefore, an accurate assessment of atmospheric escape initiated in the upper atmosphere can require consideration of deeper atmospheric layers (at least down to a few bars).

Historically, atmospheric escape mechanisms have been divided into thermal (engaged through bulk atmospheric heating) and non-thermal (driven by the energizing of atmospheric particles through other channels). It is also common to assume that thermal processes dominate atmospheric evolution at early ages ($<$1 Gyr) while non-thermal processes only matter for the older planets and do not contribute significantly to bulk atmosphere depletion. The real picture is more complicated. Both types of escape processes can occur simultaneously and be powered by the same energy sources, which complicates the evaluation of their relative contributions. For example, it has been shown that the atmospheres of some planets orbiting low-mass stars (hence, on short period orbits even in the habitable zone) may be eliminated by non-thermal erosion from stellar winds \citep{Kislyakova2014A&A...562A.116K,Modi2023MNRAS.525.5168M}, even ignoring the contribution from atmospheric heating, which is also expected to last longer in M-dwarf than G-dwarf systems. To help navigate the diversity of atmospheric escape processes, we review the most relevant ones in Sec.\,\ref{sec:escape_processes} and summarise their main drivers (energy sources) and target planets in the Solar System and beyond in Tab.\,\ref{tab:thermal_proc} and Tab.\,\ref{tab:non-thermal}.

The timescales on which different escape processes perform, as well as their efficiency, depend crucially on the properties of a specific planet and its host star. These dependencies have been extensively studied for planets in the Solar System,  both observationally and theoretically. Still, some of them remain poorly understood despite decades of observations with numerous ground and space-based facilities. Meanwhile, for exoplanets only the strongest atmospheric escape of light species (mainly H and He) may be detected observationally, and the outcomes of such observations remain inconclusive from a theoretical point of view {due to a stellar noise and many degeneracies in theoretical models} \citep[e.g.][]{Orell-Miquel2024A&A...689A.179O}. {Otherwise, the absence of atmospheres around rocky exoplanets (revealed by many recent JWST observations) can put upper limits to the atmospheric evaporation timescales.} Moreover, the exoplanets known to date (even if we only consider terrestrial-type planets) occupy a region of parameter space completely different from the planets of the Solar System, both in their masses, sizes and environments surrounding their host stars.

Our perspective on understanding atmospheric escape processes from exoplanets and their effect on planetary atmospheric evolution depends crucially on our understanding of these processes and their sources within the Solar System. Therefore, in Sec.\,\ref{sec:observations} we discuss insights into atmospheric escape processes from observations of Mars (\ref{sec:observations_mars}), Venus (\ref{sec:observations_venus}), and Earth (\ref{sec:observations_earth}). In the following sections, we discuss the main parameters defining the efficiency of different escape processes, how they perform in the Solar System, and what implications they may have for exoplanets. 
We start by discussing the relevant parameters of various host stars -- the main energy sources in their systems -- in Sec.\,\ref{sec:stellar_input}. We further discuss the role of the planetary mass and size in Sec.\,\ref{sec:planetary_mass}, the role of  atmospheric composition in Sec.\,\ref{sec:composition}, and the controversial effects of planetary magnetic field in Sec.\,\ref{sec:magnetic_field}. We summarise the key points one has to consider when studying atmospheric evolution of exoplanets in Sec.\,\ref{sec:discussion}. For the convenience of the reader, we summarise the commonly used notations and abbreviations in Tab.\,\ref{tab:notations} in the Appendix.

\section{Diversity of atmospheric escape processes}\label{sec:escape_processes}

\subsection{Thermal Escape}\label{sec:intro_thermal}
By far the strongest atmospheric mass loss is caused by so-called thermal escape processes. Unified by their main cause, which is the heating of the planetary atmosphere by external or internal sources, they span a wide range of physical regimes, evolutionary timescales, and associated mass loss rates.
Among them, one can highlight kinetic mass loss processes (considering the escape of individual particles; enchanced and classic Jeans escape regimes in Tab.\,\ref{tab:thermal_proc}) and hydrodynamic atmospheric escape (considering the bulk outflow of atmospheres; boil-off, core-powered, and XUV-driven hydrodynamic escape in Tab.\,\ref{tab:thermal_proc}). The former typically describes the escape of the atmospheres of weakly or moderately irradiated terrestrial-like planets on billion-year timescales. The latter describes extreme outflows from young planets and planets orbiting very close to their host stars.

\begin{table}[hb]
    \centering
    \caption{Thermal escape processes}    
    \begin{tabularx}{\textwidth}{X|X|X|X|X|X}
    \toprule
      {Thermal escape type} & {Main driver} & {Major species} & {Driving process} & {Affected by mag. field?} & {Relevant planets {and ages}}\\
      \midrule
      \endhead
      {Boil-off} & {Stellar $L_{\rm bol}$} & {bulk atmosphere} & {Parker-wind-like expansion} & {no} & {Close-in \& younger than $\sim$100\,Myr$^a$}\\ \midrule
      {Core-Powered Mass Loss} & {Planet's luminosity$^b$} & {bulk atmosphere} & {Parker-wind-like expansion} & {no} & {Younger than $\sim$20\,Myr$^a$ or affected by tidal/magnetic heating}\\ \midrule
      {XUV-driven hydro. escape} & {XUV} & {H, He/bulk$^c$} & {Pho\-to\-ioni\-sa\-ti\-on of H (XUV heating)} & {yes} & {Ages between $\sim$1-5\,Gyr$^a$}\\ \midrule
      {Enhanced Jeans escape} & {XUV} & {light species} & {Pho\-to\-che\-mist\-ry heating} & {yes} & {Ages between $\sim$1-5\,Gyr$^a$}\\ \midrule
      {Classic Jeans escape} & {XUV} & {light species} & {Pho\-to\-che\-mist\-ry heating} & {yes} & {Older than $\sim$3\,Gyr$^a$}\\
      \bottomrule
    \end{tabularx}
    \footnotesize{$^a$ -- values given here are indicative and actual ages of transition between different thermal escape regimes depend on the type of host star and planetary parameters. For details, see Fig.\,\ref{fig:Escape_processes} and Sec.\,\ref{sec:discussion}.\\ 
    $^b$ Post-accretion cooling luminosity or heating of planetary interior/lower atmospheric layers through tidal or magnetic interactions with the host star. Approximation by \citet{gupta_schlichting2019MNRAS.487...24G} generalises this escape type to bolometric heating; in this case, physics is the same as for boil-off.\\
    $^c$ Under strong irradiation (hence, high escape), the flow of light species (mostly hydrogen) can drag the heavier species along causing bulk outflow; otherwise, only lighter species escape.}
    \label{tab:thermal_proc}
\end{table}

\subsubsection{Kinetic Jeans Escape}
\label{sec:intro_thermal_Jeans}
Jeans escape is the longest studied type of kinetic (i.e., considering loss of individual particles) thermal escape \citep[e.g. ][]{Jeans1955dtg..book.....J,Chamberlain1963P&SS...11..901C,Oepik1963GeoJ....7..490O,Mihalas_Mihalas1984oup..book.....M}. It considers atmospheres in hydrostatic equilibrium and assumes that atmospheric species follow a Maxwell–Boltzmann distribution. Therefore, volatile atmospheric species escape from the region around the exobase, where the atmosphere transitions to collisionless. Moreover, Jeans' model implies that the peak of the Maxwell–Boltzmann distribution lies well below the escape velocity needed for a particle to leave the planet's gravitation well $v_{\rm esc} = \sqrt{\frac{2GM_{\rm pl}}{r}}$ (here, $M_{\rm pl}$ denotes the planet's mass and $r$ the radial distance from the planet's centre). Thus the atmosphere experiences no significant bulk motion and only the most energetic particles escape.

To treat this problem accurately, one has to solve the Boltzmann equation \citep[e.g.][]{Volkov2011ApJ...729L..24V}. This requires heavy numerical computations, therefore the task is often simplified to an isothermal Maxwellian atmosphere\citep{Mihalas_Mihalas1984oup..book.....M}; in this case, the problem can be addressed (semi-)analytically and the distribution of atmospheric species can be written as
\begin{equation}
    f(\vec{x}, \vec{v}) = n\left( \frac{1}{\pi v_{\rm th}^{2}}\right) ^{3/2}\exp\left( -\frac{v^2}{v_{\rm th}^2} \right) \,.\label{eq:max_jeans}
\end{equation}
Here, $v_{\rm th} = \sqrt{\frac{2k_{\rm b}T}{\mu}}$ describes the thermal velocity of particles of the mass $\mu$ and $n$ denotes the numerical density of said particles. If the velocities of the atmospheric species $v$ exceed the escape velocity at the exobase, they can escape, and the outflow flux can be described as
\begin{eqnarray}
    \Phi = n\left( \frac{v_{\rm th}^{2}}{4\pi}\right)^{1/2}(1 + \lambda_{\rm exo})\exp(-\lambda_{\rm exo})\,,\label{eq:phi_jeans}\\
    \lambda_{\rm exo} = \frac{v_{\rm esc}^2}{v_{\rm th}^{2}} = \frac{GM_{\rm pl}\mu}{k_{\rm b}T_{\rm exo}r_{\rm exo}}\,.\label{eq:lambda_exo}
\end{eqnarray}
The parameter $\lambda_{\rm exo}$ is commonly referred to as Jeans parameter and represents the ratio of the planet's escape velocity to the thermal velocity of escaping particles, squared, as calculated at exobase. This definition can also be formulated as the ratio of the planet's gravitational energy to the bulk thermal energy of its atmosphere. This parameter is widely used in studies of planetary atmospheres for the basic classification of planets' escape regimes (see Sec.\,\ref{sec:intro_thermal_classification}).
{We note, that both $T_{\rm exo}$ and $r_{\rm exo}$ can be very different from the surface/photosphere parameters. $T_{\rm exo}$ can reach $\sim$~1000-10,000\,K and $r_{\rm exo}$ can stretch up to a few or even a few tens of planetary radii \citep[e.g.][]{VanLooveren2024Trappist}.}

The simplified equations presented above have limitations  \citep[see ][ and the references therein]{Gronoff2020JGRA..12527639G}. In particular, the atmospheric particle distribution is not exactly Maxwellian even for weakly-irradiated atmospheres. The particles with highest energies constantly escape from the atmosphere \citep{Chamberlain1971P&SS...19..675C}. Furthemore, atmospheres in general can not be considered isothermal \citep[e.g.][]{Merryfield1994P&SS...42..409M,Bauer1971,johnstone2018} {and do not consist of the uniform species as implied by Eq.\,\ref{eq:max_jeans}--\ref{eq:lambda_exo}. For the mixed atmospheres, this formulation can only provide a crude approximation if particle weight ($\mu$) is substituted by the mean species weight.} Finally, particles escape not only from the exobase but from wider range of altitudes around it. Therefore, in the most general case, the analytic approach must be applied with caution.

\subsubsection{Hydrodynamic Escape}\label{sec:intro_thermal_hydro}
As the atmosphere's thermal energy budget increases (through the heating by the host star or the high internal energy of the planet), the peak of the Maxwellian distribution shifts to higher velocities and, eventually the atmospheric particles' mean energy exceeds the bounding of planet's gravity and particles' mutual collisions.
In this case, deep atmospheric layers get involved into atmospheric escape and it transforms into continuous bulk outflow \citep[e.g. ][]{watson1981Icar...48..150W,Volkov2011ApJ...729L..24V}. This pushes the exobase to higher (and depleted) altitudes such that they are no longer considered part of the planetary atmosphere. 

In this case, the classic Jeans-like prescription breaks, and the atmosphere can be considered in a fluid-like approximation known as hydrodynamic atmospheric escape. For a while, it was used to study the early evolution of terrestrial-like planets in the Solar System \citep[see e.g.][]{Dayhoff1967Sci...155..556D,Sekiya1980PThPh..64.1968S,watson1981Icar...48..150W,Tian2005Sci...308.1014T,Erkaev2016MNRAS.460.1300E,Lammer2020Icar..33913551L} and the hydrogen-dominated atmospheres of hot giant exoplanets \citep[see, e.g.][]{Lammer2003ApJ...598L.121L,Lecavelier2004A&A...418L...1L,Baraffe2004A&A...419L..13B,Yelle2004Icar..170..167Y,Erkaev2007A&A...472..329E,Erkaev2015MNRAS.448.1916E,GMunoz2007P&SS...55.1426G,GMunoz2019ApJ...884L..43G,GMunoz2023A&A...672A..77G,Penz2008A&A...477..309P,Cecchi-Pestellini2009A&A...496..863C,Murray-Clay2009ApJ...693...23M,Owen_Jackson2012MNRAS.425.2931O,kubyshkina2018A&A...619A.151K,Kubyshkina2024A&A...684A..26K,Caldiroli2021A&A...655A..30C,Schulik_Booth2023MNRAS.523..286S}. 

In the early stages of evolution, while the energy budget is high, thermal escape processes are expected to be more effective than the of non-thermal ones, with about an order of magnitude difference in the mass loss rates \citep[e.g.][]{Kislyakova2014A&A...562A.116K,Modi2023MNRAS.525.5168M}. In turn, the mass loss rates in the hydrodynamic regime can be orders of magnitude higher than typical Jeans escape values. Hydrodynamic escape can thus be seen as the main driver of bulk atmospheric losses in  early planetary evolution \citep[e.g.][]{Lopez2012ApJ...761...59L,Chen2016ApJ...831..180C,kubyshkina2019A&A...632A..65K,kubyshkina2022MNRAS.510.3039K,Pezzotti2021A&A...654L...5P,Gu_Chen2023ApJ...953L..27G}. It is expected to have contributed largely to shaping the mass-radius and radius-period distributions of low-mass ($\leq100$\,M$_{\oplus}$) exoplanets \citep[e.g.][]{Fulton2017AJ....154..109F,Fulton2018AJ....156..264F,Owen_Wu2017ApJ...847...29O,gupta_schlichting2019MNRAS.487...24G,Mordasini2020A&A...638A..52M,Affolter2023A&A...676A.119A}. 

One can describe (radially symmetric) hydrodynamic atmospheres using a set of conservation equations for the mass, momentum, and energy depending on radial distance $r$ and time $t$
\begin{eqnarray}
\frac{\partial\rho}{\partial t} + \frac{\partial(\rho v r^2)}{r^2\partial r} &=& 0\,, \label{eq:mass_cons}\\
\frac{\partial\rho v}{\partial t} + \frac{\partial[r^2(\rho v^2+P)]}{r^2\partial r} &=& - \frac{\partial U}{\partial r} + \frac{2P}{r}\,, \label{eq:mom_cons}\\
\frac{\partial E}{\partial t} + \frac{\partial[vr^2(E + P)]}{r^2\partial r} &=& Q  + \frac{\partial}{r^2\partial r}(r^2\chi \frac{\partial T}{\partial r}) - \frac{\partial(\rho U)}{r^2\partial r}\,.\label{eq:en_cons}
\end{eqnarray}
Eqs.\,\ref{eq:mass_cons}-\ref{eq:en_cons} operate on atmospheric density $\rho$, bulk flow $\rho v$, and the total energy $E$ (kinetic + thermal). Other parameters are mean (electron) temperature $T$, thermal pressure $P$, {thermal conductivity of the neutral gas}, and the gravitational potential $U$ including planet's and host star's gravity and the effects of the orbital motion \citep[e.g.][]{Erkaev2007A&A...472..329E}. The second term on the right-hand of Eq.\,\ref{eq:en_cons} accounts for the thermal conductivity of the neutral gas.

The parameter $Q\,=\,H\,-\,C$ in Eq.\,\ref{eq:en_cons} is the sum of all heating and cooling rates given by (photo-)chemical processes included in the specific model. In hydrogen-dominated atmospheres, heating is mainly controlled by the photoionisation of hydrogen {molecules} (${\mathrm H + h\nu \rightarrow \mathrm H^+ + e^*}$) occurring at higher altitudes than for heavier species, and photodissociation of hydrogen atoms.  The most typical cooling process in such atmospheres is the collisional excitation of hydrogen atoms (Ly$\alpha$-cooling, ${\mathrm H + e^* \rightarrow \mathrm H^* + e}$). For particularly hot planets (typically, hot Jupiters), the heating rate can also be affected by metal line heating \citep[e.g.][]{Fossati2021A&A...653A..52F,GMunoz2023A&A...672A..77G}. Further contributions to cooling can be provided by processes such as the free-free interaction of {hydrogenic ions} (Bremsstrahlung cooling), production of ${\mathrm H^-}$, recombination of ${\mathrm H_2^+}$, ${\mathrm H_2}$ line, and ${\mathrm H_3^+}$-cooling \citep[e.g.][]{Kubyshkina2024A&A...684A..26K}.

For secondary atmospheres, contributions to heating can also be provided by the photoionisation of heavier atmospheric species and photodissociation of various molecules. The cooling processes also become more diverse (e.g. water molecule and ${\mathrm {CO}_2}$-cooling). Accounting for all the endothermic and exothermic chemical reactions between atmospheric elements requires solving an extensive chemical network along with liquid dynamics equations Eqs.\,\ref{eq:mass_cons}--\ref{eq:en_cons}, and modelling realistic atmospheres requires large computational resources. Therefore, most hydrodynamic models employing chemistry beyond that of a hydrogen-helium mixture are restricted to 1D geometry. However, the interaction of hydrodynamic outflow with stellar winds is an essentially 3D problem. Solving the two aspects of the problem within one model is non-trivial. The same concerns apply to deeper atmospheric layers (down to $P\sim$\,1\,bar), where a broader range of photochemistry processes and a possible condensation need to be accounted for.

\subsubsection{Classification of thermal escape processes}\label{sec:intro_thermal_classification}
Direct modeling of (exo)planetary atmospheres can be a challenging task, both numerically and from the theoretical point of view, as the applicability of a specific model can often be only verified a-posteriori. Therefore, it is common to preface the detailed modelling with a diagnosis based on the basic properties of the planet (such as mass, radius, and irradiation level), allowing one to define the most likely atmospheric escape regime.

A few parameters have been suggested for such analysis. The 
aforementioned Jeans escape parameter at the exobase level $\lambda_{\rm exo}$ (Eq.\,\ref{eq:lambda_exo}) is historically applied to evaluate terrestrial-like planets and the planets of the Solar System. 
It was shown that for $\lambda_{\rm exo} \lesssim 6$, the escape transforms from a Jeans-like to a hydrodynamic regime with a smooth transition from subsonic to supersonic flow \citep{Volkov2011ApJ...729L..24V,Erkaev2015MNRAS.448.1916E}. Here the retention of the atmosphere on Gyr timescales becomes unlikely, at least under the conditions evaluated. When $\lambda_{\rm exo}$ becomes $\lesssim$ 2-3, there is no stationary hydrodynamic transition from a subsonic to a supersonic flow. In such a case, a fast non-stationary atmospheric expansion results in extreme thermal atmospheric escape rates. This is commonly referred to as a ``blow-off'' of the upper atmosphere, which is expected to erode on short timescales. Distinct from boil-off and core-powered mass loss, ``blow-off'' does not consider any specific energy sources but relies solely on the parameters of a planet. However, one should note that the position of the exobase $r_{\rm exo}$ and the value of $\lambda_{\rm exo}$ depend on a range of parameters including the molecular weight of the atmospheric species, the planet's mass, the temperature at the exobase, and atmospheric inflation. The latter two depend on the planet's internal thermal energy budget (see Sec.~\ref{sec:internal}), the planet's mass, the XUV flux of the host star, and, crucially, on atmospheric composition, as detailed in Sec.\,\ref{sec:composition}.

For the present-day terrestrial-like planets in the solar system, the exobase height can be, on astronomical scales, considered $r_{\rm exo}\,\sim\,R_{\rm pl}$ or even $r_{\rm exo}\,=\,R_{\rm pl}$ if the atmosphere consists of the exosphere only (as at Mercury). Exoplanets, especially young ones, span a much wider range of parameter space. Thus, defining the position of the exobase for any given planet can be non-trivial (in particular, for planets with thick hydrogen-helium-dominated atmospheres).
Therefore, $\lambda_{\rm exo}$ is often generalized to the parameter $\Lambda$ calculated at the planetary photosphere, where the atmosphere becomes optically thin. Namely, $r_{\rm exo}$ in Eq.\,\ref{eq:lambda_exo} is substituted by planetary radius $R_{\rm pl}$ \citep{Fossati2017A&A...598A..90F}. For compact secondary atmospheres, parameters $\lambda_{\rm exo}$ and $\Lambda$ are nearly equivalent. However, for close-in sub-Neptune-like planets, hot Jupiters, and highly irradiated secondary atmospheres, $r_{\rm exo}$ and $R_{\rm pl}$ (hence, $\lambda_{\rm exo}$ and $\Lambda$) can differ by order of magnitude.
Hydrodynamic simulations show that the transition from blow-off to XUV-driven mass loss mechanism occurs when the gravitational parameter $\Lambda$ drops below 10--30 \citep[e.g.][]{Fossati2017A&A...598A..90F,kubyshkina2018A&A...619A.151K,owen2024MNRAS.528.1615O}. Planets are expected to transition from hydrodynamic to Jeans-like escape at $\Lambda$ values well above 30; the specific value depends strongly on the orbital separation and temperature of the planets \citep[hence, also the type of the stellar host, e.g.][]{Reza2025A&A...694A..88R}.

Due to the limitations of modern observational techniques, most low-mass planets detected to date orbit low-mass M-type stars (especially, low-mass planets in the habitable zone). For such planets, tidal interactions with their host stars become crucial (see Sec.\,\ref{sec:stellar_input:tides}). Furthermore, such interactions are crucial for planets in very short orbits, where stellar gravity can largely contribute to driving the hydrodynamic outflow or even solely power the outflow \citep[e.g.][]{Koskinen2022ApJ...929...52K}. To account for these effects in the classification of exoplanets, a recent study by \citet{Guo2024NatAs...8..920G} introduces the upgraded Jeans parameter $\lambda^*\,=\,\Lambda\times K$, where parameter $K\,<\,1$ accounts for the correction of the planet's gravitational potential according to stellar gravity and orbital motion \citep{Erkaev2007A&A...472..329E}. \citet{Guo2024NatAs...8..920G} predicts that for $\lambda^*\,<\,3$, the hydrodynamic outflow is driven by the tidal forces or the internal thermal energy of a planet (which sources are hard to disentangle for planets on close orbits), while for $\lambda^*>6$ the hydrodynamic escape is XUV-driven.

\subsection{Non-Thermal Escape}\label{sec:intro_non-thermal}

In the upper atmosphere, collisions become less frequent and the suprathermal component in the velocity distribution can also contribute to the atmospheric escape. There are two major drivers of the non-thermal atmospheric escape, i.e., stellar X-rays and UV radiation (XUV) and stellar wind (SW). The non-thermal escape mechanisms are roughly categorized into two types: the collisional non-thermal escape and the stellar wind-induced escape. As shown in Tab.\ref{tab:non-thermal}, which summarizes the non-thermal escape from terrestrial planets, the photochemical escape and charge exchange ({production of energetic neutral atoms, ENA, with $v\,>\,v_{\rm esc}$}) are the major collisional non-thermal escape. The SW-induced escape includes the ion pickup, atmospheric sputtering, and ionospheric outflows for unmagnetized planets, as well as the polar wind, auroral outflows, and plasmaspheric drainage plumes (plasma elements detached from the plasmasphere and propagating outwards) for magnetized planets. {Plasma-induced erosion is an analogous mechanism to SW-induced ion pick up escape, but is driven by the magnetospheric plasma within Saturn's magnetosphere.}  The relative importance of the non-thermal escape mechanisms depends on the planetary conditions such as atmospheric composition and intrinsic magnetic field {(MF)}.
\begin{table}[ht]
    \centering
    \caption{Summary of non-thermal escape processes from terrestrial planets {and Titan}} 
    \begin{tabularx}{\textwidth}{X|X|X|X|X|X}
        \toprule
        {Non-thermal escape type} & {Main drivers} & {Major species} & {Driving processes} & {Planetary magnetisation} & {Relevant SS {bodies}} \\
        \midrule
        \endhead
        {Photochemical escape} & {XUV} & {suprathermal (hot) {H, D, C, N, O}} & {dissociative recombination, photolysis} & {both} & {Mars {(H, D, C, N, O), Venus (H, D), Titan (N)}} \\
        \midrule
        {ENA production (charge exchange)} & {SW, XUV} & {H, O, C/H} & {charge exchange} & {both} & {Earth/Venus} \\
        \midrule
        {Ion pick-up} & {SW} & {O$^+$, C$^+$} & {ionisation} & {unmagnetized} & {Venus, Mars} \\
        \midrule
        {Plasma induced erosion$^a$} & {mag. plasma$^b$} & {N$_2^+$, CH$_4^+$, H$_2^+$} & {ionisation} & {unmagnetized} & {Titan} \\
        \midrule
        {Atmospheric sputtering} & {SW{/mag. plasma}, XUV} & {H, He, O, N, CO$_2$, Ne, Ar} & {pickup ion precipitation} & {unmagnetized} & {Venus, Mars{, Titan}} \\
        \midrule
        {Cold ion outflow (Plasma instabilities)} & {SW} & {bulk ionosphere} & {plasma processes (KHI$^d$, interchange, magnetic reconnection, etc.)} & {unmagnetized} & {Venus, Mars} \\
        \midrule
        {Polar wind} & {MF$^c$} & {H$^+$, He$^+$} & {ambipolar electric field} & {magnetised} & {Earth } \\
        \midrule
        {Plasmaspheric drainage plume} & {SW} & {H$^+$, He$^+$, O$^+$, C$^+$} & {SW electric field variations} & {magnetized} & {Earth} \\
        \midrule
        {Auroral outflows} & {SW, XUV} & {O$^+$, O$_2^+$} & {plasma heating/acceleration} & {magnetised} & {Earth} \\
        \bottomrule
    \end{tabularx}
    \footnotesize{{$^a$The same escape mechanism as ion-pickup but driven by magnetospheric plasma. $^b$The interaction between the corotating magnetospheric plasma of Saturn with Titan's upper atmosphere.} $^c$Here, an intrinsic magnetic field of a planet. $d$ Kelvin-Helmholtz instability.}
    \label{tab:non-thermal}
\end{table}

On Earth-like or heavier planets, where the gravitational escape velocity is considerable (11.2 km/s for the Earth), {if irradiation is low or moderate}, thermal escape in the form of neutral atoms concerns essentially hydrogen. Heavier species, such as oxygen, carbon, and nitrogen, must be accelerated to reach escape velocities by non-thermal escape processes. Thus, atmospheric escape of these heavier species, which often constitute the major part of secondary planetary atmospheres of terrestrial planets, occurs mainly in the form of ion escape. 
Neutral species in the upper atmosphere can be ionized by the stellar XUV radiation, charge exchange interactions, or electron impact. There are various processes to supply planetary ions to the magnetospheres \citep[e.g. ][and references therein]{Seki2015SSRv..192...27S}.

\begin{figure}
    \centering
    \includegraphics[width=\linewidth]{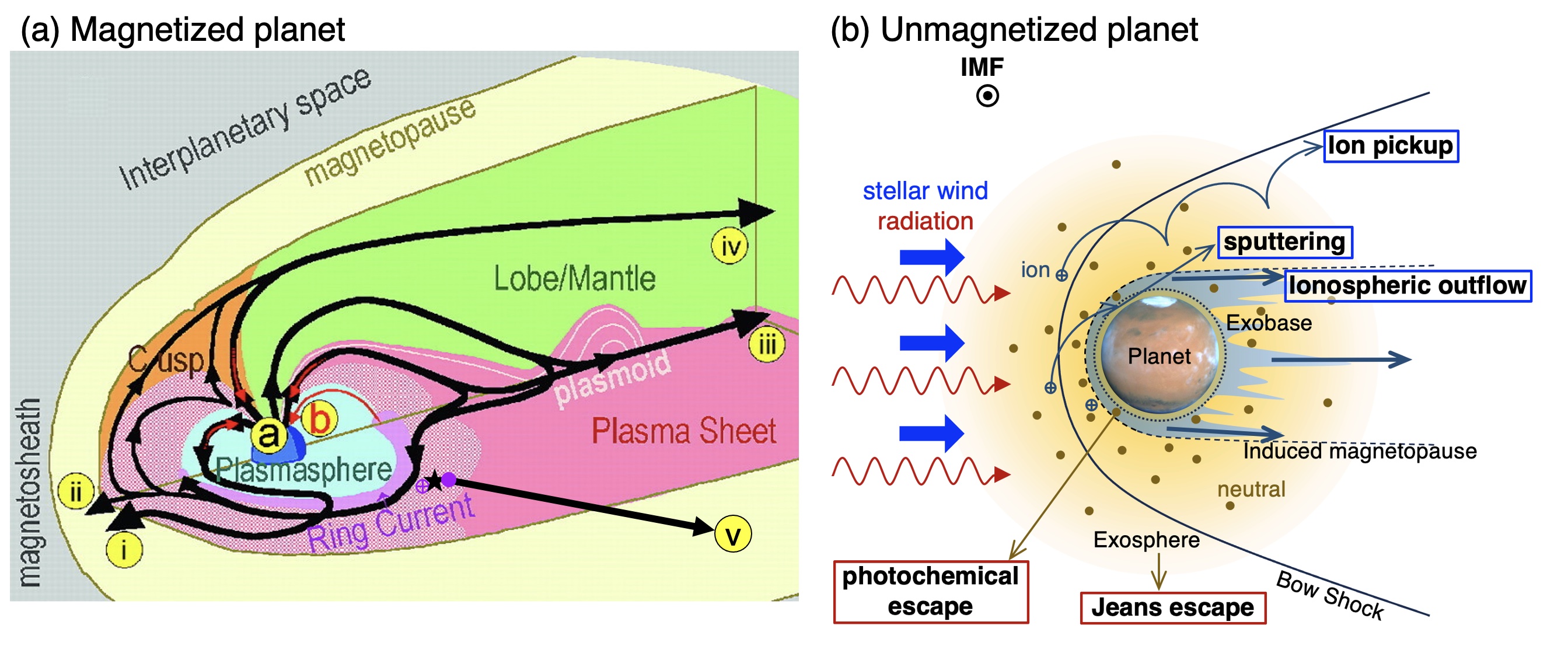}
    \caption{Schematic illustrations of (a) ion escape routes from a magnetized planet \citep[modified from][]{Seki2001Sci...291.1939S} and (b) atmospheric escape mechanisms from unmagnetized planets. In panel (a), \raise0.2ex\hbox{\textcircled{\scriptsize{a}}} and \raise0.2ex\hbox{\textcircled{\scriptsize{b}}} show outflow and return flow from/to the ionosphere, respectively. The escape routes \textcircled{ii}, \textcircled{iii}, and \textcircled{iv} in the panel (a) result from polar wind and auroral outflows, while routes \textcircled{i} and \textcircled{v} correspond to the plasmaspheric drainage plume and ENA production by charge exchange between the ring current ions and geocorona.}
    \label{fig:non-thermal_fig}
\end{figure}

In the case of magnetized planets, such as Earth, the ion escape occurs primarily from polar ionosphere, corresponding to latitudes higher than the subauroral regions as indicated with label \raise0.2ex\hbox{\textcircled{\scriptsize{a}}} in Fig.\,\ref{fig:non-thermal_fig} (a) \citep{Seki2001Sci...291.1939S}. Once the planetary ions outflow from the polar ionosphere, they can undergo various acceleration and transport in the magnetosphere (solid black arrows), and a significant part of them will eventually escape to interplanetary space (routes \textcircled{ii}, \textcircled{iii}, and \textcircled{iv}). {However}, even if the ions have energies above the planet's escape energy $E_{\rm esc}\,\sim\,v_{\rm esc}^2$, some of them return to the planetary atmosphere due to the combination of the magnetospheric convection and various plasma processes such as pitch angle scattering by wave-particle interactions as indicated by red arrows and label\raise0.2ex\hbox{\textcircled{\scriptsize{b}}}. 
{Two other mechanisms} enable the trapped ions in the planetary magnetic fields {to escape}: The plasmaspheric drainage plume (route \textcircled{i} in Fig.\,\ref{fig:non-thermal_fig} (a){)} and ENA production by the charge exchange (route \textcircled{v}). 

In the case of unmagnetized planets, the stellar wind can interact directly with their upper atmospheres as shown in Fig.\,\ref{fig:non-thermal_fig} (b). In addition to the thermal Jeans escape introduced in Sec.\,\ref{sec:intro_thermal_Jeans}, photochemical escape can be an important mechanism to cause the neutral atmospheric escape. The relative importance of the neutral escape compared to the ion escape largely depends on the stellar XUV radiation and the planetary mass. As the planet's mass increases, the relative importance of the ion escape to neutral escape generally becomes high due to the large escape energy. Three major ion escape mechanisms from the unmagnetized planet are ion pickup, atmospheric sputtering, and ionospheric outflows. Strictly speaking, the ionospheric outflow includes various plasma processes causing ion acceleration/heating. The basic characteristics of each escape mechanism are outlined in the following subsections.

\subsubsection{Collisional Non-Thermal Escape}\label{sec:intro_non-thermal_collisional} 
Photochemical escape refers to production of energetic atoms by photochemical reactions such as the dissociative recombination of molecular ions. For low-mass planets, such as Mars, the dissociative recombination itself can make atoms with $E\,>\,E_{\rm esc}$. For example, the dissociative recombination of O$_2^+$
\begin{equation}
{\rm O}_2^+  +  e^- \rightarrow  {\rm O^*  +  O^*}
\end{equation}
is an important process for oxygen escape from Mars \citep[e.g.][]{Lillis2017JGRA..122.3815L}; at more massive planets, such as Venus, {however,} the excess energy by the reaction alone is not sufficient to reach the escape energy. The resultant hot atoms of heavy species, such as O, can indirectly cause the escape of lighter species (H and D) by  elastic energy transfer collisions \citep[e.g.][]{McElroy1982Sci...215.1614M}
\begin{equation}
{\rm O^*  +  H  \rightarrow   O +  H^*}
\end{equation}

In the upper atmosphere, many photochemical reactions should be considered for the theoretical treatment of non-thermal escape \citep[e.g.][]{Fox2009Icar..204..527F}. When the atmospheric composition or host stellar conditions change, the relative importance of each reaction will also change \citep[e.g.][]{Nakamura2023JGRA..12831250N}.
Thus the efficiency of photochemical escape depends mainly on the planetary mass, atmospheric composition, and stellar XUV radiation as shown in Tab.\,\ref{tab:non-thermal}.

Charge exchange or ENA production generally refers to collisional processes of neutral exospheric species with energetic ions to produce translationally energetic and incompletely thermalized atoms. If the resultant ENAs have energies greater than $E_{\rm esc}$, they can cause atmospheric escape. 
If the planet has a global intrinsic field strong enough to form a substantial magnetosphere like Earth, magnetospheric dynamics facilitate various plasma processes accelerating planetary ions. During disturbed periods, such as large magnetic storms at Earth, planetary ions (e.g. O$^+$) contribute significantly to the formation of the ring current -- the high-energy ions trapped by the intrinsic dipole magnetic field \citep[e.g.][ and references therein]{Yue2019JGRA..124.7786Y}. The charge exchange between the ring current O$^+$ ions and geocorona form ENA by untrapping planetary ions from the magnetic field{, which then} contribute to {the escape of heavy species (oxygen) from the atmosphere} \citep[e.g.][ and references therein]{Keika2006JGRA..11111S12K,Ilie2013JASTP..99...92I}. Thus, the ENA production escape efficiency depends both on the stellar XUV radiation, which affects the corona, and stellar wind, which determines the ring current formation as shown in Tab.\,\ref{tab:non-thermal}.

It should be noted that the charge exchange process also indirectly affects atmospheric escape from unmagnetized planets by contributing to  upper atmospheric heating through precipitation of stellar wind origin ENA (mainly H and He) \citep[e.g.][]{Halekas2017JGRE..122..901H}. In some cases, it has been pointed out that the precipitation of ENA can cause non-thermal hydrogen escape from low-mass planets like Mars \citep[e.g.][]{Gregory2023JGRE..12807802G}.

\subsubsection{Stellar Wind Induced Escape from Unmagnetized Planets}\label{sec:intro_non-thermal_sw_unmag} 

The ion pickup process is caused by ionization of the neutral corona.  Photochemical reactions, such as dissociative recombination discussed above, contribute not only to the direct escape of neutrals but also to the formation of a hot neutral corona around the planet. The composition of the {hot} corona can include H, O, C, and N, depending on the {planetary mass,} atmospheric composition, and stellar XUV conditions\footnote{{Apart from H, O, C and N, sulphur is the only additional element that could theoretically escape photochemically. However, this can only happen from low-mass bodies with masses below roughly 0.5\,M$_{\mercury}$ such as Io. The only photochemically relevant reaction, i.e., the dissociative recombination reaction \ce{SO+ + e -> S + O}, produces S with an excess energy of 1.66\,eV, which is slightly above the escape energy needed for S to escape from Io but already far below the energy needed to escape from Mercury. In addition to S, SO also escapes photochemically from Io via \ce{SO2+ + e -> SO + O} with an excess energy of 1.66\,eV for SO (1.63\,eV are needed for SO to escape from Io)  \citep{Huang2023}.}}. Once the neutral atoms of the corona are ionised by stellar photon, electron impact, or charge exchange, they can be rapidly accelerated {to energies larger than $E_{\rm esc}$} by the {SW-induced} electric field. 
This escape mechanism is referred as the ``ion pickup'' \citep[e.g.][]{Luhmann2006P&SS...54.1457L,Kislyakova2014A&A...562A.116K,Curry2015P&SS..115...35C}. Since the number of pickup ions increases with decreasing distance from the planet, it can cause the deceleration of the stellar wind and is sometimes referred to as ``mass loading'' of the stellar wind. Phenomenologically, pickup ions contribute to a part of polar plumes, where planetary ion flow is primarily accelerated in the direction of the stellar wind electric field \citep[e.g.][]{Dong2015GeoRL..42.8942D,Sakakura2022JGRA..12729750S}. Since the trajectory of the ions depends on the location of the ionization, the distribution and efficiency of the ion pickup escape have been studied based on a global simulation of interaction between the stellar wind and the planetary atmosphere either with a hybrid simulation \citep[e.g.][]{Jarvinen2018JGRA..123.1678J} or a combination of a global {magnetohydrodynamic (MHD)} simulation and a statistical trajectory tracings of test particles \citep[e.g.][]{Curry2015P&SS..115...35C}. The ion pickup escape efficiency depends both on the {filling of} hot neutral corona influenced by the stellar XUV and the stellar wind conditions, especially on its electric field \citep[e.g.][]{Masunaga2017JGRA..122.4089M} as shown in Tab.\,\ref{tab:non-thermal}.

Atmospheric sputtering is caused by precipitation of the pickup ions into the atmosphere. A significant part of the pickup ions accelerated by the stellar wind electric field can re-enter the atmosphere, where they can sputter neutrals from the region near the exobase \citep[e.g.][]{Luhmann1991JGR....96.5457L,Curry2015P&SS..115...35C}. The spatial distributions of precipitating pickup ions have a hemispheric asymmetry in terms of the stellar wind electric field direction \citep[e.g.][]{Hara2013JGRA..118.5348H,Hara2017JGRA..122.1083H}. The spatial distribution may also depend on crustal magnetic fields \citep[][]{Hara2018JGRA..123.8572H} in a case like Mars. To estimate the escape rates due to  atmospheric sputtering caused by pickup ion precipitation, it is necessary to have a description of not only the atmosphere target gas and the exospheric source of the precipitating pickup ions, but also of the stellar wind properties that determine the precipitating pickup ion fluxes and energies. Since atmospheric sputtering efficiently facilitates the loss of neutral atmospheric constituents close to the exobase, it can cause the escape of various species such as H, He, C, O, CO$_2$, Ne, and Ar, contributing to mass fractionation of light noble gases \citep[][]{Jakosky1994Icar..111..271J}. While some models predict a strong increase of atmospheric sputtering loss with increasing stellar XUV flux, observations of heavy ion precipitation did not show an expected XUV dependence \citep[][]{Martinez2019GeoRL..46.7761M} and further studies are needed for clarification.

The cold ion or ionospheric outflows from unmagnetized planets is the general term for phenomena in which planetary-origin plasma is transported from the ionosphere into the induced magnetotail due to some bulk acceleration processes. These processes are mainly related to the momentum transfer from the SW to the upper ionosphere. It leads to energization and, thus, an outward flow of ionospheric ions throughout the planetary tail \citep{Lundin2007}. From spacecraft observations of  ions in the tail region of Venus, it is expected that polarized electric fields are also related to the energization process of the ions \citep{Hartle1990, Lammer2006}.

Plasma instabilities can also contribute to the escape of cold ionospheric ions. The observation of wave-like structures and plasma clouds or bubbles are indications that Kelvin-Helmholtz \citep[KH; e.g.][]{Wolff1980, Penz2004} and interchange \citep{Arshukova2004} plasma instabilities could be relevant to ion loss processes, especially around unmagnetized planets like Venus and Mars. These instabilities are generated due to disturbances of the interface between the solar wind and the ionopause layer, where the velocity shear between the two separated plasma layers and the curvature of the magnetospheric field can lead to a so-called detachment of ionospheric clouds or bubbles from the ionosphere.

Other energization processes include flux ropes formed by magnetic reconnection, magnetic tension force in the MHD regime, magnetic tension force with unmagnetized ions, ponderomotive force, and combinations of some heating and magnetic mirror effects \citep[e.g.][ and references therein]{Inui2019JGRA..124.5482I}. 
These processes are sometimes referred {to }as cool ion outflow, {momentum transfer}, cold ion escape or tailward escape \citep[e.g.][]{Fraenz2015P&SS..119...92F,Dong2017JGRA..122.4009D,Inui2018GeoRL..45.5283I}.
As summarized in Tab.\,2 in \citet{Inui2019JGRA..124.5482I}, each process has different characteristics such as mass dependence and the direction of acceleration against local magnetic fields, which are often useful to identify the acting process based on observations.
However, the relative contribution of each ion energization process is far from understood and upcoming multipoint observations with continuous solar wind monitoring at Mars will be essential for a quantitative understanding.

\subsubsection{Stellar Wind Induced Escape from Magnetized Planets}\label{sec:intro_non-thermal_sw_mag} 
{In this section, we focus on the effects of the dipolar magnetic field, which is expected to be the most relevant for (exo)planets. Dipole potential declines with distance slower than higher harmonics, hence, it is expected to cause largest-scale effects and therefore dominate the magnetic field effects even if the planetary magnetic field is not strictly dipolar \citep[e.g.][]{AndreevaTsyganenko2016JGRA..121.2249A}. If the planet has no dipolar magnetic field, the contribution from the higher harmonics/local magnetisation (as e.g. crustal field of Mars) can become non-negligible, but is still expected to have weaker effect on the escape processes. We discuss such cases in more detail in Sec.\,\ref{sec:magnetic_field_unmag}. }

The most basic escape mechanism associated with magnetised planets is the polar wind, which is caused by the ambipolar electric field in open magnetic field regions.
The polar wind is an ambipolar outflow of thermal plasma from the high-latitude ionosphere of planets with a significant global dipole intrinsic magnetic field like Earth \citep[e.g.][ and references therein]{Yau2007JASTP..69.1936Y}. In the original concept of the polar wind, it consists primarily of electrons and light (H$^+$ and He$^+$) ions, and \citet{Axford1968JGR....73.6855A} coined the term ``polar wind'' to describe the supersonic nature of the thermal plasma expansion and outflow, in analogy to the solar wind plasma from the solar corona into interplanetary space. However, observations at Earth indicate that the polar wind also consists of O$^+$ ions in addition to H$^+$, He$^+$, and electrons, which often have small velocities compared to the escape velocity. Various mechanisms accelerate the polar wind ions further at higher altitudes and help them to escape \citep[e.g.][]{Delcourt1993JGR....98.9155D}. There have been many efforts to model the polar wind theoretically \citep[e.g.][ and references therein]{Lemaire2007JASTP..69.1901L,Tam2007JASTP..69.1984T,Glocer2012JGRA..117.4318G}. As the polar wind is accelerated by the ambipolar electric field, in general, its velocity correlates with the local electron temperature. The escape rate due to the polar wind depends less on the stellar wind conditions than the auroral outflows described below. The dependence on the stellar XUV through the photoelectron effects is also small, since the escape flux is regulated by a net production of ions \citep[e.g.][]{Kitamura2015GeoRL..42.3106K}.

Auroral outflows are also prominent atmospheric escape, when a planet has a global intrinsic magnetic field like Earth.
High-energy particle precipitations from space are primarily concentrated on the auroral oval (including the cusp, Fig.\,\ref{fig:non-thermal_fig} (a)), a ring-shaped region around the magnetic poles where auroras frequently occur, caused by interactions between the solar wind and Earth's magnetosphere \citep[e.g.][]{Newell2009JGRA..114.9207N}. Both the particle kinetic energy and electromagnetic energy inputs are enhanced in the auroral oval and significant ion escape occurs through various plasma processes \citep[see details, e.g.][ and references therein]{Gronoff2020JGRA..12527639G}. %

There are many acceleration and heating mechanisms of the ionospheric ions \citep[e.g.][ and references therein]{Yau1997SSRv...80....1Y,Yau2021GMS...259..207Y}. 
To efficiently cause the outflow of heavy ions such as O$^+$ and O$_2^+$, upward transport of heavy ions from the low- to high-altitude ionosphere is important. The upward transport is often referred to as ion upflows, and both electron precipitations \citep[e.g.][]{Ogawa2008JGRA..113.5306O} and enhanced Joule heating by a strong electric field \citep[e.g.][]{Takada2021JGRA..12628951T} drive the upflows.

The ions need to be further accelerated or heated at high altitudes to escape, outflowing to the magnetosphere \citep[e.g.][ and references therein]{Yau1997SSRv...80....1Y,Yau2021GMS...259..207Y}. Morphologically, there are two types of auroral outflows: ion beams accelerated by the parallel potential drop along the magnetic field \citep[e.g.][]{Hull2003JGRA..108.1007H} and conics caused by perpendicular plasma heating and following acceleration by the magnetic mirror force \citep[e.g.][]{Miyake1996JGR...10126961M}. Wave-particle interactions with various waves including broadband extremely low frequency (BBELF), electromagnetic ion cyclotron (EMIC), electrostatic ion cyclotron (ESIC), lower hybrid (LH), and kinetic Alfven waves can contribute to ion heating. Since the effects of the stellar XUV are important for the source ionosphere, the efficiency of the ion escape by the auroral outflows depends both on the stellar wind conditions and XUV radiation.%

\begin{figure}
    \centering
    \includegraphics[width=0.55\linewidth]{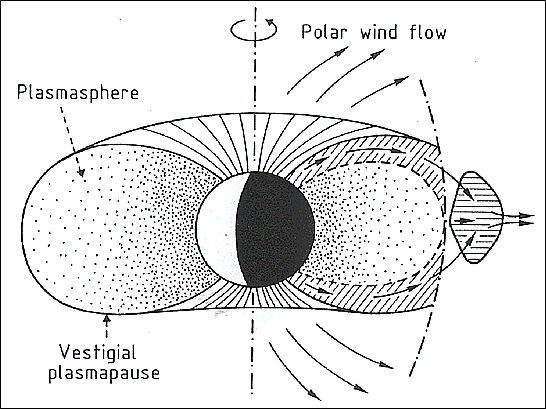}
    \caption{Plasma element being detached from the plasmasphere through the enhanced magnetospheric convection. From \citet{Lemaire2001JASTP..63.1285L}.}
    \label{fig:2.3.4_plasmasphere}
\end{figure}
%
Another path for the cold ion escape is through the plasmaspheric drainage plumes. As shown in Fig.\,\ref{fig:non-thermal_fig} (a), the plasmasphere is the region where the cold planetary plasma is trapped by the global planetary magnetic field and its size is determined by the balance between the corotation and stellar wind induced electric fields. Thus variations of the SW-induced electric field, which depends on the stellar wind velocity and interplanetary magnetic field, can cause the outflow of the plasmaspheric plasma (see Figure\,\ref{fig:2.3.4_plasmasphere}), which has important effects on the inner magnetospheric dynamics \citep[e.g.][]{Glocer2020JGRA..12528205G,Yamakawa2023JGRA..12831638Y}.
The stream of the cold planetary plasma from Earth's inner magnetosphere to the dayside magnetopause especially during geomagnetic storms is often referred to as plasmaspheric drainage plumes and is expected to contribute to ion escape from a magnetized planet \citep[e.g.][ and references therein]{Borovsky2008JGRA..113.9221B}. The efficiency of ion escape by plasmaspheric drainage plumes depends both on the strength of the planetary magnetic field and the stellar wind conditions.

\subsection{Contribution from the internal energy sources}\label{sec:internal}
The formulation of many escape processes discussed above assumes that the upper atmosphere's thermal energy budget is only supplied by external sources (XUV, SW). Thus, the contribution from the heat sources within the planet's interior or lower atmosphere has not been considered. However, such heating sources can affect atmospheric mass loss in multiple ways. 
This ``internal'' heating can be sustained by a few mechanisms, such as accretion of solids and gravitational contraction during the formation stage \citep[e.g.][]{ginzburg2018}, decay of the radioactive elements in the planetary core \citep[e.g.][]{Kamland2011NatGe...4..647K,Mordasini2012A&A...547A.112M}, stellar bolometric heating from \citep[e.g.][]{owen_wu2016boil-off,gupta_schlichting2019MNRAS.487...24G}, or tidal and magnetic interactions of a close-in planet with its host star (see details in Sec.\,\ref{sec:stellar_input}). 

Suppose that the heating is strong and the planetary gravity is low. In that case, the planet's internal thermal energy can power a hydrodynamic outflow on its own, without any contribution from photochemical heating (boil-off or core-powered mass loss). Such a regime is expected to be short-lived, as it is associated with extreme atmospheric mass loss rates that can overcome the XUV-driven escape by orders of magnitude. This leads to rapid cooling and contraction of the planet \citep[e.g][]{owen_wu2016boil-off,kubyshkina2020MNRAS.499...77K}, and quick cessation of the process.
Such escape mechanisms can, therefore, lead to the total evaporation of the atmosphere on a short timescale ($\sim$1--100\,Myr) or a swift reduction in its size and thermal budget. This mechanism is mainly relevant for primordial hydrogen-dominated atmospheres.

Furthermore, while core-powered mass-loss dominates the escape, the classic XUV-driven escape can be suppressed. The XUV heating, normally occurs in a narrow range of altitudes near the planetary photosphere ($R_\mathrm{pl}$), and its specific position and extension \citep[reduced to a single radius $R_{\rm eff}$ in common approximations, e.g.][]{watson1981Icar...48..150W,Erkaev2007A&A...472..329E} is defined by atmospheric opacity and the local pressure gradient. The heat sources located below the photosphere lead to atmospheric inflation, the pressure gradient becomes shallow, and the atmosphere remains dense \citep[$\rho>\sim10^{12}$\,${\rm cm^{-3}}$][]{kubyshkina2018A&A...619A.151K,owen2024MNRAS.528.1615O} and opaque up to altitudes order of a few planetary radii (which, for very short orbits, can be about as far as the planet's Roche lobe). Thus, high-energy stellar photons cannot penetrate the deeper atmospheric layers, making XUV heating ineffective.%

As the internal heat and the inflation decrease, the contribution from XUV heating increases. Eventually, it overcomes the core-powered mass loss, which soon becomes negligible. Still, the remaining heat can be sufficient for inflating the atmosphere to some extent. In such a case, the XUV heating occurs in an otherwise hydrostatic atmosphere. Yet at higher altitudes it increases the interaction surface $\pi R_{\rm eff}^2$ with stellar photons. The XUV heating is roughly proportional to $R_{\rm eff}$ to the power of 3 \citep[][]{watson1981Icar...48..150W} or higher \citep{kubyshkina2018ApJ...866L..18K}, and in the case of inflated atmospheres can be considered as enhanced XUV heating \citep[e.g.][]{owen2024MNRAS.528.1615O}.

Atmospheric inflation is further relevant for {both} non-hydrodynamic {thermal and some of} non-thermal escape processes, as atmospheric inflation also forces the exobase ($r_{\rm exo}$) to higher altitudes. {For the Jeans-like escape, increasing $r_{\rm exo}$ leads to the decrease of $\lambda_{\rm exo}$ and hence, increase of the mass loss (see Eq.\,\ref{eq:phi_jeans}).}
{For} the non-thermal escape processes, {the expansion of the exosphere implies that} the interaction surface with stellar XUV and SW increases; it is particularly relevant for photochemical escape, ENA production by SW, and ion pick-up.

\section{Solar System Observables in the context of atmospheric loss mechanisms}\label{sec:observations}

\begin{figure}
    \centering
    \includegraphics[width=1.0\linewidth]{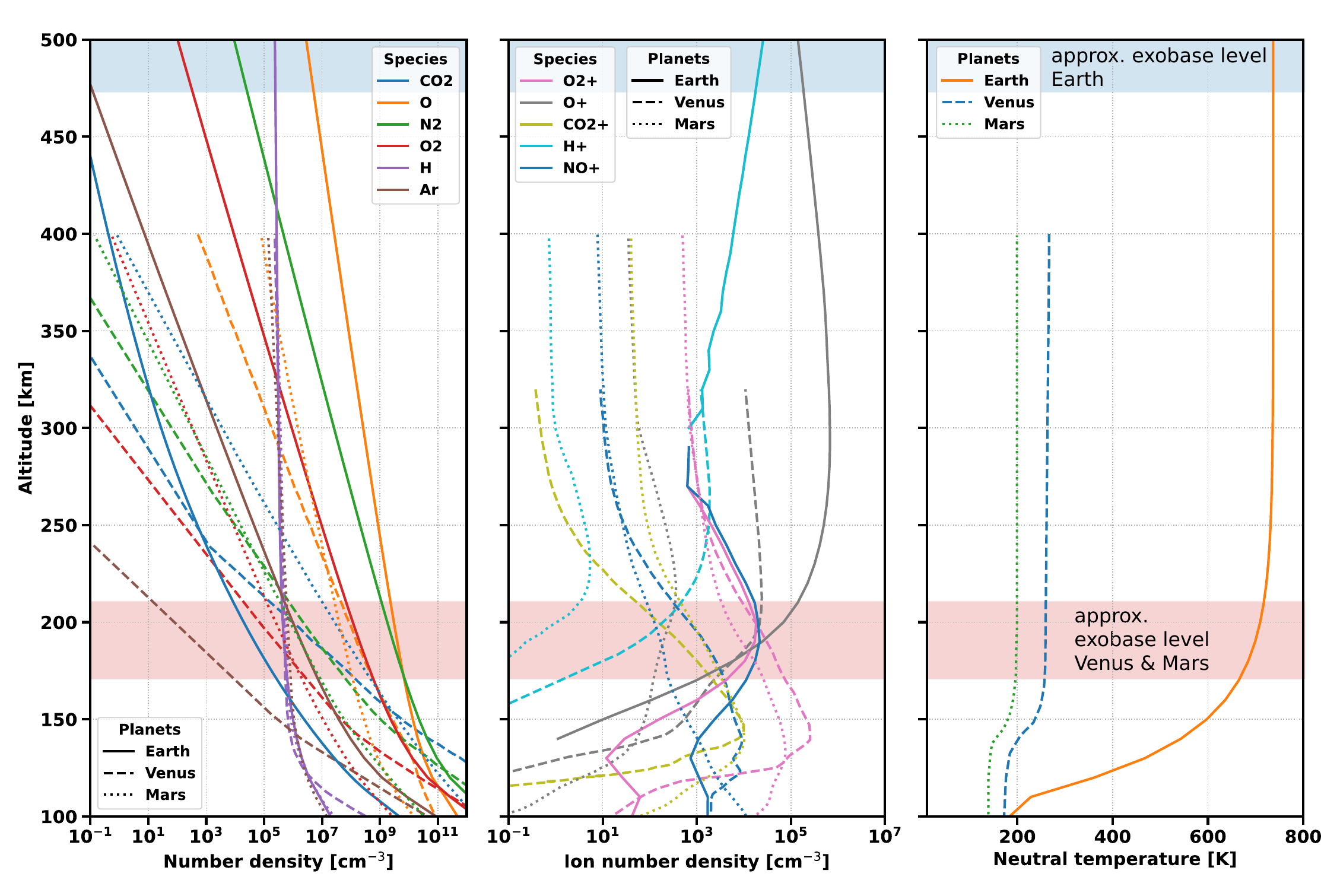}
    \caption{{Upper atmospheric profiles for neutrals (left panel), ions (middle panel), and the neutral temperature (right panel) for Earth (solid), Venus (dashed), and Mars (dotted). The profiles for the Earth are from the reference atmosphere model NRLMSIS-2.0 \citep{Picone2002} and the reference ionosphere model IRI \citep{Bilitza2001}, both for solar minimum (December 02, 2015), except for CO$_2$, which was modelled with the 1D upper atmosphere model Kompot \citep{johnstone2018} and taken from \citet{Scherf2025}. For Venus and Mars, all profiles are based on solar minimum conditions and were taken from \citet{Fox2001} and \citet{Fox2009Icar..204..527F}, respectively. We note that for the solar maximum, temperatures and atmospheric densities can be slightly higher for all three planets. The solar activity-dependent exobase levels for Venus, Mars, and Earth are schematically highlighted in the figure.}}
    \label{fig:profiles}
\end{figure}

{Three of the four terrestrial planets in the solar system, namely, Venus, Earth, and Mars, host a collision-dominated atmosphere. Their surface pressure and atmospheric composition, however, vary significantly. Whereas Venus hosts a CO$_2$-dominated atmosphere with almost 100\,bar surface pressure, Earth's atmosphere is dominated by N$_2$ and O$_2$ with only minor amounts of CO$_2$ and has a surface pressure almost two orders of magnitude lower ($\sim1$\,bar). The Martian atmosphere, dominated by CO$_2$, has a surface pressure more than two orders of magnitude lower than that of Earth (i.e., $\sim$5\,mbar). The difference in atmospheric composition between Earth, on the one hand, and Venus and Mars, on the other, implies a substantial difference in their upper atmospheric structures. Figure~\ref{fig:profiles} compares the thermospheric profiles of some neutral (left panel) and ion species (middle panel) on Earth (solid), Venus (dashed), and Mars (dotted), and their neutral temperature profiles.} 

{Earth's upper atmosphere is relatively hot, reaching temperatures of up to $\sim$700-1200\,K \citep[depending on solar activity, e.g.][]{Picone2002} in the thermosphere, where the incident XUV surface flux from the Sun is absorbed. Its exobase level can be as high as roughly 500 to 1000\,km and both temperature and exobase level depend on the solar activity. The upper atmospheres of Venus and Mars, on the other hand, are compact and cold. Besides the higher molecular weight of their atmospheres compared to the Earth's, the infrared-coolant CO$_2$ efficiently re-emits the incident XUV surface flux back into space, which cools their thermosphere temperatures to about 200 to 250\,K and reduces their exobase altitudes to roughly 200\,km \citep[see, e.g., Fig.~10 in][]{Way2022}. These are notable differences since Venus experiences twice as much irradiation from the Sun as the Earth, and Mars has a mass that is only a tenth of the Earth's. Atmospheric composition, therefore, matters for investigating and understanding atmospheric escape processes.}

{Finally, we note that among the three terrestrial planets with collisional atmospheres, the Earth is the only planet that hosts an intrinsic magnetic field, whereas Venus and Mars do not. This significantly alters the dominant processes responsible for atmospheric escape on these planets (see also Table~\ref{tab:non-thermal} for a summary on the various non-thermal loss channels). For Venus, atmospheric escape is dominated by ion outflow and ion pickup escape for O and C, and by photochemical escape for H and D. Sputtering of various species constitutes a minor loss channel, whereas thermal escape is negligible for all of the species. In the case of the Earth, ion outflows originating from high-latitude regions such as the polar wind and auroral outflows of its magnetosphere are the dominant process for most of the species (H$^+$, He$^+$, N$^+$, C$^+$, O$^+$, O$^{2+}$), with H also being lost photochemically. For Mars, O, N, and C are predominantly lost through photochemical processes and via ion escape. Thermal escape, however, is the major loss channel for hydrogen. In addition, Mars is the only planet for which sputtering can be an important driver of escape for various species ranging from the light H to the heavy Ar isotopes. For Titan, photochemical escape and ion escape induced via Saturn's magnetospheric plasma are at present the most important loss channels for N, H, and \ce{CH4}. Summarizing the above, the major loss channels for Mars, Venus, and Titan are photochemical and ion escape, whereas it is polar outflow for the Earth. For details on the predominant loss channels for each planet, we also refer the interested reader to \citet[][this topical collection]{Steinmeyer2025}. In the following subsections, we discuss how atmospheric structure, composition, and the presence or absence of an intrinsic magnetic field influence atmospheric loss mechanisms and their observable features.}

\subsection{Mars: Small, cool, and weakly magnetised planet with CO$_2$ rich atmosphere}\label{sec:observations_mars}
%
Presently, Mars has no global intrinsic magnetic field, and the solar wind interacts directly with the upper atmosphere. As introduced in Sec.\,\ref{sec:intro_non-thermal}, in addition to the thermal (Jeans) escape, there are four major atmospheric escape mechanisms, i.e., photochemical escape, ion pickup, atmospheric sputtering, and cold ionospheric outflows (Fig.\,\ref{fig:non-thermal_fig} (b) and Tab.\,\ref{tab:non-thermal}). In the past decade, there have been significant advances in observations of atmospheric escape from Mars based on Mars Express and Mars Atmosphere and Volatile Evolution (MAVEN) spacecraft observations \citep[e.g.][and references therein]{Jakosky2018Icar..315..146J,Nilsson2023Icar..39314610N}. Below, we summarize how each mechanism is identified by observations.

The mass loss rate by Jeans escape can be estimated from observations of the density and temperature of neutral species at the exobase. %
In-situ observations of the altitude profile of neutral densities have been used to determine the exobase height \citep[e.g.][]{Jakosky2017Sci...355.1408J}.

Concerning photochemical escape, the direct measurement of the suprathermal component of the neutral atmosphere at high altitudes is often challenging, although important for estimating the photochemical escape rate and the exosphere (especially neutral corona). Therefore, exosphere observations have been primarily conducted by using optical instruments, such as UV spectrographs \citep[e.g.][]{Chaffin2018JGRE..123.2192C,Chirakkil2024JGRA..12932342C}. \citet{Lillis2017JGRA..122.3815L} observationally estimated the oxygen photochemical escape rate by combining in-situ density observations of electrons, ions, and neutrals. They found a power law exponent of 2.6 for the EUV dependence of the photochemical escape rate. However, such an exponential dependence was not confirmed by other studies \citep{Zhao2015,Zhao2017,Amerstorfer2017,Dong2018,Scherf2021} for larger EUV fluxes, as two competing mechanisms influence the photochemical escape rates. Whereas an increase in the EUV flux tends to increase the photochemical production of suprathermal O (and C) atoms, the accompanied EUV-induced expansion of the upper atmosphere above the main photochemical production zone leads to more collisions, and hence to larger thermalization rates \citep{Amerstorfer2017,Lichtenegger2022Icar..38215009L}.  Although photochemical losses still increase toward larger EUV fluxes, this effect reduces photochemical escape rates compared to the exponential assumption by orders of magnitude for EUV fluxes $\gtrsim$10 times the Sun's present-day flux \citep[see also Fig.~4a in][]{Lichtenegger2022Icar..38215009L}.
\begin{figure}
    \centering
    \includegraphics[width=0.6\linewidth]{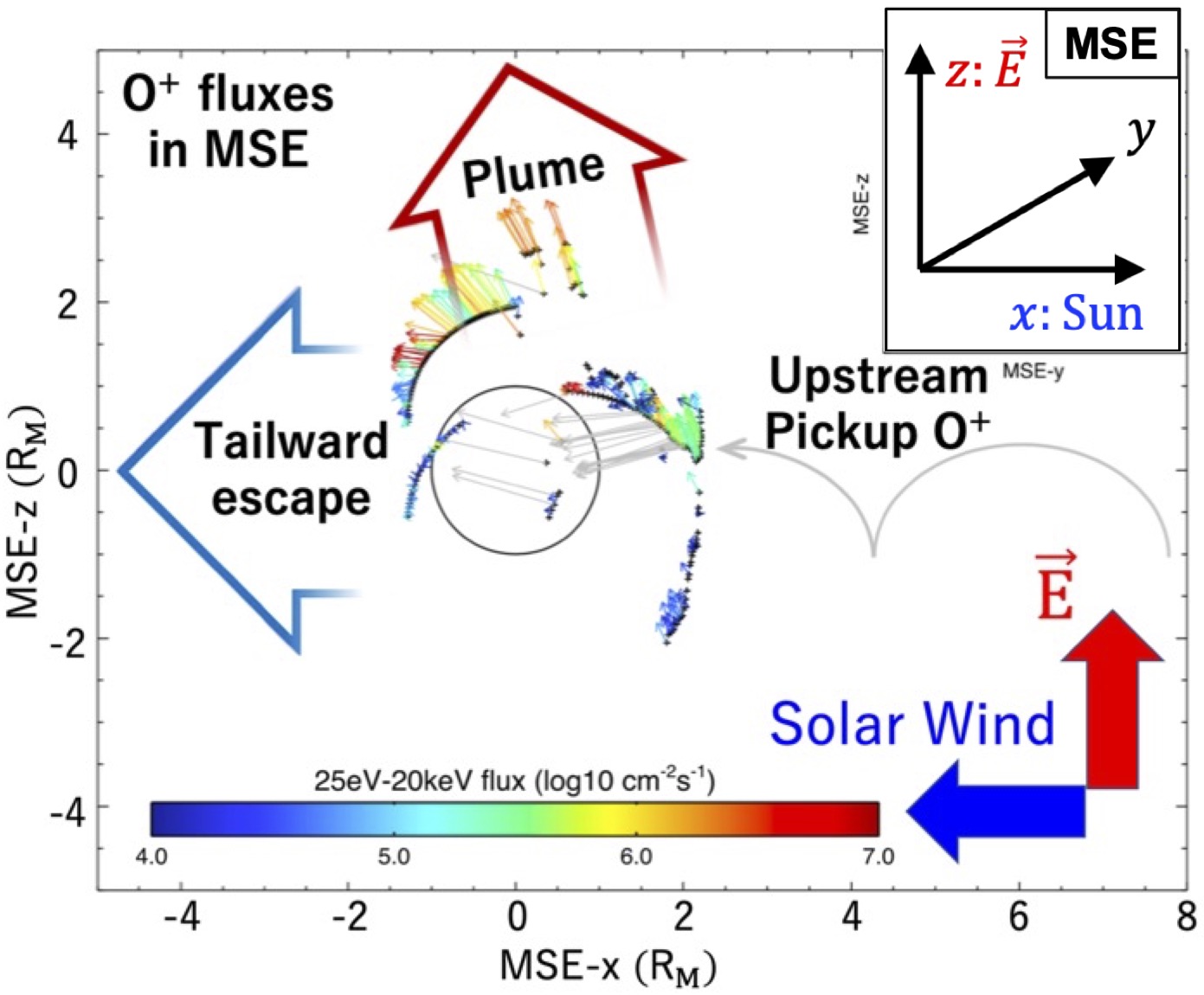}
    \caption{Two major ion escape channels from Mars observed by MAVEN, i.e., polar plume (red empty arrow) and tailward escape (blue empty arrow), in MSE (Mars-centered Solar Electric) coordinates \citep[modified from][]{Dong2015GeoRL..42.8942D}.}
    \label{fig:3.1.1}
\end{figure}

\begin{figure}
    \centering
    \includegraphics[width=\linewidth]{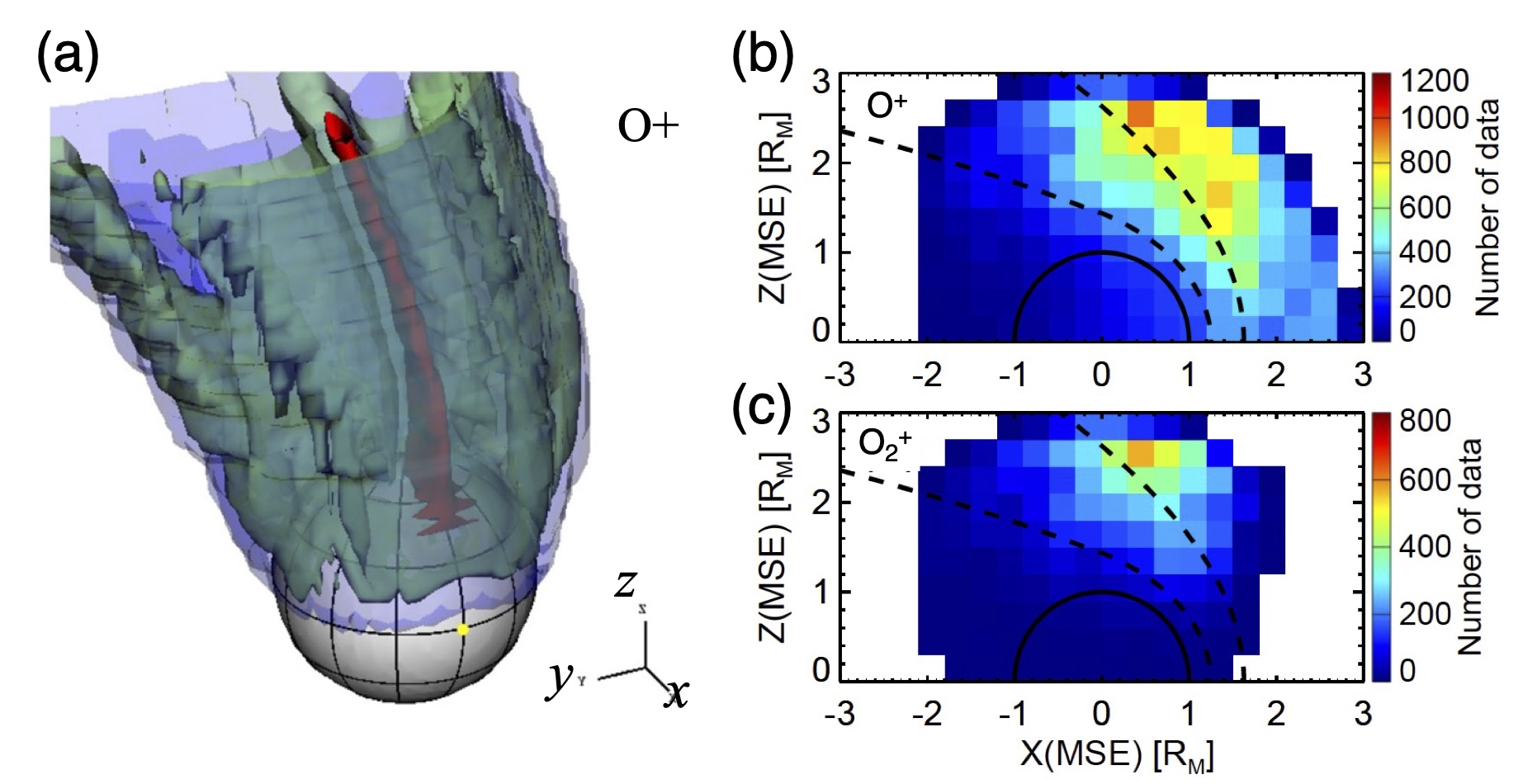}
    \caption{(a) Spatial distribution of the O$^+$ flux of polar plumes originating from the ionosphere. (b) Observed spatial distributions of O$^+$ and (c) O$_2^+$ polar plumes in the MSE coordinates \citep[modified from ][]{Sakakura2022JGRA..12729750S}.}
    \label{fig:3.1.2}
\end{figure}

Significant part of the escape from Mars occurs through ion pickup. Observationally, there are two major escape channels of ions, i.e., polar plumes and tailward flows. The polar plume is a permanent plasma structure in the Mars-centered Solar Electric (MSE) coordinates, with the x-axis pointing to the Sun, the z-axis in the direction of the upstream solar wind convection electric field, and the y-axis completing a right-handed system \citep[][]{Dong2015GeoRL..42.8942D}. As shown with the red empty arrow in Fig.\,\ref{fig:3.1.1} the polar plume is the planetary ion flow with a significant velocity component along the solar wind electric field direction. The O$^+$ plume includes a significant contribution from the ion pickup from the oxygen corona. Since the solar wind electric field can penetrate the ionosphere, acceleration by the electric field can occur down to the penetration altitude. The polar plume also includes the bulk outflow of ionospheric ions toward the electric field direction as shown in Fig.\,\ref{fig:3.1.2} (a). It should be noted that polar plumes of molecular ions such as O$_2^+$ and CO$_2^+$ have also been observed and their spatial distributions are more confined than that of O$^+$ ions, since they originate only from the ionosphere (no source in the corona) as shown in Fig.\,\ref{fig:3.1.2} (b) and Fig.\,\ref{fig:3.1.2} (c) \citep[][]{Sakakura2022JGRA..12729750S}. 

Since the spatial distribution of the pickup ions strongly depends on the location in the MSE coordinates, the ambiguity in the MSE coordinate determination and limited spatial coverage make it difficult to derive an accurate total escape rate via ion pickup. One of the solutions is to estimate the rate from the source corona observations. The total escape rate due to ion pickup originating from the corona has also been estimated from a corona retrieval based on observations of the ring-shaped {distributions} formed by pickup ions \citep[e.g.][and references therein]{Masunaga2024PSJ.....5..180M}.

Escape by atmospheric sputtering has been estimated based on observations of high-energy ion precipitation onto the atmosphere. Using the observed energy distributions of the precipitating ions, the exospheric structure induced by the precipitation needs to be modelled to estimate the expected escape rate \citep[e.g.][ and references therein]{Leblanc2018GeoRL..45.4685L}. For present-day Mars, the contribution of atmospheric sputtering is negligible for heavy ions. However, it could have been significant in ancient times, when the solar wind flux and XUV radiation were much stronger than the present day values \citep[e.g.][]{Luhmann1991JGR....96.5457L}.

\begin{figure}
    \centering
    \includegraphics[width=0.9\linewidth]{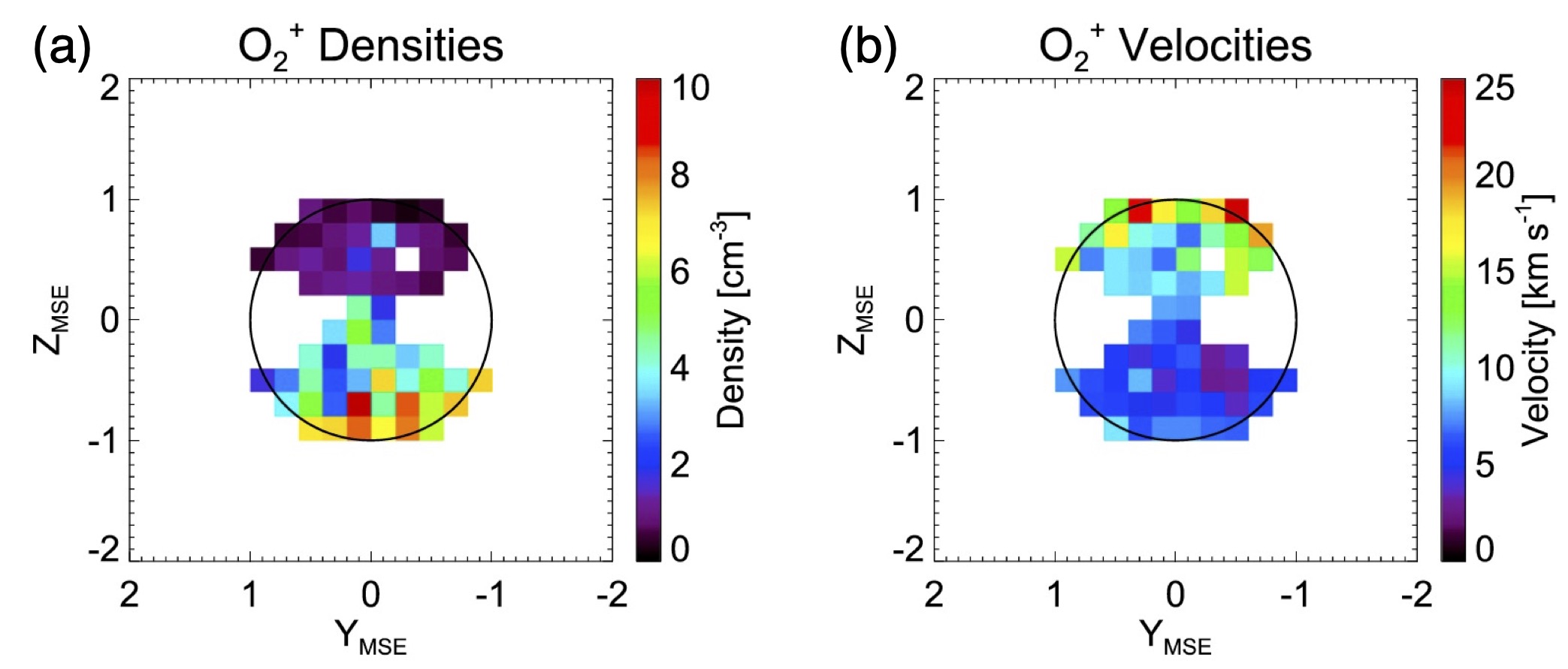}
    \caption{Spatial distribution of O$_2^+$ density (a) and velocity (b) in the Martian induced magnetotail observed by MAVEN. Flux of polar plumes originated from the ionosphere in MSE coordinates \citep[][]{Sakakura2022JGRA..12729750S}. }
    \label{fig:3.1.3}
\end{figure}

\begin{figure}
    \centering
    \includegraphics[width=0.5\linewidth]{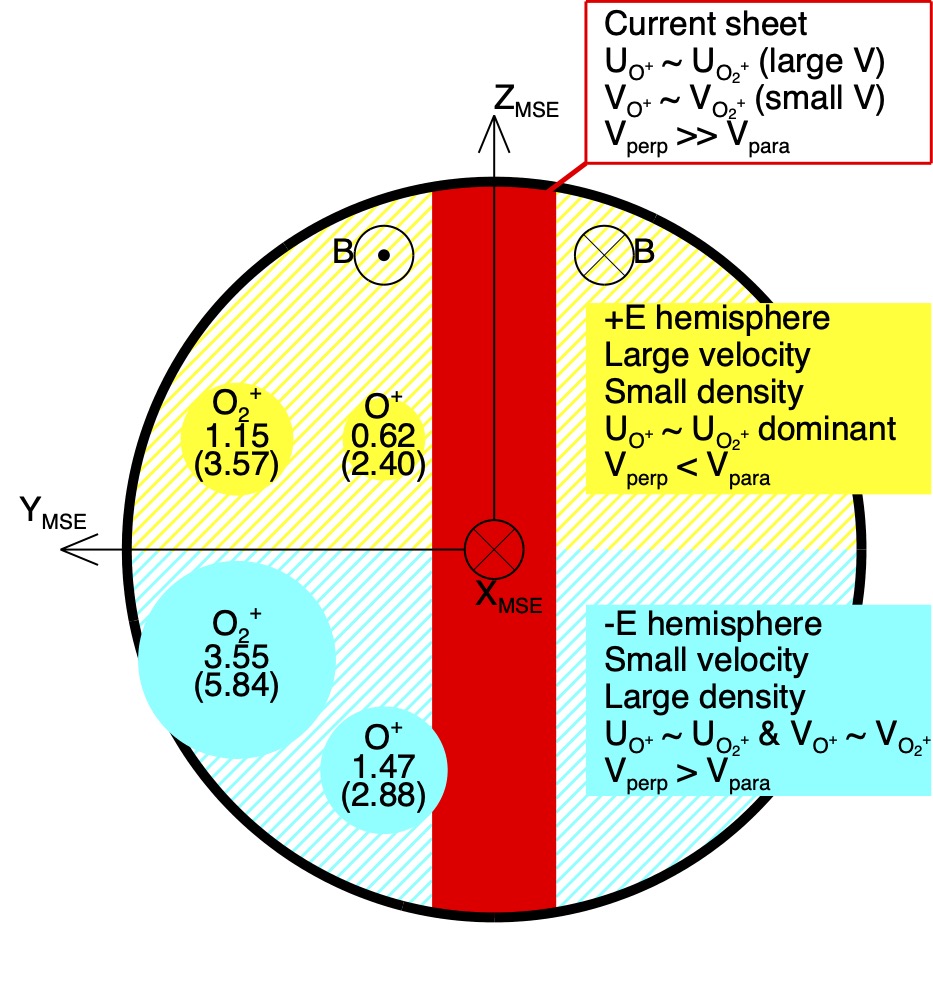}
    \caption{Summary of plasma characteristics observed in the magnetotail, which are important to identify relevant plasma processes to the ion acceleration and heating adopted from \citet{Inui2019JGRA..124.5482I}.}
    \label{fig:3.1.4}
\end{figure}

The cold ionospheric outflow is also an important contributor to the ion escape.
The Martian induced magnetotail is filled with planetary cold ions often flowing in the anti-sunward direction \citep[e.g.][]{Nilsson2012JGRA..11711201N,Inui2019JGRA..124.5482I}. This tailward escape, shown with a blue empty arrow in Fig.\,\ref{fig:3.1.1}, observationally includes the most of ionospheric outflows. As introduced in Sec.\,\ref{sec:intro_non-thermal}, the cold ionospheric outflows can be caused by various plasma processes. In-situ observations of planetary ions are essential to identify the specific acceleration and heating processes causing their escape. Observations have revealed systematic structures in the ionospheric outflows in the magnetotail: evolution of heavy ion distribution functions with distance from Mars show the gradual continuous acceleration of the ions flowing tailward in the current sheet \citep[][]{Nilsson2012JGRA..11711201N}. As shown in Fig.\,\ref{fig:3.1.3}, the velocity (density) is higher in the +E (-E) hemisphere than the -E (+E) in the MSE coordinates. As summarized in Fig.\,\ref{fig:3.1.4}, observations suggest that major acceleration/heating mechanisms causing the tailward ion flows differ depending on the location of the induced magnetotail. Detailed classification can be found in the literature \citep[e.g.][ and references therein]{Inui2019JGRA..124.5482I}. 
In addition to the identification of the escape mechanisms, estimation of the total escape rate based on observations has been intensely studied \citep[e.g.][]{Jakosky2018Icar..315..146J,Ramstad2021SSRv..217...36R}. 

\subsection{Venus: Hot Earth-size planet with induced magnetosphere and CO$_2$ rich atmosphere}\label{sec:observations_venus}
Currently, Venus has a thick CO$_2$-dominated atmosphere and lacks a strong intrinsic magnetic dipole field. {Therefore, the interaction with the solar wind creates an induced magnetosphere, similar to that of Mars as described above.} The interaction with the solar wind is the main cause of atmospheric escape from Venus today \citep[e.g.,][]{Lammer2006,gillmann2022}, along with the photochemical escape of hydrogen (with escape rates $>10^{25}$\,s$^{-1}$ \citep{Lammer2006}). Starting at present-day Venus, we can first elaborate on the presence of the atmospheric escape channels described in the previous sections. A full review of the escape processes at Venus and its potential consequences for the Venusian atmospheric evolution is provided by \cite[][]{gillmann2022}.

The thermal escape at Venus today is minuscule, as a consequence of two combined factors. Firstly, the temperature of its exosphere is only around 270 K \citep[][]{limaye2017}, cooled by the infrared emissions from CO$_2$, which stands in stark contrast to its extremely hot surface temperature of $\sim$735 K. Secondly, the escape velocity at the exobase is around 10 km s$^{-1}$, which is significantly higher than at Mars. This leads to only a tiny portion of the assumed Maxwellian distribution at the exobase being above the escape velocity that can escape the atmosphere.  Therefore, thermal escape can only account for a very small percentage of the total escape rate today \citep[e.g.,][]{Lammer2006}. Hydrodynamic escape is not currently taking place at Venus but is considered to have been an important factor during the runaway greenhouse phase earlier in Venus' history \citep[][]{ingersoll1969}.

Therefore, non-thermal escape processes are the most important channels for the atmospheric escape at Venus today. {One should note that the processes that dominate at Venus are different from those at Mars.} This may in part be explained by the difference in mass between the two planets, hence, in their escape velocities. %

At Venus, photochemical escape is only important for  lighter species, such as hydrogen, and can be considered negligible for heavier species, such as O and O$_2$  \citep[][]{Lammer2006}. Sputtering, similarly to Mars, has never been directly measured at Venus and can only be estimated through modelling and indirect measurements of related properties. Estimates of the sputtering rate at Venus contribute less than 30\% to total atmospheric escape \citep[][]{Luhmann1991JGR....96.5457L,Lammer2006}. Consequently, except for H, the main escape occurs in the form of ion escape caused by the direct interaction between the solar wind and the Venusian ionosphere. Here several processes are important, as elaborated upon in Sec.\,\ref{sec:intro_non-thermal_sw_unmag}. 

Similarly to Mars, an ion pickup plume forms in the direction of the solar wind motional electric field, which carries with it an estimated 30\% of the total escaping ions from Venus \citep{masunaga2019}. Due to the larger mass of Venus compared to Mars, and the higher interplanetary magnetic field strength around the orbit of Venus, the ion pickup plume is less spectacular in its shape than Mars’. The smaller ion gyroradius in relation to the planetary mass causes the pickup plume to form closer to Venus than Mars, where it follows along the stream of the shocked solar wind plasma as it is diverted around the induced magnetosphere \citep[c.f. Fig. 6 \& 9 in][]{jarvinen2016}. 

Besides the loss of suprathermal H, the largest channel of ion escape today occurs in the induced magnetotail of Venus, where a plethora of escape mechanisms contribute to the total escape, ranging from ion cloud formations \citep[e.g.][]{brace1982}, plasma sheet acceleration \citep[e.g.][]{barabash2007}, electric field acceleration \citep{collinson2016}, etc. (see Sec.\, \ref{sec:intro_non-thermal_sw_unmag}). The estimated total ion escape rates from Venus today lie in the range of (3-6)$\times$10$^{24}$ s$^{-1}$ \citep{futaana2017}. However, it has been shown to vary depending on solar cycle and upstream conditions \citep[e.g.][]{persson2018, persson2020, masunaga2019}, including extreme conditions such as Interplanetary Coronal Mass Ejections (ICMEs) and Corotation Interaction Regions (CIRs) \citep[e.g.][]{luhmann2007, edberg2011, mcenulty2010}.

As indicated above, the presence of these different escape channels at Venus is dependent on the current conditions at Venus. Other work \citep[e.g.][]{Way2020Venus} has shown that Venus' atmospheric composition and density could have been drastically different at various times in its past. If we {neglect} the post-accretion magma ocean atmosphere, and assume a path to temperate conditions like that of early Earth, the early Venusian atmosphere was likely N$_2$--CO$_2$ dominated, {as it is also suggested for} the late-Hadean/early-Archean Earth \citep[e.g.][]{Charnay2017}. In that case, the atmosphere {had a substantially lower} surface pressure{, presumably around a few 100 mbar to several bar, compared} to the current 92 bar surface pressure, with CO$_2$ partial pressures (PPs) ranging from roughly 100\% (like today) to less than 20\% in an otherwise N$_2$ dominated atmosphere. In such scenarios, smaller CO$_2$ PPs alongside higher N$_2$ PPs would influence exospheric composition, temperature, and height with corresponding consequences for escape (see Sec.\,\ref{sec:composition_Known-Atm-Types}).

Note that a lower CO$_2$ PP in the Venusian atmosphere would decrease the radiative cooling, which consequently increases the exospheric temperature and causes an expansion of the upper atmosphere. How much the exospheric temperatures, and subsequent exospheric expansion are affected depends on the interplay between the chemical heating/cooling and the radiative cooling in the atmosphere \citep[for more details see][]{Gronoff2020JGRA..12527639G}.  \cite{johnstone2018} showed that increasing the CO$_2$ mixing ratio of Earth's atmosphere decreases the exobase altitude and temperature from today's value. They found that a mixing ratio of around 30\%, which is approximately a factor of three lower than the PP in Venus' today atmosphere today, causes a decrease of the exobase temperature to around 300 K and an altitude of less than 200 km, which is not far from the current exobase at Venus of around 270 K and 180 km. However, when decreased to less than 10\% \cite{johnstone2018} showed that both the temperature and the altitude of the exobase increased by a factor of two from a Venusian-like mixing ratio atmosphere. %

In such scenarios, we may expect several things to take effect. An increase in exospheric temperatures would lead to an increase in {Jeans escape} \citep[][]{johnstone2018}, which in combination with a higher XUV flux from the Sun at about 4.4-4.5 Gyrs ago \citep[][]{Tuetal15} could result in hydrodynamic escape that would cause a significant portion of the atmosphere to be lost \citep[e.g.][]{Lammer2006,tian2008,Johnstone2021Atmosphere}. An expansion of the atmosphere would also lead to a larger interaction area between Venus and the Sun, affecting the size and structure of the induced magnetosphere. A larger interaction area leads to an increase of energy and momentum transfer to the atmosphere and higher ion escape rates \citep{persson2021, Gronoff2020JGRA..12527639G}. However, the induced magnetosphere boundary does not linearly increase in size with an increased exobase altitude as it is controlled by the pressure balance between the ionospheric thermal pressure and the solar wind dynamic pressure \citep{futaana2017}. One should also account for the different properties of the wind at young ages, it is expected to be denser and hotter {\citep[e.g.][]{Vidotto21}}. An increase in the size of the induced magnetosphere that is smaller in relation to the inflation of the exosphere would lead to an increased portion of the expanded atmosphere ending up outside of the induced magnetosphere boundary, which could also lead to a significant increase in atmospheric loss due to an increase in charge exchange, ion pickup, and sputtering \citep[e.g.][]{Gronoff2020JGRA..12527639G, kulikov2006}. A change in the atmospheric composition may also affect the conductivity of the ionosphere if it is combined with a change of the ionospheric composition and structure, which will have further effects on the structure of the induced magnetosphere \citep{cravens1980}.

Additionally, the configuration of the Interplanetary Magnetic Field (IMF) plays an important role for unmagnetized planets such as Venus and Mars. For example, it was recently shown that an IMF that is aligned with the solar wind flow will cause the induced magnetospheres to degenerate, and will affect the total ion escape from the planet \citep[e.g.][]{fowler2022, zhang2024}. Conditions for the flow-aligned IMF may have been more common during the earlier stages of our solar system when the Sun had a faster rotation rate \citep[e.g.][]{Tuetal15}. This may be more common for exoplanets that lie closer to their host stars than Venus in which the aberrated solar wind is more inclined to be aligned with the nominal Parker Spiral angle of the star’s magnetic field \citep{zhang2024}.%

Another important factor for Venus is the question of an intrinsic magnetic field. It is not known if Venus ever had an intrinsic magnetic field \citep{orourke2018}. If it did it would have changed the interaction with the solar wind significantly and made it more Earth-like (which is further described in Sec.\,\ref{sec:observations_earth} below).

All these scenarios must also take into consideration the increased solar XUV radiation levels in the past \citep[e.g.][]{Tuetal15}. In combination with any atmospheric composition changes, this would lead to additional heating effects on the Venusian upper atmosphere and its atmospheric escape \citep[e.g.][]{Johnstone2021Atmosphere}. Therefore, it is fair to say that it is not straightforward to extrapolate escape rates observed in the present day backwards in time to understand the past Venusian atmosphere and its atmospheric escape.


\subsection{Earth: Moderate HZ planet with intrinsic magnetosphere and N$_2$-O$_2$ dominated atmosphere}\label{sec:observations_earth}
\begin{figure}
    \centering
    \includegraphics[width=0.6\linewidth]{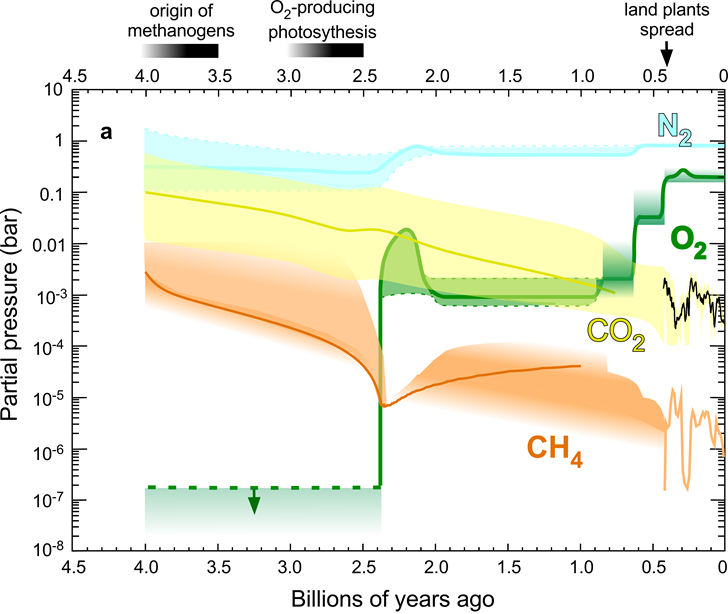}
    \caption{A synopsis of the evolution of the terrestrial atmosphere during the last 4 billion years. From \citet{Catling2020}.}
    \label{fig:2.3.1_evo-earth-atmo}
\end{figure}

The present composition of the terrestrial atmosphere is quite different from that of Mars and Venus, though the initial atmospheric compositions of these three planets are believed to have been similar \citep{Lammer2018,Lammer2021SSRv..217....7L}. The long-term development of the Earth’s atmosphere is shaped by its interactions with the planet’s interior, surface, living organisms (hence, it is highly connected to the development of habitable conditions), incoming material from space such as meteoroids, and the loss of atmospheric particles into space \citep{Yamauchi2007AsBio...7..783Y,Lammer2008SSRv..139..399L,Lammer2020SSRv..216...74L,Avice2020SSRv..216...36A,Gronoff2020JGRA..12527639G}. The present atmospheric composition is thus quite different from that of the primordial Earth \citep{Stueken2020SSRv..216...31S,Catling2020}, compare Fig.\,\ref{fig:2.3.1_evo-earth-atmo}.

Given the value of the gravitational escape velocity from our planet (11.2 km s$^{-1}$), present-day (kinetic) thermal escape by means of neutral atoms or molecules is mostly limited to hydrogen. All other heavier elements, as for example oxygen and nitrogen, have to go through a series of acceleration mechanisms in order to attain escape velocities. These elements, which presently represent over 99\% of the mass of the terrestrial atmosphere, need thus to be first ionised (see Sec.\,\ref{sec:intro_non-thermal}).

Although the magnetosphere of the Earth deflects most of the solar wind flow away from the upper atmosphere operating as a shield, at the same time it increases the cross-sectional area between the solar wind and the Earth by a factor of more than 200. Hence the terrestrial magnetic field increases the amount of solar wind kinetic energy that is intercepted by the magnetosphere and which can then be channelled down to the high-latitude ionosphere, a portion of it ending up in driving ion outflow and acceleration \citep{Li2017JGRA..12210658L,Gunell2018A&A...614L...3G,Maggiolo2022JGRA..12730899M}. At the equatorial latitudes the trapped cold plasma population of the plasmasphere (Fig.\,\ref{fig:2.3.4_plasmasphere}) constitutes another source of plasma transport to the outer magnetosphere, which can then eventually escape to the interplanetary medium \citep{Lemaire2001JASTP..63.1285L,Dandouras2013AnGeo..31.1143D,Borovsky2014JGRA..119.6496B}.

\begin{figure}
    \centering
    \includegraphics[width=0.6\linewidth]{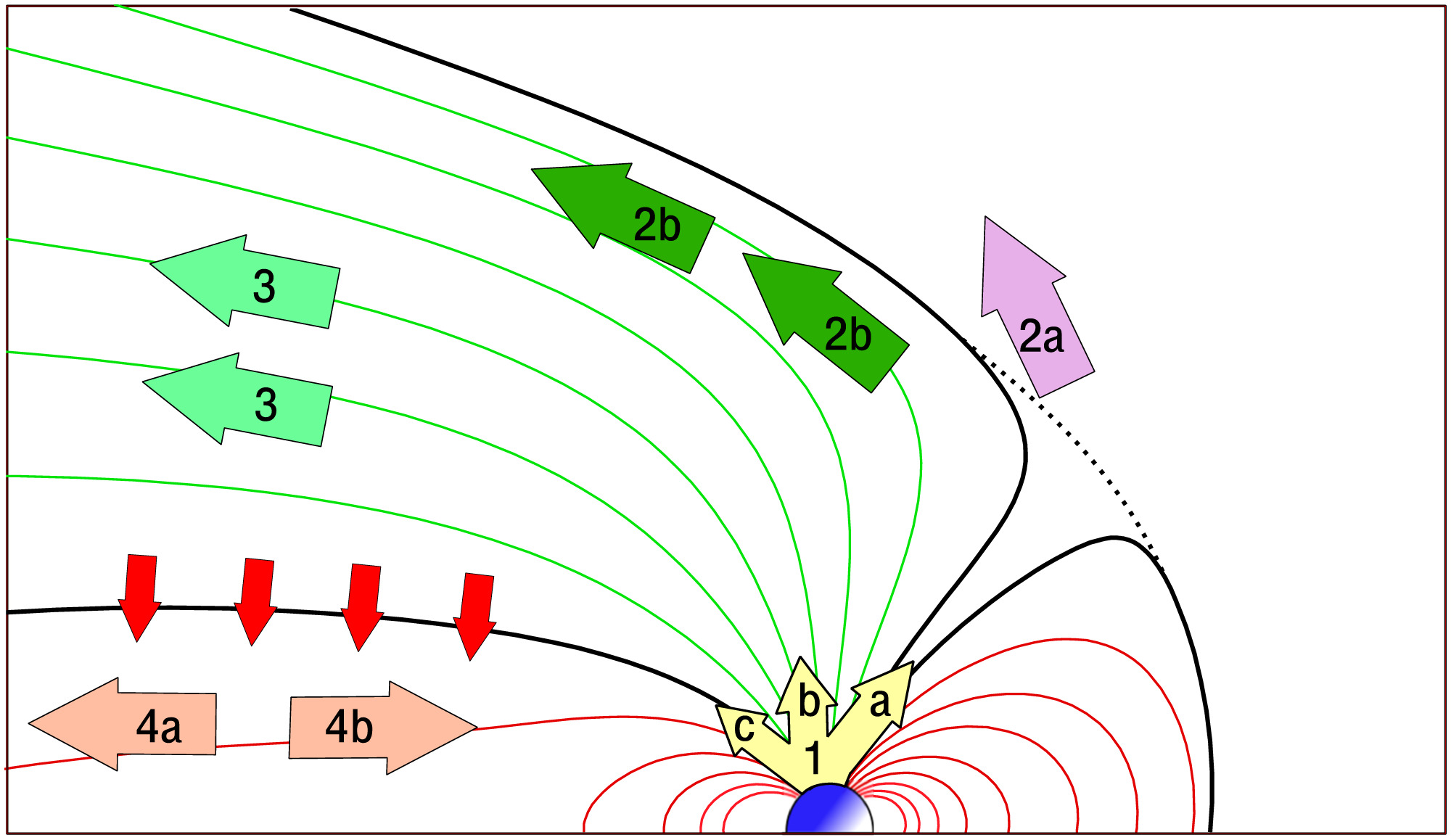}
    \caption{Schematic of the terrestrial magnetosphere (meridian cut, Sun is on the right) showing the three main source regions of outflowing ions from the ionosphere (yellow arrows) and their pathways into the magnetosphere (purple, green, red and beige arrows). 1a, 1b and 1c relate to the cusp, polar wind and nightside auroral zone outflow respectively (see text). The red and green curved lines represent in their turn the closed and open magnetic field lines. Upwelling ions can either directly escape through the cusp (2a), or along the plasma mantle (2b). Polar wind ions get into the magnetotail lobes (3), where magnetospheric convection (red arrows) can bring them into the plasma sheet. There, they can be either injected tailward (4a) or Earthward (4b). From \citet{Slapak2018GeoRL..4512669S}.}
    \label{fig:2.3.2_msphere-ion-flows}
\end{figure}

The ions upwelling from the high-latitude ionosphere can originate either from (see Fig.\,\ref{fig:2.3.2_msphere-ion-flows})
\begin{itemize}
    \item {The polar cap, which is connected to the open magnetic field lines and where ions are extracted from the ionosphere by the ambipolar electric field and then form the polar wind \citep[e.g.][]{Schunk2000JASTP..62..399S}, or}
    \item {The funnel-shaped polar cusp and cleft, where the ions are accelerated upwards following multiple step processes \citep{Andre1990JGR....9520809A}, or}
    \item {The auroral zone, where energy is provided by the energetic electron precipitation originating from the magnetosphere \citep{Yau1997SSRv...80....1Y}.}
\end{itemize}

\begin{figure}
    \centering
    \includegraphics[width=0.8\linewidth]{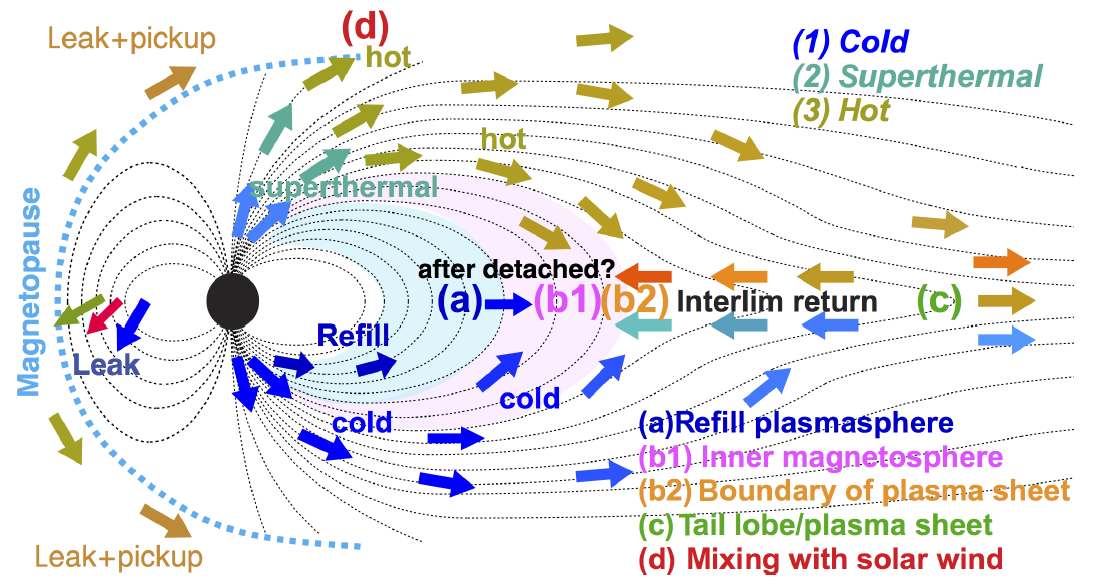}
    \caption{Simplified diagram showing the various transport routes in the magnetospheres for the ions originating from the terrestrial ionosphere. From \citet{Yamauchi2019AnGeo..37.1197Y}.}
    \label{fig:2.3.3_msphere-circulation}
\end{figure}

Depending on the original location of the outflowing ions in the ionosphere, the type of ion species, the magnetospheric convection, and the conditions of interplanetary magnetic field, these ions, as shown schematically in Fig.\,\ref{fig:2.3.3_msphere-circulation}, can have very different trajectories in the magnetosphere \citep{Yamauchi2019AnGeo..37.1197Y}. The ions outflowing from the high-latitude ionosphere generally follow tailward bended trajectories. They can then get out of the magnetosphere through the cusp, or can move along the plasma mantle, or can get transported anti-Earthward along the open field lines of the magnetotail lobes, or from the lobes they can eventually get into the plasma sheet. There, depending on the local magnetic configuration and the state of the magnetospheric convection, they can be re-injected into the inner magnetosphere, or can directly escape tailwards \citep{Ebihara2006JGRA..111.4219E,Haaland2012JGRA..117.7311H,Chappell2015SSRv..192....5C,Yamauchi2019AnGeo..37.1197Y,Dandouras2021JGRA..12629753D}. We note that, even for the ions injected into the inner magnetosphere, a large number will finally escape in the form of energetic neutral atoms, following charge-exchange collisions with the neutral hydrogen of the exosphere.

\begin{figure}
    \centering
    \includegraphics[width=0.6\linewidth]{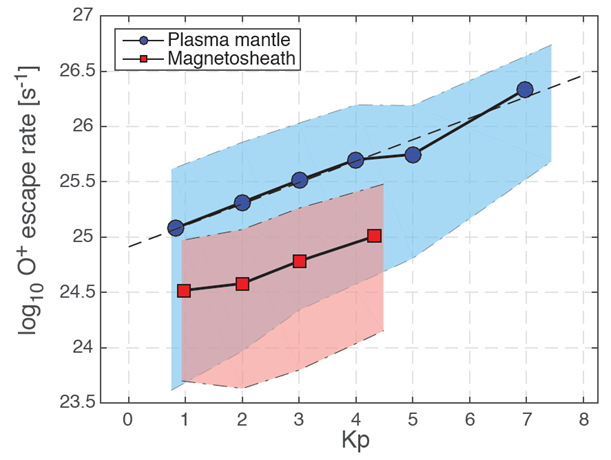}
    \caption{Observed escape rates of ${\rm O^+}$ (ions s$^{-1}$) through the plasma mantle (blue circles) and through the dayside magnetosheath (red squares) as a function of the geomagnetic activity index Kp (which quantifies disturbances in the horizontal component of Earth’s magnetic field with an integer in the range 0–9, with 1 being calm and 5 or more indicating a geomagnetic storm). From \citep[][ and Corrigendum]{Slapak2017AnGeo..35..721S}.}
    \label{fig:2.3.5_Oplus}
\end{figure}

Earth’s ionosphere and magnetosphere have been studied in-situ or by remote sensing techniques by several missions, providing information on the ionospheric source, the transport and acceleration through the magnetosphere, and eventual leakage of these particles to the interplanetary medium \citep[e.g.][]{Chappell2015SSRv..192....5C,Dandouras2020SSRv..216..121D}. Some of the most prominent results are
\begin{itemize}
    \item {Low-energy (few eV) ions coming from the polar caps fill the lobes of the magnetotail almost continuously. This upflow is sensitive to the Solar EUV intensity and can show a twofold growth at the solar cycle maximum \citep{Engwall2009NatGe...2...24E,Andre2015JGRA..120.1072A};}
    \item {The cusps, where energy supplied from the solar wind is concentrated due to the funnel-shaped local magnetic field geometry, are the main source outflowing high-energy ions \citep[mostly ${\rm O^+}$ and ${\rm N^+}$, but also some molecular ions ${\rm N_2^+}$, ${\rm NO^+}$, ${\rm O_2^+}$; see][]{Kistler2010JGRA..115.3209K,Kronberg2014SSRv..184..173K,Yamauchi2024SSRv..220...82Y};}
    \item {The energetic ($\sim$100\,eV to several keV) heavy-ion outflow is closely tied to the solar wind parameters. A growth in the outflow by a factor of up to $\sim$100 has been observed as a function of the dynamic pressure of the solar wind\citep{Schillings2019EP&S...71...70S};}
    \item {The energetic heavy-ion outflow is also highly dependent on the geomagnetic activity, as expressed by the Kp activity index, and follows a logarithmic trend (Fig.\,\ref{fig:2.3.5_Oplus}). At peak activity levels, the escape rate of heavy ions is almost 100 times as high as in normal conditions \citep{Slapak2017AnGeo..35..721S};}
    \item {The total loss of heavy ions during the last four billion years, considering the sensitivity of the outflow on the geomagnetic and solar activity conditions and the highly active Sun in its early stages \citep{guedel2020SSRv..216..143G}, could be roughly equivalent to 40\% of the current atmospheric oxygen content \citep{Slapak2017AnGeo..35..721S,Kislyakova2020JGRA..12527837K};}
    \item {Plasmaspheric outflows, in the form of plasmaspheric plumes (plasma elements detached from the plasmasphere and propagating outwards) and plasmaspheric wind (continuous slow outwards plasma transport), constitute the main ion outflow mechanisms from the equatorial latitudes \citep{Dandouras2013AnGeo..31.1143D,Borovsky2014JGRA..119.6496B}. Similar outflows should also exist around other planets, quickly rotating and having an ionised atmosphere and an intrinsic magnetic field.}
\end{itemize}

However, there are issues concerning the outflow and escape of terrestrial ions that have to be addressed. The most important of them concerns the accurate composition of the escaping ion populations and the way it responds to the external conditions, as the solar and geomagnetic activity conditions \citep{Dandouras2021JGRA..12629753D,Yamauchi2024SSRv..220...82Y}. Most of the ion mass spectrometers flown in the magnetosphere and operating in the few keV energy range do not have enough resolution to separate the nitrogen ions from the oxygen ions \citep{Reme2001AnGeo..19.1303R,Young2016SSRv..199..407Y}. 

\section{Role of Stellar Input}\label{sec:stellar_input}
\noindent
Different physical processes ruled by the host star contribute to the energy input to a planetary atmosphere according to the planetary orbit, the stellar parameters, and the atmosphere's chemical composition. 

\subsection{Stellar magnetic activity and high-energy radiation}
In late-type stars with an internal structure  similar to our Sun, we observe  non-radiatively heated outer stellar atmospheres showing spatial and temporal variability owing to the complex geometry and evolution of stellar magnetic fields \citep[e.g.,][]{Schrijveretal24}. 
The X-ray and EUV (XUV) radiation comes from the corona, with temperatures $T$ of the order of $10^{6}$~K, and from the thin transition region between the chromosphere ($T \sim 10^{4}$~K) and the corona itself. The coronal plasma is mainly responsible for the emission below $\sim 45$~nm, dominated by He~II, Mg~X, and highly ionized Fe lines in the Sun. In contrast, at longer wavelengths, the spectrum is dominated by the He~I continuum and the hydrogen Lyman continuum up to the hydrogen ionization threshold at 91.2\,nm. Beyond that limit, the transition region lines emitted by O~III, O~VI, Si~IV, N~V are the most prominent features over a very low continuum, up to the 121.5 nm of the strong Ly-$\alpha$ line  \citep{Linsky19,LinskyRedfield24}. Lower energy chromospheric lines, for example, C~II at 133.4~nm \citep{Franceetal18}, are usually not as important as the XUV spectrum.

The part of the XUV spectrum  most relevant for atmospheric escape depends mainly on the planetary atmosphere's chemical composition. In  a primordial atmosphere dominated by hydrogen and helium, the radiation below 91.2~nm is responsible for photoionisation heating of H and He (see Sec.\,\ref{sec:intro_thermal_classification}).  On the other hand, in the case of secondary atmospheres containing molecules such as CO$_{2}$, H$_{2}$O, N$_{2}$, CO, and O$_{2}$, photons with wavelengths longer than $\sim 80$~nm are the most relevant for their photodissociation. The strong chromospheric Ly-$\alpha$ line is especially important because it photodissociates water and methane \citep[cf. Fig.\,10 of][]{LinskyRedfield24}. 

The XUV spectrum of late-type stars depends on their magnetic fields, produced by the stellar hydromagnetic dynamo. X-ray emission is well correlated with the Rossby number $R{\rm o}$, that is, the ratio of the stellar rotation period $P_{\rm rot}$ to the convective turnover time $\tau_{\rm c}$. This characterizes internal stellar convection and is estimated by means of stellar models \citep[e.g.][]{Spadaetal13}, $R{\rm o} \equiv P_{\rm rot}/\tau_{\rm c}$. 
The best correlation {between X-ray emission and stellar rotation} is obtained by considering the ratio of the stellar X-ray luminosity to its bolometric luminosity, $R_{\rm X} \equiv L_{\rm X}/L_{\rm bol}$. This can be expressed as a broken power law with the exponent decreasing below a critical value of $Ro^{\rm sat}$, that indicates saturation of the stellar coronal emission in the most rapid rotators \citep[e.g.][]{wright2011ApJ...743...48W,Johnstoneetal21} 

\begin{equation}
    \bf{R_{\rm X} =} \begin{cases}
        & \bf{R_{\rm X}^{\rm sat},}${ if }$ \bf{R{\rm o}\leq R{\rm o}^{\rm sat},} \\
        & \bf{CR{\rm o}^{\beta},}${ if }$ \bf{R{\rm o} > R{\rm o}^{\rm sat},}
    \end{cases}
\end{equation}
{where parameters C, $\beta$, $R{\rm o}^{\rm sat}$, and $R_{\rm X}^{\rm sat}$ can be defined empirically.}

The greatest advantage of the $R_{\rm X}-R{\rm o}$ empirical correlation is that it is valid for stars of all spectral types between F and M,  independently of the stellar intrinsic luminosity,  due to the $L_{\rm bol}$ normalization {\citep[e.g.][]{wright2011ApJ...743...48W,Jackson2012MNRAS.422.2024J}}. Similarly, the Rossby number threshold for $R_{\rm X}$ saturation is independent of the spectral type and the rotation period. Main-sequence stars with a mass smaller than $\sim 0.35$~M$_{\odot}$ are  considered fully convective. 
The dependence of $R_{\rm X}$ on the Rossby number is generally regarded as similar to that established for more massive stars \citep[cf. Sect.~3.1 of][]{Johnstoneetal21}, although some very-low-mass stars, such as Trappist-1 ($M \sim 0.09$~M$_{\odot}$), show a significantly lower XUV luminosity than predicted by that relationship \citep{Wheatley2017MNRAS.465L..74W}. However, the present-day XUV surface flux ($F_{\rm XUV}$) in the habitable zone (HZ) of Trappist-1 is still roughly two orders of magnitude higher than the present-day flux at Earth \citep[e.g.,][]{Wheatley2017MNRAS.465L..74W,Birky2021}. 

The intrinsic variability of the X-ray emission on timescales from minutes to decades produces a remarkable dispersion around the mean $R_{\rm X}-R{\rm o}$ correlation. Nevertheless, it can still be used to estimate the total flux received by a planet over evolutionary timescales \citep{Johnstoneetal21}, as they are much longer than those characteristic of stellar variability. 
The estimate of the total flux received in the EUV bandpass is uncertain within at least a factor of two \citep{Franceetal18} due to strong absorption by the interstellar medium between the ionization threshold of hydrogen at 91.2 nm and $\sim 35$~nm. Therefore, all parameterizations of the stellar EUV flux as a function of the Rossby number or of the stellar age {\citep[e.g.,][]{Sanz-Forcadaetal11,Johnstoneetal21,King2018MNRAS.478.1193K}} are based on indirect methods  \citep{Franceetal18,Linsky19,LinskyRedfield24}. Similarly, the Ly-$\alpha$ flux is strongly absorbed by interstellar hydrogen, thus empirical correlations are based on a reconstruction of the line profile starting from its extended wings. These limitations must be considered when simulating the evolution of planetary atmospheres.

A cumulative effect by stellar flares is included in the $R_{\rm X}-R{\rm o}$ correlation because they are a kind of short-term variability \citep{Johnstoneetal21}. However, the temporary intensification of the short-wavelength continuum and spectral line radiation by up to two orders of magnitude can be relevant for planetary atmospheric escape because it increases the exobase radius remarkably, thus enhancing  atmospheric mass loss rates \citep{doAramaletal22}. Moreover, flares are often accompanied by slower propagating CMEs (though the latter are not necessarily directed in a way that allows interaction with the planet) which can interact with already ``excited'' atmospheres \citep[e.g.][]{Hazra2025MNRAS.536.1089H}.%

\subsection{Co-evolution of stellar rotation and high-energy radiation}\label{sec:stellar_input:rotation}
Late-type stars are born with a wide range of rotation rates: $\Omega$ between $\sim 1$ and $\sim 50$~$\Omega_{\odot}$. Their rotation stays constant during the pre-main-sequence phase, as long as they are dynamically locked to their circumstellar accretion discs. This phase is short -- the mean disc lifetime is of the order of a few Myr  \citep[e.g.][]{Mamajek09} -- and is followed by a phase of contraction, while the star moves towards the zero-age main sequence (ZAMS). This leads to a rapid acceleration of the rotation owing to the reduction of the moment of inertia, making $\Omega$ maximal on the ZAMS. Afterwards, the effect of the stellar magnetized wind leads to a steady decrease of the stellar angular velocity, until all the stars of spectral types F, G, and early K converge towards the same evolutionary sequence where $\Omega$ is roughly proportional to $t^{-1/2}$. Where $t$ is the age of the star, a dependence called the Skumanich law \citep[after][]{Skumanich72}. Such a convergence of the rotational evolution tracks takes about 0.6~Gyr for F and G stars, and longer for later spectral types. It takes Gyrs for M dwarfs, which exhibit a wide range of rotation periods even in clusters of 2.5 Gyr of age or older. 

\begin{figure}
    \centering
    \includegraphics[width=0.6\linewidth]{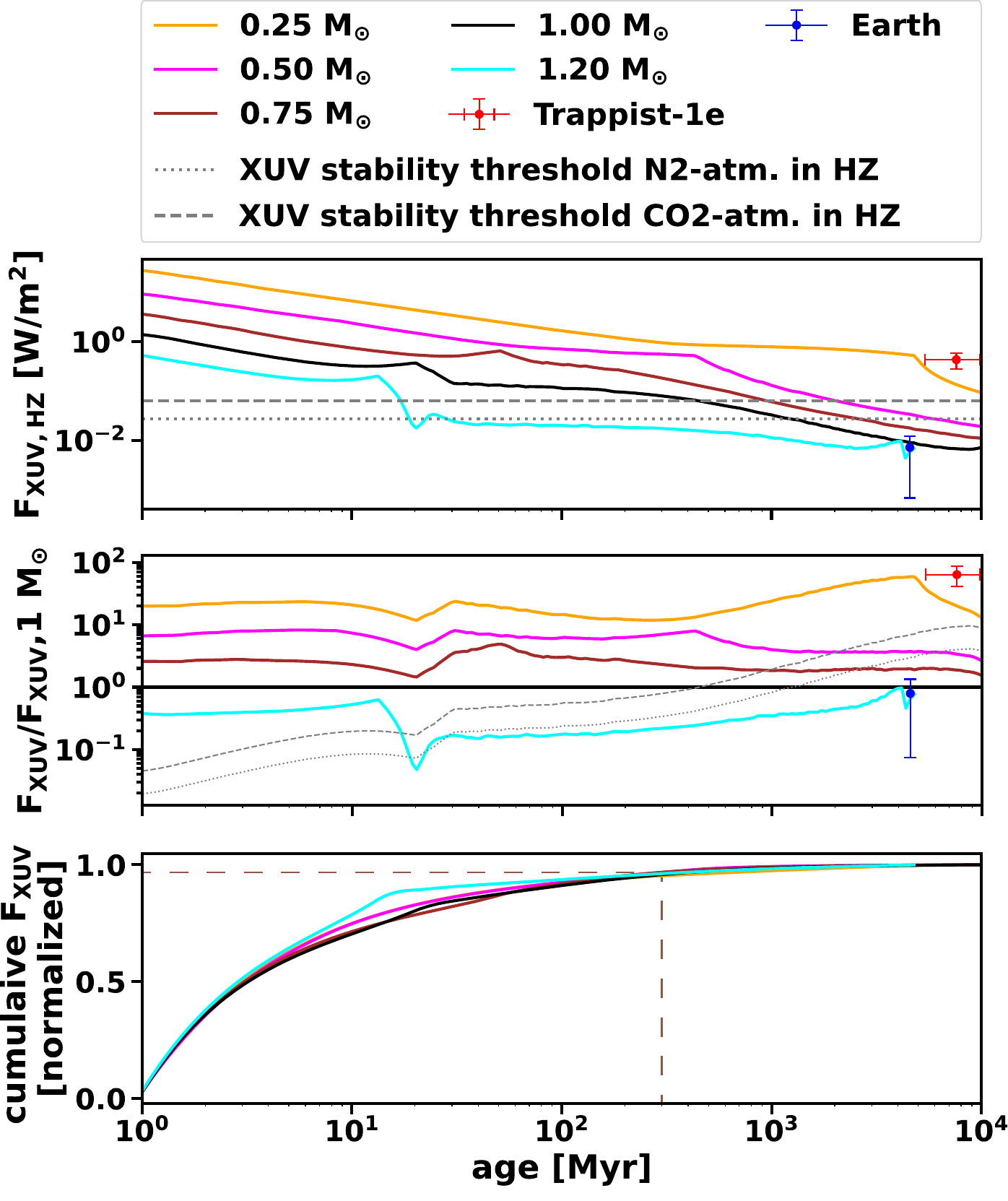}
    \caption{The $F_{\rm XUV}$ evolution of 0.25-1.2\,$M_{\odot}$ stars according to the model of \citet{Johnstoneetal21} (solid lines). The upper panel shows the $F_{\rm XUV}$ evolution of moderate rotators of the given masses in the middle of their respective HZ. The {middle} panel shows the same, but normalized to a moderately rotating Sun-like star (the black line in both panels). {The lower panel shows the cumulative flux of moderate rotators normalized to 1 and highlights an age of 300\,Myr (dashed lines) at which $>$95\% of the cumulative flux is already emitted for all evolutionary tracks}. The red dot with error bars shows $F_{\rm XUV}$ at Trappist-1\,e \citep[from][]{Wheatley2017MNRAS.465L..74W} (whose host star Trappist-1 has a mass of $\sim$0.09$M_{\odot}$) and the blue dot shows $F_{\rm XUV}$ for the Earth \citep[based on][]{Tuetal15}. The dashed and dotted grey lines illustrate the thermal stability threshold of secondary CO$_2$-dominated (99\% CO$_2$, 1\% N$_2$) and N$_2$-dominated (90\% N$_2$, 10\% CO$_2$) atmospheres in the HZ according to \citet{VanLooveren2024Trappist}; see also \citet{Scherf2024AsBio..24..916S}.}
    \label{fig:XUVEvo}
\end{figure}

Since the convective turnover time $\tau_{\rm c}$ stays {almost} constant during the main-sequence evolution, the XUV fluxes of those stars steadily decrease with age and the decrease speeds up after a star leaves the saturation regime. {It turns out that, even though most of the cumulative XUV flux received by their planets is likely to be contributed after the saturated phase during the first 0.3-0.5 Gyr on the main sequence, as illustrated in the bottom panel of Fig.\,\ref{fig:XUVEvo} \citep[see][for details]{Johnstoneetal21}, the XUV flux becomes the dominant contributor to atmospheric evaporation starting from a few tens of Myr of evolution and then remains dominant on Gyr timescales \citep[][]{king2021MNRAS.501L..28K}.} Stars having short-lived discs on the pre-main sequence can reach the ZAMS with a rotation rate exceeding 50~$\Omega_{\odot}$ and are characterized by a much higher level of XUV radiation than slower ZAMS rotators. {As a consequence, their cumulative XUV radiation is higher than for slower rotators, especially during the first Gyr of evolution \citep[cf.][ Fig. 18]{Johnstoneetal21}. This may lead to the complete evaporation of the primordial atmospheres of close-in Neptune-sized planets \citep{Tuetal15}.}

There is an indication that the angular momentum loss rate decreases below the value predicted by the Skumanich law in solar-like stars older than the Sun \citep{vanSadersetal16,Halletal21,Davidetal22}. This should keep their XUV at higher values than predicted based on the Skumanich law. Other effects, such as the stalled rotational evolution observed in stars less massive than the Sun between 0.6 and 1.5-2~Gyr, \citep[e.g.,][]{Curtisetal20}, may also affect the XUV cumulative flux received by their planets, although not significantly compared to the XUV fluence received within the first 0.3-0.5~Gyr of their evolution. Second-order effects can be related to the stellar metallicity affecting the rotational evolution \citep{Amardetal20} and the coronal emission of the stars \citep{Poppenhaegeretal22}.

Although the XUV flux decreases at least by an order of magnitude during the first few 100 Myr for most M-F stars \citep[e.g.][]{Johnstoneetal21}, the $F_{\rm XUV}$ received by a planet can remain high enough to rapidly erode an atmosphere over longer timescales. As the upper panel of Fig.~\ref{fig:XUVEvo} illustrates for various stellar masses, the $F_{\rm XUV}$ at the middle of the HZ of lower-mass stars takes significantly longer to drop below a certain threshold than the $F_{\rm XUV}$ of higher-mass stars. This figure shows the XUV surface flux evolution of so-called moderate rotators (i.e., stars belonging to the 50th percentile of the initial $P_{\rm rot}$-distribution of their stellar mass) calculated with the stellar evolution model Mors \citet{Johnstoneetal21} and the stellar isochrones from  \citet{Spadaetal13}, respectively. %
The dashed and dotted grey lines in Fig.~\ref{fig:XUVEvo} show the thermal stability thresholds for CO$_2$-dominated (99\% CO$_2$, 1\% N$_2$) and N$_2$-dominated (90\% N$_2$, 10\% CO$_2$) atmospheres hosted by a 1\,M$_{\oplus}$-planet \citep{VanLooveren2024Trappist}. As long as an evolving star lies above one of these thresholds, the respective atmosphere is not thermally stable in its HZ due to XUV heating of the upper atmosphere (Sec.~\ref{sec:intro_thermal}). Only stars below these thresholds can hence host temperate planets with these certain types of secondary atmospheres (see Sec.~\ref{sec:composition}). The HZ planet Trappist-1\,e, for instance, lies above both thresholds \citep[values from][]{Wheatley2017MNRAS.465L..74W} although its host star has a mass of $\sim$0.09$M_{\odot}$ and is $\sim$8\,Gyr old. It indicates that Trappist-1 may never reach an $F_{\rm XUV}$ in its HZ that allows for the retention of a secondary atmosphere (as CO$_2$-atmosphere is expected to be the most thermally stable among them).

In contrast, another HZ planet included in Fig.~\ref{fig:XUVEvo} -- the Earth{ --} is clearly below both thresholds. The Sun is a 4.56\,Gyr old quiescent G star that appears anomalously weakly active compared to other solar-type stars \citep{Reinhold2020}. Its XUV is below that of the average G-type star of similar age \citep{Johnstoneetal21}, as well as super-flaring rates. Whereas the average Sun-like star produces super-flares with energies $<10^{34}$\,erg roughly once per century \citep{Vasilyev2024}{, no} such energetic super-flares have been observed at the Sun -- with the largest flare observed to date being the Carrington event with estimated energies of $~\sim 5\times10^{32}$\,erg \citep[e.g.,][]{Hayakawa2023}. Assuming the Sun can produce super-flares, the occurrenc rate is estimated to be $\sim 10^{-3}$ per year \citep{Shibata2013}, about an order of magnitude lower than the average for solar type stars. If the Sun is indeed anomalously weakly active, this has far-reaching implications for habitability as secondary atmospheres are highly susceptible to atmospheric erosion around active stars \citep{Scherf2024AsBio..24..916S}.

Finally, the bottom panel of Fig.~\ref{fig:XUVEvo} shows the same data as the top panel but normalized to the XUV surface flux evolution of a moderately rotating solar-like star. This plot is therefore valid for any orbital separation, for which the respective stars induce the same total incident stellar surface flux, $S_{\rm eff}$. Fig.~\ref{fig:XUVEvo} clearly shows that stellar mass and age are crucial parameters for the retention and loss of planetary atmospheres. See also \citet{Scherf2024AsBio..24..916S} for an in-depth discussion on this issue.

\subsection{Stellar magnetized winds}
Stellar winds interact with planetary atmospheres, especially those not shielded by a magnetosphere (see sec.\,\ref{sec:intro_non-thermal}). Unfortunately, we only have detailed information for the solar system and mostly for the present-day Sun close to the plane of the ecliptic. However, a stellar wind is a complex 3D MHD flow  that depends on the level of activity of the star and is perturbed by transients produced by flares and CMEs. 

Stellar wind mass loss rates (a proxy of the wind's speed and density) have been measured only for about a dozen stars, based on different indirect techniques. The dependence of the wind mass loss flux on the coronal X-ray flux has been investigated by \citet{Vidotto21,Vidotto23} providing information to study the  wind impact on the evolution of planetary atmospheres. 
Even less understood is the impact of CMEs associated with major stellar flares on exoplanetary atmospheres because of the difficulties in observing them \citep[cf. Sect. 1 of][]{Xuetal24}. An extrapolation based on  solar CMEs  to more active stars may not be appropriate because the mechanism producing and accelerating CMEs can be less efficient in more active stars than in the Sun \citep{Alvarado-Gomezetal18}. %
{However, some works attempt to estimate the CME frequencies at stars of different types through measurable parameters, such as temperatures and stellar spot occurrence \citep[e.g.][]{herbst2021ApJ...907...89H}.} 

{Though the impact of CMEs is well studied for solar system planets \citep[e.g.][]{luhmann2007,Edberg2010GeoRL..37.3107E,edberg2011}, no direct observations are available for exoplanets. Theoretical modeling suggests that CMEs can lead to some enhancement of the atmospheric escape \citep[e.g.][]{Lammer2007AsBio...7..185L,Lynch2019ApJ...880...97L,Alvarado-Gomez2022ApJ...928..147A,Hazra2025MNRAS.536.1089H} but also to an enhancement of prebiotic atmospheric chemistry \citep[e.g.][]{Kay2019ApJ...886L..37K,Airapetian2016NatGe...9..452A}.}

\subsection{Tides in planets and stars}\label{sec:stellar_input:tides}
Tides have been reviewed by \citet{Ogilvie14}, while a number of other works have been published in the context of star-planet interactions, e.g., \citet{VanLaerhoven2014,Blackledge2020,Lanza22}. Tidal dissipation inside a planet produces heating in its interior, tends to decrease the eccentricity of its orbit, and synchronizes its rotation with the orbital motion, except when the planet has a permanent quadrupolar deformation as in the case of Mercury, in which case its rotation can be captured in a spin-orbit resonance \citep[e.g., Ch.~5 of][]{MurrayDermott99}. On the other hand, thermal atmospheric tides are considered to account for the retrograde and very slow rotation of Venus \citep[][]{Dobrovolskis1980,Dobrovolskis1980b,CorreiaLaskar10}. 
By secularly modifying the star-planet separation, tides in the star influence the amount of flux received by the planet. The rotation of the star can also be affected, if the planet is sufficiently massive, thus affecting the XUV flux received by the planet \citep{Attiaetal21}. 

The dissipation of the tides inside a planet raised by the host star is usually parameterized by the modified tidal quality factor $Q^{\prime}_{\rm p}$ \citep[e.g.,][]{Zahn08}. 
In most applications, given our uncertainty on tidal dissipation,  $Q^{\prime}_{\rm p}$ is assumed to be constant, thus greatly approximating the modeling of tidal effects. The power $P_{\rm tide}$ dissipated inside a (pseudo)-synchronized planet on a slightly eccentric orbit ($e < 0.2$) can be expressed as \citep[cf. Eq.~(4) of][]{Milleretal09}
\begin{equation}
    P_{\rm tide} = \frac{63}{4} \left[ (GM_{\rm s})^{3/2} \left(\frac{M_{\rm s} R_{\rm pl}^{5} e^{2}}{Q^{\prime}_{\rm p}} \right)\right] a^{-15/2},
    \label{tidal_diss}
\end{equation}
where $G$ is the gravitational constant, $M_{\rm s}$ the mass of the star, $R_{\rm pl}$ the radius of the planet, $e$ the eccentricity of the orbit, and $a$ the orbit's semimajor axis.  For a highly eccentric orbit,  the $Q^{\prime}_{\rm p}$ parametrization breaks down and a different approach must be used that allows one to include  the effects of non-synchronous rotation \citep{Leconteetal10}. 
For terrestrial planets, typical values of $Q^{\prime}_{\rm p}$ range from a few tens to a few thousands, while for planets with a sizeable gaseous envelope or giant planets, values from a few $10^{4}$ up to $10^{5}-10^{6}$ are usually adopted \citep[cf.][in the case of the solar system]{Laineyetal16}. However, given our limited understanding of tidal dissipation inside planets and stars \citep{OgilvieLin07,Ogilvie14}, predicting an accurate value of $Q^{\prime}_{\rm p}$ for a specific planet is highly challenging. The physical mechanism of tidal heating is discussed in more detail in \citet{chapter4}.

Tidal heating in a terrestrial planet can produce relevant volcanic activity {\citep[for details see e.g.][this topical collection]{Lourenco2025}}, thus influencing the composition of its secondary atmosphere (see sec.\,\ref{sec:composition}). It can also lead to the enhancement of thermal and non-thermal escape processes (sec.\,\ref{sec:internal}). An average non-zero orbital eccentricity can be maintained by the perturbations of other planets in a compact system, especially in the case of a resonant chain such as in Trappist-1 \citep[e.g.,][]{Barretal18}. 

\subsection{Star-planet magnetic interactions}
The interaction between the stellar magnetic field and a close-in planet produces phenomena that can indirectly probe the planet's magnetic field  \citep[e.g.,][]{Shkolniketal08,Cohenetal11,Sauretal13,Cauleyetal19}. In addition to tides, the angular momentum loss of the star can be affected by the field of a closely orbiting giant planet \citep{Lanza10,Tejadaetal21}, thus modifying the evolution of its magnetic activity. In some cases, stellar flare activity could be affected as well {\citep{Loydetal23,Ilinetal24,Ilin2025Natur.643..645I}. The latter can, in turn, increase the planet's atmosphere evaporation rate \citep[][]{Ilin2025Natur.643..645I}. }

If the planet has no intrinsic magnetic field, its motion through the magnetic field of the star (closed corona or open stellar wind structures) produces an electric induction in its interior and ionosphere, provided that the magnetic flux is time-dependent. The amount of heating in the different layers -- ionosphere,  salty ocean, or interior -- depends critically on their electric conductivity \citep{Kislyakovaetal17,Kislyakovaetal18,KislyakovaNoack20}. 
This is relevant only for very close-in planets orbiting strongly magnetized stars, such as young late-type stars or late M-type stars such as Proxima Centauri or Trappist-1. The magnetic fields associated with coronal mass ejections may also produce induction heating \citep{Grayveretal22}.

The strong and varying magnetic field along the close-in orbit of a planet around an M star additionally induces an electric current in its ionosphere whose dissipation results in strong Joule Heating (JH) of the upper atmosphere. For Trappist-1\,e, the JH energy flux is likely larger than the incident XUV surface flux of $\sim$0.3\,W\,m$^{-2}$ \citep[$\sim$300 erg/s/cm$^2$, e.g.,][see also Fig.~\ref{fig:XUVEvo}]{Wheatley2017MNRAS.465L..74W} and may even approach up to a few percent of the entire stellar energy flux received by the planet in the case that its ionosphere thickness reaches $\sim$10,000\,km \citep{Cohenetal24}. Such extended ionospheres can indeed be expected for secondary atmospheres on close-in planets like Trappist-1e \citep[e.g.,][]{Cohenetal24,Nakayama2022}, thereby implying intense heating of the upper atmosphere and strongly enhanced atmospheric escape. 

When the planet has an intrinsic magnetic field, magnetic reconnection, and particle acceleration occur at the boundary of its magnetosphere \citep[e.g.][]{Lanza09,Lanza12,Buzasi13} and energy can be conveyed towards the star and the planetary atmosphere. The  maximum power available in the case of a hot Jupiter interacting with a Sun-like star can reach  $10^{15}-10^{18}$~W \citep[or $10^{22}-10^{25}$~erg/s][]{Lanza09,Lanza12}. A remarkably larger amount of power is predicted when a loop interconnecting the planetary field  with the stellar field is formed, reaching a maximum of $10^{20}-10^{21}$~W \citep[or $10^{27}-10^{28}$~erg/s][]{Lanza13}.

\section{Role of Planetary Mass (Escape Velocity)}\label{sec:planetary_mass}
From Sec.\,\ref{sec:escape_processes}, one can derive that thermal escape processes, if effective at the planet, are generally expected to dominate planetary atmospheric outflow. Hence, the escape from closer-in planets is expected to be stronger. However, besides the level of incoming stellar radiation, other planetary parameters are at play. To first order the most important one is planetary mass or, for planets subject to inflation (e.g., due to a combination of high internal thermal energy and low mass), the interplay between the mass and the radius of the planet, as discussed in Sec.\,\ref{sec:internal}.%

Even planets located far from their host star, hence less subject to intense thermal escape processes, can lose the entirety of their atmospheres if their mass is too small. In the Solar System, a good example of such a planet is Mars, which is significantly farther from the Sun, but about ten times less massive than Earth.

\begin{figure}[hb]
    \centering
    \includegraphics[width=0.8\linewidth]{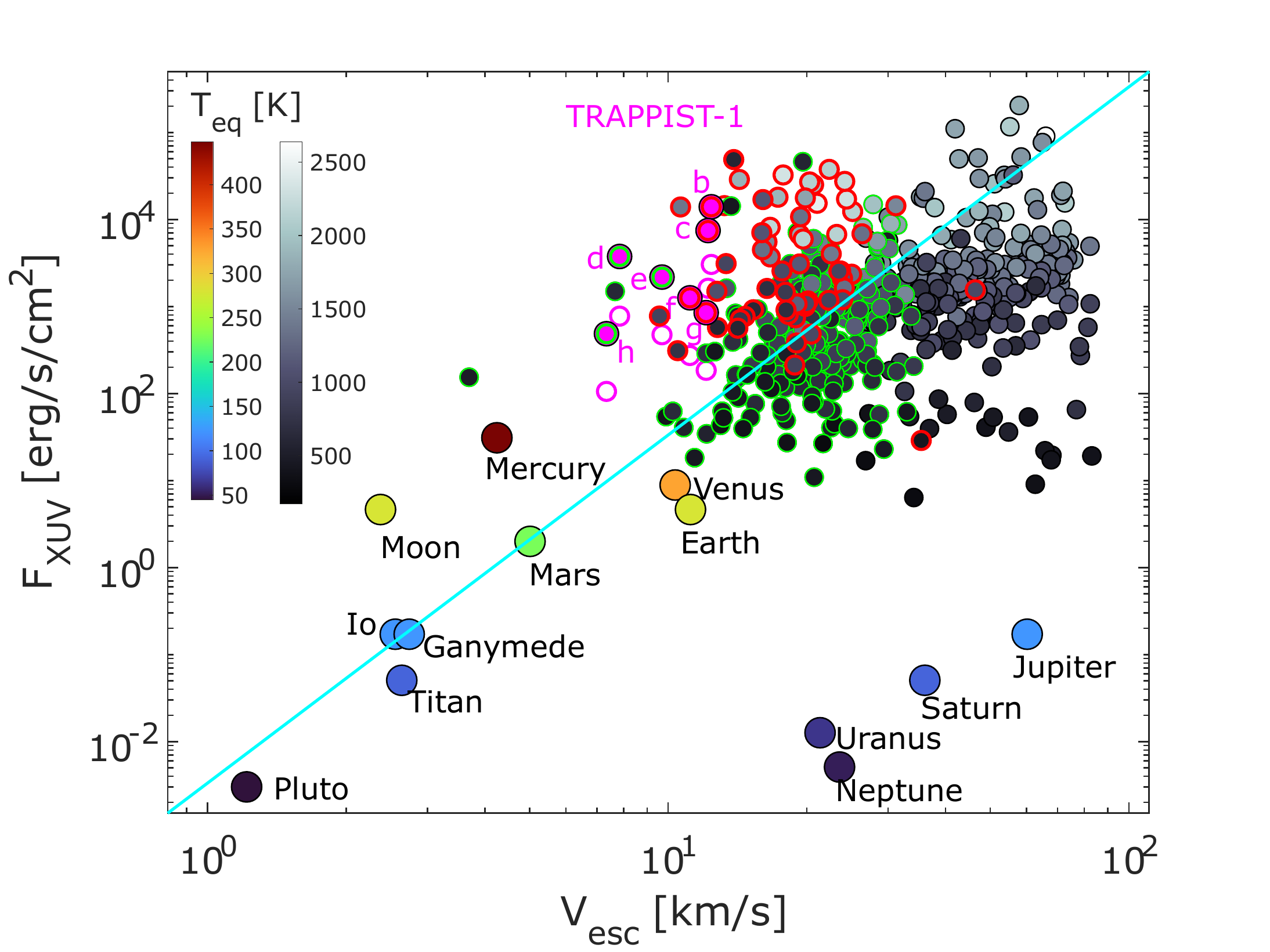}\\
    \caption{The Cosmic Shoreline following \citet{Zahnle2017} and \citet{Gronoff2026}. The XUV flux for a given planetary orbit was estimated by employing Mors stellar evolution code \citep[for planets orbiting stars with masses 0.1-1.2\,$M_{\odot}$][]{Johnstoneetal21} or the analytical approximations by \citet{wright2011ApJ...743...48W} and \citet{Sanz-Forcadaetal11} (for lighter and heavier stars). The colour of the dots reflects planetary equilibrium temperature (assuming zero albedo), as indicated by the colourbars in the plot; we additionally highlight the planets in Trappist-1 system with filled magenta markers. For comparison, empty magenta markers show the XUV flux estimates of Trappist-1 system from \citet{Wheatley2017MNRAS.465L..74W}. {Green and red outlines denote the exoplanets lighter than 40\,$M_{\oplus}$ with H-He atmosphere fraction larger and smaller than $10^{-5}$ according to the model by \citet{johnstone2015ApJ...815L..12J}, respectively.}}
    \label{fig:cosmic_shore}
\end{figure}

To illustrate the point, in Fig.\,\ref{fig:cosmic_shore} we present the distribution of the known exoplanets in the escape velocity ($v_{\rm esc}$) -- XUV flux ($F_{\rm XUV}$) plane following \citet{Zahnle2017}. Here, we use the data from the 
NASA Exoplanet Archive\footnote{\href{https://exoplanetarchive.ipac.caltech.edu/}{https://exoplanetarchive.ipac.caltech.edu/}} and the NASA Planetary Fact Sheet\footnote{\href{https://nssdc.gsfc.nasa.gov/planetary/factsheet/}{https://nssdc.gsfc.nasa.gov/planetary/factsheet/}} for the Solar System bodies. We only consider planets with masses smaller than $2\,M_{\rm jup}$, mass uncertainties smaller than 45\%, and radius uncertainties smaller than 15\%. 
For exoplanets, we used the following approach. To estimate the $F_{\rm XUV}$ at planetary orbits consistently for all planets, we employed the Mors stellar evolutionary code \citep{Johnstoneetal21,Spadaetal13} to calculate the fluxes according to the mass and age of their host stars, assuming that all stars have evolved as moderate rotators. For the planets orbiting stars outside the applicability limit of the Mors code (which is, $0.1-1.2$\,$M_{\odot}$), we used the empirical approximations by \citet{wright2011ApJ...743...48W} and \citet{Sanz-Forcadaetal11} as described in \citet{kubyshkina2019A&A...632A..65K}, which are  expected to provide average values. If the age of the star was not available in the Exoplanet Archive, we assigned it to the average value of 5\,Gyr, except for the Trappist-1 system where the age was set to 7.6\,Gyr \citep{Burgasser2017ApJ...845..110B}. %
The cyan line in Fig.\,\ref{fig:cosmic_shore} ($F_{\rm XUV}\,\sim\,v_{\rm esc}^4$) is expected to separate the planets capable of keeping {some type of} an atmosphere (below the line, where $F_{\rm XUV}$ is low and $v_{\rm esc}$ is high) from the bare cores (above the line); following \citet{Zahnle2017}. We set the proportionality coefficient more or less arbitrarily (as in \citet{Zahnle2017}), assuming that Mars was on the borderline of keeping its atmosphere.

The first thing one notices when analysing Fig.\,\ref{fig:cosmic_shore} is that the Solar System planets and the exoplanets known to date (at least, those with well-constrained masses and radii) occupy different regions both in the $F_{\rm XUV}$-$v_{\rm esc}$ diagram and in terms of equilibrium temperature. Furthermore, the exoplanets' population does not seem to align with the $F_{\rm XUV}\,\sim\,v_{\rm esc}^4$ {as lower-mass planets of the Solar System}. Instead, {they follow a shallower slope, in particular at high planetary masses.}
A few contributors seem to be at play in smoothing the dependence: generally higher equilibrium temperatures (e.g. a larger contribution from interior heat, and infrared heating from the planetary core), tidal forces of the host star, and radius inflation due to star-planet interaction.

{If one assumes that, for exoplanets, division between planets with and without atmospheres occurs along a line (power law) in $F_{\rm XUV}$-$v_{\rm esc}$ diagram, the shallower dependence implies that, for a given $v_{\rm esc}$, the transition occurs at lower XUV level that predicted by $F_{\rm XUV}\,\sim\,v_{\rm esc}^4$ dependence shown in Fig.\,\ref{fig:cosmic_shore}. However, this statement is currently impossible to verify, as we do not know if there is an atmosphere at any given exoplanet, unless it is a substantial H-He dominated atmosphere. Even in the latter case, defining the actual atmospheric mass fraction from observable parameters is tricky, as all the available internal structure models are strongly degenerated, in particular for small atmospheric mass fractions \citep[e.g.][and references therein]{Egger2025A&A...696A..28E}. To illustrate this point, in Fig.\,\ref{fig:cosmic_shore}, we include predictions on the presence/absence of H-He atmospheres for planets lighter than 40\,$M_{\oplus}$ based on simple structure model from \citet{johnstone2015ApJ...815L..12J} (assuming pure H-He atmospheres and fixed core parameters). Red and green markers' outlines show planets without and with H-He atmospheres (where the boundary atmospheric mass fraction value is taken at $10^{-5}$). Though the transition between the two groups of planets occurs near $F_{\rm XUV}\,\sim\,v_{\rm esc}^4$ line, they strongly overlap, not allowing to retrieve a clear boundary.}

Besides controlling the critical level of irradiation leading to atmospheric evaporation, planetary mass can affect specific escape channels. This can affect the composition of material leaving the planet. For non-thermally escaping atmospheres, the larger mass means a higher fraction of ion escape compared to neutrals (see Sec.\ref{sec:intro_non-thermal}). For thermally escaping atmospheres, the fraction of ions in the escaping material at high altitudes is also expected to increase: at a given $T_{\rm eq}$, ionisation fronts are more narrow for more massive planets. Thus, for heavy planets, the atmosphere is fully ionised at the exobase, while for low-mass planets this is not necessarily the case \citep[in particular, if the planet is hot; e.g.][]{kubyshkina2018A&A...619A.151K,Guo2024NatAs...8..920G}.
%

We highlight the fact that atmospheric escape could exhibit a complex dependence on the planet's size, through a simplified analysis that parallels \citet{Chin2024ApJ...963L..20C}. Suppose that we have a planet of radius $R_{\rm pl}$. As the size (or radius) of the planet is expanded, its cross-sectional area would roughly increase as $R_{\rm pl}^2$, if we assume a purely geometric scaling. In this case, a higher fraction of the electromagnetic radiation (extreme UV and X-rays) and stellar wind emitted by the host star would be intercepted by the planet's atmosphere. Hence, this would translate to an increase in the atmospheric escape rate. On the other hand, if we continue to increase the planet’s size further (and assume that the atmosphere does not contribute to $R_{\rm pl}$), we run into the issue that the escape velocity of the planet also increases, which obeys
\begin{equation}
v_\mathrm{esc} = \sqrt{\frac{G M}{R_{\rm pl}}} \propto R_{\rm pl}^{1.64},
\end{equation}
where we have used the mass-radius relationship for rocky planets $M_{\rm pl}\,\propto\,R_{\rm pl}^{3.7}$ \citep[][]{Zeng2016ApJ...819..127Z}. Hence, as the planet’s size increases, the corresponding rise in the escape velocity implies that fewer particles would possess the requisite energy to escape the planet’s gravitational well. 

To summarize the previous discussion, we expect the atmospheric escape rate to exhibit a potentially non-monotonic dependence on the planet size, as the planet transitions from a source-limited regime (influenced by the planet’s cross-sectional area) to the energy-limited regime (modulated by the planet’s escape velocity). As we have not specified what the driver of atmospheric escape is, it may be anticipated to hold true for different types of atmospheric escape, albeit to varying degrees and with different scalings.

In the case of ion escape, the trend delineated in the preceding paragraph appears to be consistent with empirical data from Mars and Venus, which may be source- and energy-limited, respectively \citep[][]{persson2020,Ramstad2021SSRv..217...36R}. Moreover, the above prediction has been corroborated through state-of-the-art simulations of putative exoplanets performed by \citet{Chin2024ApJ...963L..20C}, who found that ion escape peaked at the radius of $\sim 0.7\,R_{\oplus}$ for selected stellar wind parameters.

Turning our attention to late-type hydrodynamic escape powered by XUV radiation \citep[considering an energy-limited approximation][]{Erkaev2007A&A...472..329E}, a similar trade-off may be manifested \citep[e.g.][]{Lingam2021lcfb.book.....L}. If we take the size of the exobase to be loosely comparable to the planetary radius -- which is not strictly valid even for terrestrial planets \citep[e.g.][]{Lammer2013oepa.book.....L} -- the escape rate in the energy-limited regime obeys the simple scaling $\mathbf{R_{\rm pl}^3/M_{\rm pl} \propto R_{\rm pl}^{-0.7}}$ \citep[][]{Lingam2019RvMP...91b1002L}, where we have used the mass-radius relationship from earlier. In contrast, if we consider the photon-limited regime (analogous to the source-limited regime), the atmospheric escape scales as $R_{\rm pl}^2$ \citep[][]{Owen_Alv2016ApJ...816...34O}. Therefore, we can explicitly see how an increase in the radius would permit an increase in the atmospheric escape rate in the photon-limited regime. This is followed by a decrease in the energy-limited regime, mirroring our earlier statements. This trend is compatible with the results of numerical simulations \citep[e.g.][]{Chen2016ApJ...831..180C}.

\section{Role of atmospheric composition in escape}\label{sec:composition}
\subsection{Known atmospheric types}\label{sec:composition_Known-Atm-Types}
The types of (collision-dominated) atmospheres of terrestrial planets can broadly be divided into (i) primordial atmospheres, (ii) steam atmospheres, (iii) secondary atmospheres, and (iv) silicate atmospheres. In addition, (v) so-called airless bodies host a thin, collision-less atmosphere commonly called an exosphere \citep{Lammer2022}. Primordial atmospheres are the only ones that are not degassed from the planet's interior but are instead accreted during the planetary formation stage 
from the nebula surrounding the newly formed star \citep[e.g.][]{Hayashi1979}. Since the stellar nebula is compositionally similar to its host star, pristine primordial atmospheres consist mostly of hydrogen with a smaller amount of helium, and metals (elements heavier than He) being present only in trace amounts. The more massive a planet grows within the protostellar nebula, the more hydrogen and helium it will accrete. The size of the atmosphere is further {constrained} by the amount of gas available around a forming planet, hence, its position within the disk. For a planet in the HZ of solar-like stars, a mass around 1\,$M_{\oplus}$ seems to be the limit below which the entire primordial atmosphere can be lost subsequently \citep{Erkaev2023MNRAS.518.3703E,Lammer2024}. For lower-mass host stars and shorter orbits, this mass limit will be higher and vice versa. 
Furthermore, if the irradiation from the host star and the planet's mass are at a certain sweet spot, a hydrogen-dominated primordial atmosphere can become helium-dominated through fractionation: most hydrogen is lost into space, while part of the heavier He remains at the planet \citep[e.g.,][]{Hu2015Helium}.

Steam atmospheres, on the other hand, {which were initially assumed to be purely dominated by H$_2$O and minor amounts of CO$_2$,} form either through (i) degassing from a planet's solidifying magma ocean \citep[e.g.,][]{ElkinsTanton2011}, (ii) the evaporation of an ocean, or (iii) through migration of an ice- and/or water-rich planet toward its star \citep[e.g.,][]{Burn2024}. For case (i), the steam atmosphere can subsequently condense and form oceans if the planet is within the HZ. Closer to the star, the planet can evolve differently. In case of Venus, for instance, it is currently disputed whether the degassed steam atmosphere was initially able to condense to form early oceans \citep[e.g.,][]{Hamano2013,Lebrun2013,Salvador2017,Way2020Venus,Salvador2023} or the temperature at its orbit was always too hot for condensation \citep[e.g.,][]{Hamano2013,Lebrun2013,Salvador2017,KrissansenTotton2021Venus,Turbet2021,Salvador2023}. If it does not condense, a steam atmosphere can also escape into space. In this case, oxygen provided by the dissociation of H$_2$O can accumulate at the planet, as it escapes less efficiently than H due to its weight. In addition, steam atmospheres generally escape less efficiently than primordial atmospheres, not only because of their higher molecular weight but also because H$_2$O acts as an efficient coolant \citep[e.g.][]{Yoshida2022}. However, recent {studies} \cite[e.g.,][]{Bower2022,Maurice2024} {have} put into question the entire foundation of what the constituents are for such atmospheres. They suggest that steam atmospheres are not viable for some magma ocean redox states and C/H ratios (Earth's C/H$\sim$1). In the case that Earth did not have a steam atmosphere over its magma ocean, there are few ideas on exactly how the Earth obtained its initial Hadean era liquid water ocean \citep[e.g.,][]{Miyazaki2022}.

Secondary atmospheres may be initially degassed by volcanic activity after a primordial atmosphere was lost and the subsequently degassed hypothesized steam atmosphere either condensed or was also lost into space. These atmospheres can have very different compositions, structures, and densities and are inclined to change substantially over a planet's history due to various abiotic, and potentially biotic, processes. The present-day Solar System illustrates part of the diversity secondary atmospheres can exhibit. For example, atmospheres have {likely} ranged from N$_2$-CO$_2$-dominated (Archean Earth), N$_2$-O$_2$-dominated (modern Earth) and N$_2$-CH$_4$-dominated (Titan) toward thick (Venus) and thin (Mars) CO$_2$-dominated atmospheres.
However, all these bodies likely had a different atmospheric compositions through time. Additional types of secondary atmospheres can be envisioned to exist in the Universe such as CO- \citep{Zahnle2008Mars} and O$_2$-dominated ones \citep{Wordsworth2014,Luger2015}. Crucially, their compositions lead to very different 
upper atmosphere structures, as different molecules have different molecular weights and interact differently with the incoming XUV flux of the host star. CO$_2$, for instance, is a strong infrared coolant whereas N$_2$ is not. Together with their different molecular weights, this is an important reason why the upper atmospheres of Earth and Mars/Venus look so different \citep[see, e.g.,][]{Way2022}.

Planetary bodies that are close enough to their host star, such as CoRoT-7~b \citep{Leger2011Corot7b}, may experience surface temperatures high enough to melt and vaporize rock, which then either escapes into space or forms a silicate atmosphere consisting of species such as Na, K, and SiO \citep[e.g.,][]{Ito2015Silicate}. Although silicate atmospheres have a relatively high molecular weight, only massive planets can prevent their escape, as they are genuinely exposed to very high XUV fluxes. Although recent observations of the Super Earth 55 Cancri e \citep{Hu2024Cancri55e} demonstrate that it may have an atmosphere dominated by volatiles, not silcates.
If a rocky body such as Mercury is far enough from the star for its mantle to solidify but still close and small enough to lose any (collision-dominated) atmosphere, it will still keep an exosphere dominated by, e.g., O, K, and Na. This exosphere is balanced by processes that release particles from its surface \citep[e.g., sputtering, photodesorption, and micrometeorite impacts; see][]{Wurz2022} and by their escape into space\footnote{In principle, volcanic degassing could be another atmospheric source. However, volcanic activity on small terrestrial bodies (e.g. Mercury) ceases relatively fast as long as tidal heating can be neglected. Otherwise, volcanic activity can persist as is the case for Io, which hosts an SO$_2$-dominated exosphere. For highly active stars even Earth-mass planets can lose their secondary atmospheres and form an exosphere balanced by volcanic degassing and escape into space.}. For small icy bodies, the exosphere can be dominated by interaction with their UV and particle environments (e.g., Europa, Callisto, and Ganymede) and/or the equilibrium vapor pressure of its constituents at the respective surface temperatures. The latter can be observed at Pluto and Triton whose exospheres are dominated by N$_2$ with minor amounts of CH$_4$ and CO \citep[e.g.,][]{Scherf2020}.

\subsection{Primordial (Nebular) atmosphere}
\label{sec:composition_nebular}
Formation models predict that the majority of planets, including low-mass ones, accrete hydrogen-dominated atmospheres with compositions resembling that of their host stars, while embedded in their protoplanetary discs \citep[e.g.][]{Jin_Mordasini2018ApJ...853..163J,Morbidelli2020A&A...638A...1M,Venturini2020A&A...643L...1V,Burn2021A&A...656A..72B}. For terrestrial-like planets, such atmospheres contribute less than $\sim1\%$ to the total planetary mass \citep[e.g.][]{Mordasini2020A&A...638A..52M}, and their lifetime after the protoplanetary disk dispersal is short \citep[$\lesssim1-100$\,Myr; e.g.][]{Lammer2008SSRv..139..399L,owen_wu2016boil-off}. %
However short the lifetimes of primordial atmospheres are, the extreme outflow of hydrogen during this early phase of planetary evolution can have long-term consequences. First, if the atmospheric escape rate exceeds a certain limit, hydrogen outflow can drag along some heavier elements (the critical mass, {i.e., the maximum mass of the species that can escape alongside hydrogen,} is defined by the outflow parameters). This leads to the fractionation of atmospheric species \citep[e.g.][]{Oepik1963GeoJ....7..490O,Zahnle1986Icar...68..462Z,Hunten1987Icar...69..532H,Pepin1991Icar...92....2P,Odert2018Icar..307..327O,Lammer2020SSRv..216...74L}. 
Second, the fractionation of radioactive elements can adjust the planet's thermal budget \citep{Erkaev2023MNRAS.518.3703E}. Furthermore, as long as a hydrogen-dominated atmosphere is present, it can interact with the surface and  facilitate water production \citep[e.g.][]{Ikoma2006ApJ...648..696I,Ikoma2018SSRv..214...76I,Lammer2021SSRv..217....7L,Salvador2023,Rogers2024ApJ...970...47R} and, hence, alter the interior composition \citep[e.g.][]{Rogers2024ApJ...970...47R}.
Besides the direct impact on atmospheric composition, the lifetime of the primordial atmosphere sets the timescale of the formation of the steam and secondary atmospheres dominated by elements heavier than hydrogen. Hence, it can affect whether, when, and under which external conditions the secondary atmosphere is formed. %

The early phase (first $\sim$100\,Myr for Sun-like stars and up to $\sim$1\,Gyr for late M-dwarfs) of atmospheric evaporation is expected to occur in the hydrodynamic regime, while the main driving mechanism and duration of this phase for a specific planet can vary depending on primordial planetary parameters. As discussed in sections\,\ref{sec:intro_thermal_classification} and \ref{sec:planetary_mass}, the type and strength of the outflow depends strongly on the planet's mass and irradiation level set by the planetary orbit and properties of the host star. For low-mass (lighter than $\sim10\,-\,30$\,${\rm M_{\oplus}}$) planets on short orbits, early escape is likely to occur on account of stellar (bolometric) heating, i.e., in the boil-off/core-powered regime. It will later transition into an XUV-driven regime, while for more massive and distant planets, the escape can be XUV-driven from the start \citep[e.g.][]{Guo2024NatAs...8..920G}.
This type of escape is particularly sensitive to stellar input: while the bolometric luminosity of a star (especially for Sun-like stars) does not change dramatically over the main sequence, the XUV luminosity at young ages (see Sec.\,\ref{sec:stellar_input:rotation}) can be orders of magnitude above that of Gyr-old stars.

Whether the hydrogen-dominated atmosphere survives this phase of extreme mass loss depends on the planet's mass and orbit: increase{ in the planetary} mass (increasing escape velocity) allows a low-mass planet to keep its atmosphere with decreasing orbital separation \citep[e.g.][]{Lopez2014ApJ...792....1L,Chen2016ApJ...831..180C,Kubyshkina_Vidotto2021MNRAS.504.2034K}. %
{The renowned observational confirmation of this dependency is the negative slope of the radius valley \citep[which is a lack of planets with intermediate radii $\sim1.5$\,$R_{\oplus}$ in sub-Neptunes' population within $\sim$100 days orbit][]{Fulton2017AJ....154..109F} in exoplanets' distribution in the radius-period plane \citep[e.g.][]{Owen_Wu2017ApJ...847...29O,Martinez2019ApJ...875...29M}. The position of the centre of the valley (minimum in $R_{\rm pl}$ distribution at a given orbital period $P_{\rm orb}$) changes with orbit as $R_{\rm valley} \sim (P_{\rm orb}/10$ days$)^{0.1}$.}
The origin of the valley is commonly attributed to the presence of H-dominated atmospheres at some planets. At high equilibrium temperatures, structure models \citep[e.g.][]{Stoekl2015A&A...576A..87S} predict that even tiny ($\lesssim 0.1\%$ of $M_{\rm pl}$) H-dominated atmospheres can lead to the significant inflation of planetary radii. In this context, both core-powered and XUV-driven escape mechanisms can reproduce the valley \citep[e.g.][]{Affolter2023A&A...676A.119A}. With increasing orbital separation, both the radius inflation and the {maximum bare core (MBC, which is roughly equivalent to the minimum planetary mass needed to keep the primordial atmosphere at the given orbit)} mass decrease, and the radius valley closes. However, the {MBC} mass remains an important parameter to assess a planet's evolution. Thus, for the Earth, it was estimated to be slightly below the present-day mass, indicating that the planet reached its final mass after protoplanetary gas disk dispersal {\citep[otherwise, it would keep its primordial atmosphere; on the other hand, the mass had to be high enough to accrete some of it, to explain the noble gases inventory;][]{Lammer2020Icar..33913551L,Erkaev2023MNRAS.518.3703E,Lammer2024}}.
%

Besides the dependence upon basic planetary parameters, the atmospheric evaporation timescale of H-dominated atmospheres from low-mass planets depends on their primordial atmospheric mass fraction. Models predict that the lifetime of atmospheres maximizes at intermediate initial atmospheric mass fractions{, that is typically 1-3\% of the planetary mass for the sub-Neptune-mass planets subject to significant atmospheric losses.} The specific value increases with increasing planetary mass and orbital separation, and (for a given $T_{\rm eq}$) with decreasing mass of the host star \citep[e.g.][]{Chen2016ApJ...831..180C,Kubyshkina_Vidotto2021MNRAS.504.2034K}. The effect is most pronounced for planets with masses near the {MBC} one for their respective orbit and host star.

The discussion above concerns mainly low-mass and/or close-in planets. In the case of cool giants, the loss of primordial atmospheres is dominated by Jeans-like or non-thermal escape processes, such as charge exchange \citep[e.g.][]{Parkinson2002PhDT........13P,Mauk2020JGRA..12528697M}.

\subsection{Venus and Mars-type CO$_2$-dominated atmospheres: implications for exoplanets}\label{sec:composition_CO2}
Although Venus is comparable in size and mass to the Earth, the planet’s surface shows no indications of modern Earth-style subductive plate tectonics. {The same applies to the Martian surface}. Today Venus and Mars are surrounded by CO$_2$-dominated atmospheres ($\approx~92$ bar and $\approx~6.5$ mbar) that consist of 96.5\% CO$_2$ with 3.5\% N$_2$ \citep{Oyama1980} and $\approx~95.97$\% CO$_2$ with $\approx~1.89$\% N$_2$ \citep{Mahaffy2013}, respectively. 
\citet{Weller2023} modelled secondary atmospheres and compared them with modern Venus' atmosphere. They found that volcanic outgassing in an early phase of plate-tectonic-like activity during the first billion years, followed by a stagnant-lid-phase that lasts until today, can best explain Venus’ dense CO$_2$ atmosphere and especially its present-day high N$_2$ content.
As addressed in detail in \citet{Scherf2021}, it is not likely that Mars could build up a dense secondary CO$_2$ atmosphere during its first 400 Myr. Afterward, it may have had a denser atmosphere from sporadic asteroid impacts and large volcanic events. These could have resulted in sporadic warmer climate periods $\approx 3-4$ Gyr ago \citet{Scherf2021,Lichtenegger2006,Schmidt2022circumpolar,kamada2020coupled,wordsworth2016climate,turbet2019paradoxes,forget20133d,guzewich20213d} with a denser CO$_2$ or CO-dominated atmosphere. This atmosphere must have been lost later by CO$_2$ sequestration into the soil and non-thermal atmospheric escape processes \citep{Lammer2018,Lichtenegger2022Icar..38215009L,wordsworth2016climate}.

Because Venus and Mars have a similar atmospheric composition up to the exobase level, one can estimate their average exobase temperatures by a scaling method that can also be applied to exoplanets as long as the atmosphere remains under hydrostatic conditions. From the effective heat production below the exobase level, which is balanced by the divergence of the conductive heat flux of the XUV radiation, one gets the following simplified expression for the exobase temperature $T_{\rm exo}^j$ \citep{Bauer1971,Gross1972}
\begin{eqnarray} 
    T_{\rm exo}^j\approx\frac{\eta_{j}\; F_{\rm XUV}\; k_{\rm B}
    \; \sigma_{{\rm c}_{j}}} {\alpha\; K_{0_{j}} \; m_{j} \; g \;
    \sigma_{{\rm a}_{j}}}+T_{0},
    \label{eq:T-exo}
\end{eqnarray} 
where $\sigma_{{\rm c}_{j}}$ and $\sigma_{{\rm a}_{j}}$ are the collision and absorption cross-sections of species $j$, $K_{0_{j}}$ is the thermal conductivity of species $j$ at the reference temperature, $\alpha$ is a factor related to the planetary rotation ($\approx~0.25$ for fast-rotating planets like Earth and $\approx~0.5$ for tidally locked planets). $T_{0}$ is the temperature at the base of the thermosphere, which can also be approximated with the homopause temperature. If the homopause temperature is unknown one can adopt $T_0\simeq T_{\rm eq}$, which has a negligible effect if $T_{\rm exo}^j \gg T_{0}$.

For Earth's N$_2$/O$_2$-dominated upper atmosphere with its very low atmospheric CO$_2$ abundance, the present Sun's XUV flux results in a $T_{\rm exo}\approx\,1000\,-\,1200$\,K \citep{Jacchia1977,Crowley1991}. Although Venus is closer to the Sun, due to its 96\% CO$_2$ atmosphere the average dayside $T_{\rm exo}$ is only $\approx285$\,K \citep{vonZahn1980,Hedin1983,Mueller-Wodarg2006}. A main reason for cooler Venus' thermosphere-exosphere environment compared to Earth is a very efficient cooling of the upper atmosphere in Infra-Red (IR) emission of the 15\,$\mu$m CO$_2$ band \citep{Gordiets1982,Gordiets1985,Mueller-Wodarg2006}. 

If two planets have similar atmospheres, their planetary parameters are known {and the atmospheres of both planets can be characterized by high Jeans escape parameters (with $\lambda >> 6$),} one can use Eq.\,\ref{eq:T-exo} in the case that $T_{\rm exo}$ is known for one planet to estimate the unknown exobase temperature of the second. Since Venus and Mars have a similar atmospheric composition $j {\rm =CO_2}$ in their thermosphere, one obtains the following scaling relation \citep{Bauer1971}, 
\begin{eqnarray}
    \frac{\left(T_{\rm exo}^{\rm CO_2} - T_{0}\right)_{\rm Venus}}
    {\left(T_{\rm exo}^{\rm CO_2}- T_{0}\right)_{\rm Mars}} =
    \frac{F_{\rm XUV_{\rm Venus}} g_{\rm exo}^{\rm Mars}}{F_{\rm XUV_{\rm Mars}} g_{\rm exo}^{\rm Venus}}.
    \label{eq:T-exoscaleing}
\end{eqnarray}
One can see from Eq.\,\ref{eq:T-exoscaleing} that $T_{\rm exo}^{\rm CO_2}$ depends on $F_{\rm XUV}$, which decreases with distance from the star, and on the planet's gravitational acceleration at the exobase level $g_{\rm exo}$, which is related to the exobase position $r_{\rm exo}$ and the planetary mass. 
Using the average exobase temperature of Venus and the corresponding values for $F_{\rm XUV}$ at the orbits of Venus and Mars, the corresponding $g$ values for both planets at the exobase level of 200 km and $T_{\rm 0}$ of 220 K and 160 K for Venus and Mars, one obtains an average $T_{\rm exo}^{\rm CO_2}$ for Mars of $\approx$200 K, which is in agreement with the Martian exobase temperature inferred by 
MAVEN spacecraft of $\approx 239.2\pm27.6$ K near perihelion and $\approx 162.4 \pm 19.9$ K near aphelion \citep{Qin2020}. One can also make the same estimate the other way around by getting the exobase temperature at Venus from $T_{\rm exo}^{\rm CO_2}$ of Mars. Similarly, the same equation can be utilized to estimate the exobase temperature of exoplanets as long as these are hydrostatic {and the exobase level is known, as $g_{\rm exo}$ can diverge significantly from the gravitational acceleration on the surface of a planet for extended atmospheres. If we, we for instance, take the simulation for a planet with $1.0\,M_{\oplus}$, an atmospheric composition of 10\%\,N$_2$ and 90\%CO$_2$, and an incident XUV flux of $8\,F_{\rm XUV,\oplus}$ (i.e.,8 times the XUV surface flux received by the Earth) by \citet{VanLooveren2024Trappist} with an exobase level of $\sim$240\,km, Equation~\ref{eq:T-exoscaleing} gives an exospheric temperature of $\sim820\,K$ by comparing it Venus. This is very close to the $\sim$800\,K from their simulation. However, at higher XUV fluxes, Equation~\ref{eq:T-exoscaleing} deviates from the simulations and yields lower temperatures due to the induced expansion and increasing divergences in the upper atmosphere chemistry resulting from the increased XUV surface flux. Equation~\ref{eq:T-exoscaleing} should therefore be applied with caution.}

\begin{figure}
    \centering
    \includegraphics[trim=130 100 110 90,clip,width=0.8\linewidth]{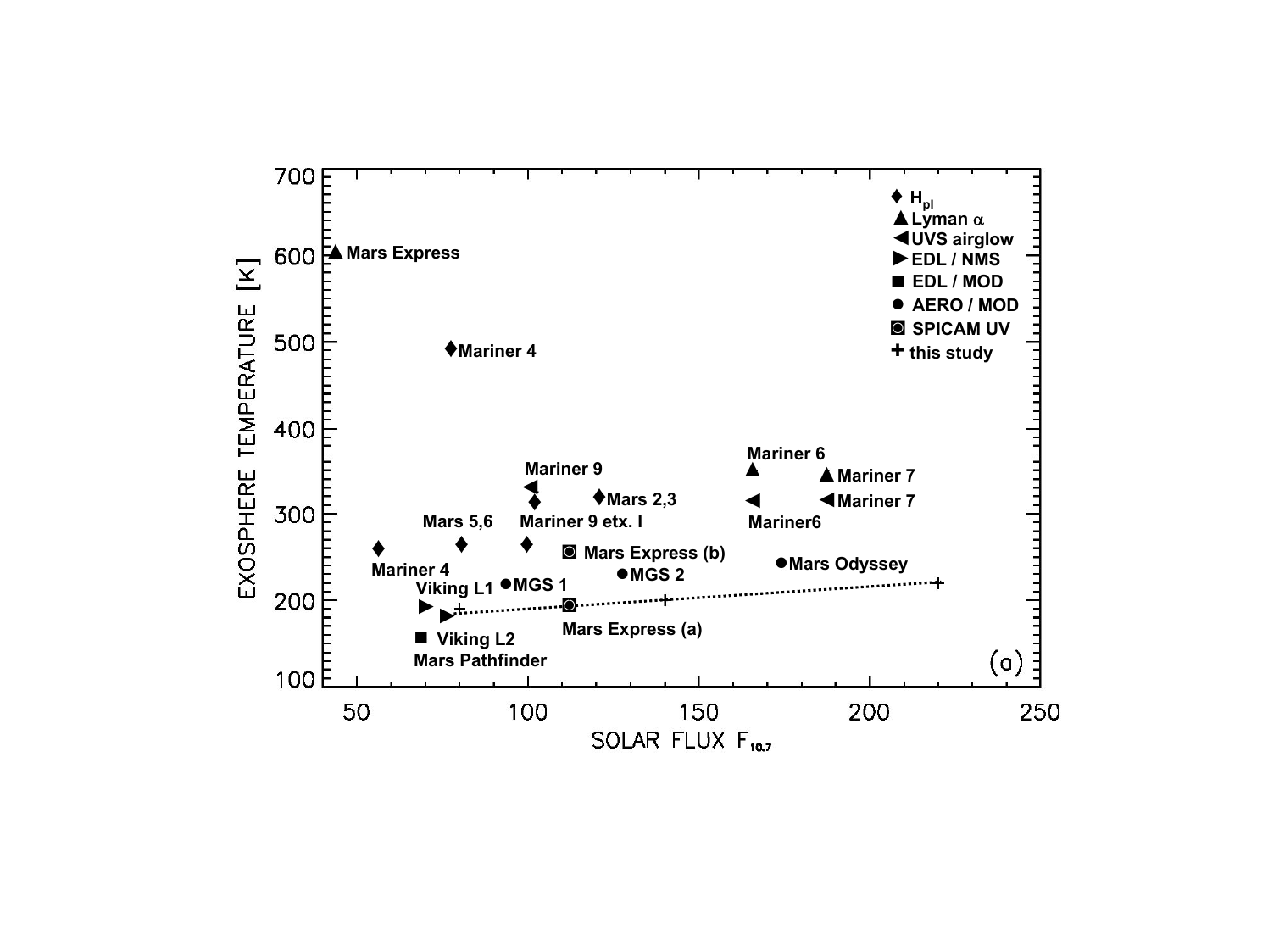}
    \caption{Exosphere daytime temperature estimates based on various methods, data, and space missions during the solar cycle at Mars. The horizontal dashed lines correspond to photochemically produced suprathermal H atoms and CO molecules \citep[adopted from][]{Lichtenegger2006}. The x-axis is the 10.7 cm radio flux (F$_{\rm 10.7}$), a commonly used proxy for solar activity.}
    \label{fig:Fig_3a}
\end{figure}

As one can see in 
Fig.~\ref{fig:Fig_3a}, the estimate from Eq.\,\ref{eq:T-exoscaleing} is more accurate than the early exobase temperature estimates inferred from plasma measurements (Mariner 4, Mariner 9, Mars 2, Mars 3, Mars 4, Mars 6), UVS airglow observations (Mariner 6, Mariner 7), and Lyman-$\alpha$ observations of the Martian hydrogen corona by Mariner 6, Mariner 7 and Mars Express \citep{Lichtenegger2006}. The reason for these high exosphere temperatures between $\approx$ 300 – 600 K obtained by the various methods listed above is related to the photochemically produced suprathermal H, O, and CO populations in the exosphere, while the scaling method above focuses on the thermal neutral atmosphere around the exobase level.  

\begin{figure}
    \centering
    \includegraphics[width=0.7\linewidth]{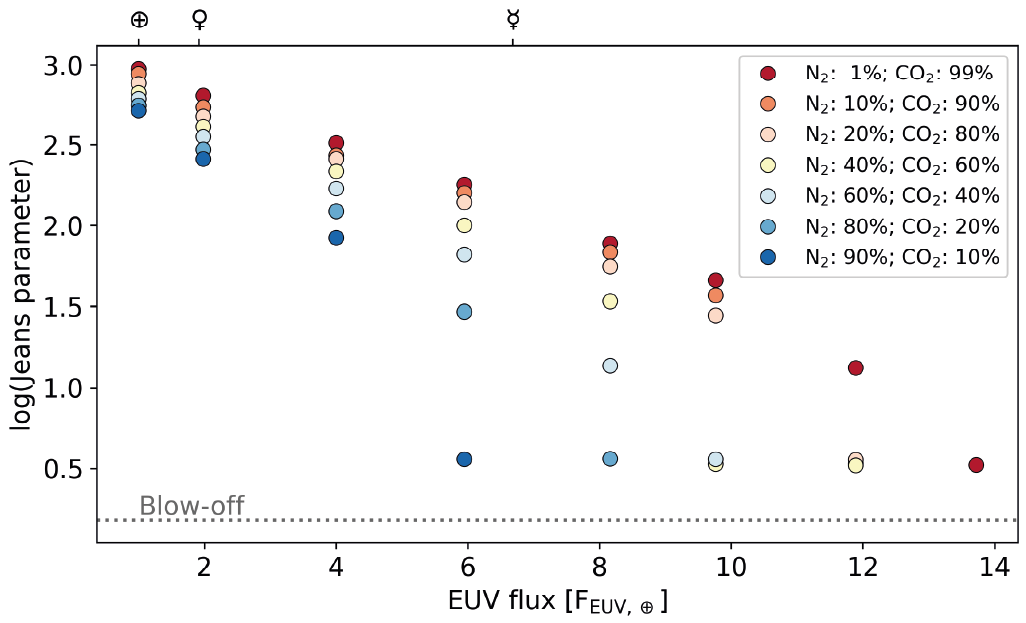}
    \caption{The Jeans parameter $\lambda_{\rm exo}$ for an Earth-mass planet with various initial atmospheric N$_2$/CO$_2$ mixing ratios. The horizontal dotted line corresponds to atmospheric blow-off conditions and the x-axis shows the EUV flux in Earth units \citep[adopted from][]{VanLooveren2024Trappist}. }
    \label{fig:Fig_3b}
\end{figure}

For high XUV fluxes, however, CO$_2$ molecules will dissociate, which results in less IR cooling, and a hotter thermosphere. In such cases, the exobase will expand to higher altitudes and $g_{\rm exo}$ will decrease. Under such conditions, Eq.\,\ref{eq:T-exo} will underestimate the $T_{\rm exo}^{\rm CO_2}$ value, if the exobase level, and the resulting $g$ is unknown. Therefore, one will not obtain accurate exobase temperatures if one applies Eq.\,\ref{eq:T-exoscaleing} to exoplanets such as those in the Trappist-1 system. 

\citet{VanLooveren2024Trappist} applied a thermosphere model including photochemistry to Earth-mass planets in the Trappist-1 system and found that CO$_2$ molecules photodissociate below an extended exobase level for XUV fluxes $> 2 \times$\ that of the Earth. They do so very efficiently for XUV fluxes $> 10 \times$ that of the Earth, producing atomic O and C, amongst others. These dissociation products are much lighter than CO$_2$, and are lost more easily. Additionally, the reduction of IR-cooling in the thermosphere yields the aforementioned expansion of the exobase level to higher altitudes. 

As shown in 
Fig.~\ref{fig:Fig_3b}, \citet{VanLooveren2024Trappist} calculated the Jeans parameter $\lambda_{\rm exo}$ (Eq.\,\ref{eq:lambda_exo}) at the exobase level for an Earth-mass planet with different atmospheric mixing ratios of N$_2$ and CO$_2$. Depending on the mixing ratio, one can see that $\lambda_{\rm exo}$ reaches the conditions for hydrodynamic blow-off (no stable upper atmosphere; see sec.\,\ref{sec:intro_thermal_classification}) for XUV fluxes $>$ 6 times that of today's Sun at 1 AU. %
An Earth-mass planet with a 99\% CO$_2$ atmosphere will reach blow-off escape conditions beyond 14 times present day XUV fluxes at Earth's orbit \citep{VanLooveren2024Trappist}. The findings of \citet{VanLooveren2024Trappist} also agree with \citet{Tian2009}, who found that early Mars could not build up a dense CO$_2$-atmosphere due to the dissociation of CO$_2$ molecules at comparably high XUV fluxes (assuming low--moderate outgassing rates). Based on the escape rate evolution from \cite{Tian2009}, volcanic degassing cannot counterbalance thermal escape at Mars during the first $\sim$450\,Myr \citep{Scherf2021}. %

Since the XUV surface fluxes at the planets in the Trappist-1 system\footnote{The EUV surface fluxes for the Trappist-1 planets derived from the lower X-ray flux values of \citet{Wheatley2017MNRAS.465L..74W} are 68 (planet h), 120 (planet g), 177 (planet f), 306 (planet e),
529 (planet d), 1050 (planet c), and 1982 (planet b) times the present-day EUV flux, $F_{\rm EUV}$, at the Earth's orbit with $F_{\rm EUV} = 4.77$\,erg\,s$^{-1}$\,cm$^{-2}$ in the wavelength range 10-121\,nm \citep[as calculated by][]{VanLooveren2024Trappist}. Here we adopted the ``EUV'' definition used in \citet{VanLooveren2024Trappist}, which includes wavelengths up to Ly-$\alpha$ line instead of 91.2\,nm.} are much higher than the values studied by \citet{VanLooveren2024Trappist} stable dense secondary atmospheres could not build up on the Trappist-1 planets (assuming moderate outgassing; see \citealt{Tian2009}){, at least according to their simulations\footnote{{At least for N$_2$-O$_2$-dominated atmospheres, \citet{Nakayama2022} find much lower thermal escape rates due to the atomic line cooling of O. However, these authors point out that the structure of highly irradiated atmospheres in their model would instead favour increased non-thermal escape rates compared to less strongly irradiated planets. Since non-thermal escape is severe around M dwarfs even without considering expanded atmospheres \citep[e.g.,][]{Airapetian_2017,Dong2018trappist,GarciaSage2017,RodriguezMozos2019,Scherf2024AsBio..24..916S}, it can therefore be expected that the existence of dense secondary atmospheres on the Trappist-1 planets remains unlikely even if the simulations by \citet{VanLaerhoven2014} overestimate thermal loss rates. Non-thermal escape can still be critical.}}. This agrees with the tentative} non-detections of CO$_{\rm 2}$ atmospheres at Trappist-1 b \citep{Greene2023} and Trappist-1 c \citep{Zieba2023} and other low mass M-star planets (although issues have been raised about the veracity of these observations; see \citealt{Fauchez2025}). {Note, however, that the existence of atmospheres on the Trappist-1 planets is still disputed \citep[e.g.,][]{Lincowski2023,Ducrot2025,Gillon2025}; see also \citet{Ducrot2026}, this topical collection, for details.} The aforementioned findings indicate that early Venus and Earth may have encountered similar problems beyond 4\,Gyr ago.

\subsection{Earth-Type Atmospheres}\label{sec:composition_earth-type}
A modern Earth-type atmosphere is a secondary atmosphere composed of N$_2$ and O$_2$ as the major atmospheric species, with CO$_2$ as an important trace molecule and potential additional trace species such as O$_3$ and N$_2$O \citep[see, e.g., definition of Earth-like atmosphere in][]{Lammer2024EEI}. Earth is the only planet known so far that hosts such an atmosphere, and it is important to note that Titan's N$_2$-dominated atmosphere cannot be categorized as Earth-type, since (i) the processes that led to its formation are completely different, (ii) its upper atmosphere structure diverges significantly, and (iii) O$_2$ and CO$_2$ are missing in Titan's atmosphere in Earth-like abundances {\citep[e.g.,][]{Scherf2020,Spross2021}}. If one were to put Titan into Earth's orbit today, its atmosphere would likely escape into space \citep{Spross2021}, which highlights that planetary mass and the stellar environment strongly matter for the structure and thermal stability of nitrogen-dominated atmospheres of any kind.

\begin{figure}
    \centering
    \includegraphics[width=\linewidth]{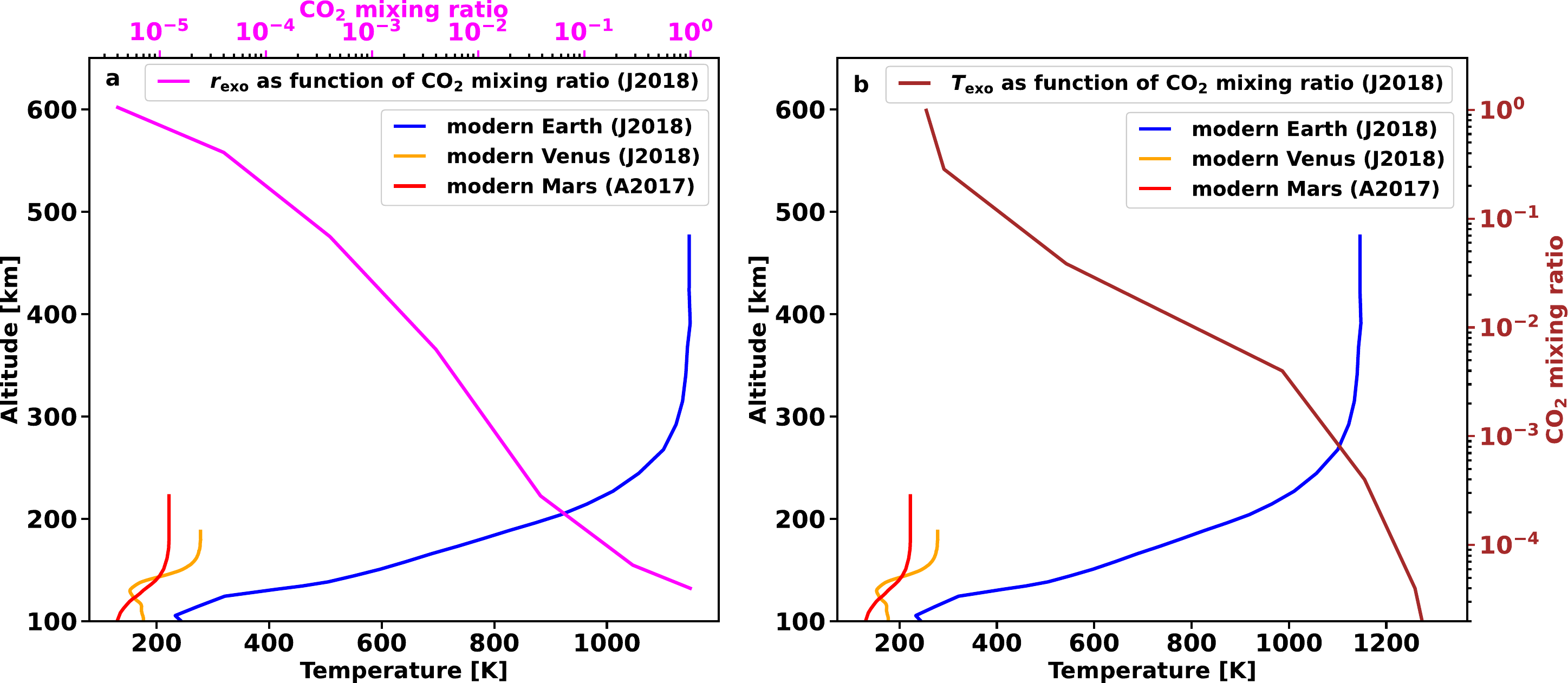}
    \caption{Both panels show the present upper atmosphere temperature profiles of Venus (orange), Earth (blue), and Mars (red), which stop at the exobase for all planets. Panel (a) additionally shows the exobase level, $r_{\rm exo}$ (magenta), as a function of the CO$_2$ to N$_2$ mixing ratio for Earth-like conditions (i.e., for the present $F_{\rm XUV}$ at the Earth's orbit and its planetary mass of 1.0\,M$_{\oplus}$). Panel (b) shows the exobase temperature, $T_{\rm exo}$ (brown), again as a function of CO$_2$ to N$_2$ mixing ratio for the same Earth-like conditions. One can see how $r_{\rm exo}$ and $T_{\rm exo}$ decrease for increasing CO$_2$ mixing ratios. Data from \citet{johnstone2018} and \citet{Amerstorfer2017}.}
    \label{fig:VEMcomp}
\end{figure}

Earth-type atmospheres are generally regarded to be a product of life interacting with the planet as both N$_2$ and O$_2$ are not simultaneously stable in an atmosphere \citep[e.g.,][]{Lovelock1982,Stueeken2016,Spross2021}. A high N$_2$ to CO$_2$ mixing ratio may be an additional hint of tectonic activity and a working carbon-silicate cycle being present at the respective planet \citep[e.g.,][]{Mikhail2014,Lammer2019}, which makes the potential detection of such an atmosphere at an exoplanet -- in contrast to CO$_2$-dominated atmospheres -- not only a potential bio- but also a geo-signature. 
Another important difference between Earth-type and CO$_2$-dominated upper atmospheres is their thermal structure since CO$_2$ is an IR-coolant and N$_2$ is not. While the CO$_2$-dominated atmospheres are relatively compact and cool under hydrostatic conditions, Earth-type atmospheres are more extended and exhibit much higher thermospheric temperatures with $r_{\rm exo}> 400$\,km and $T_{\rm exo}>1000$\,K. This is illustrated in Fig.~\ref{fig:VEMcomp}, which compares the upper atmosphere thermal structures of Venus, Earth, and Mars. The left panel additionally shows how the exobase radius, $r_{\rm exo}$, changes as a function of atmospheric CO$_2$ mixing ratio whereas the right panel further illustrates the effect of the CO$_2$ mixing ratio on the exobase temperature, $T_{\rm exo}$. It is important to note that these exobase temperatures and radii were modelled by \citet{johnstone2018} assuming Earth-like conditions, namely, for a planet with a mass of 1.0\,M$_{\oplus}$, an XUV surface flux identical to today's surface flux at 1\,AU, and an Earth-type atmosphere but with varying CO$_2$ mixing ratios. Therefore, $r_{\rm exo}$ and $T_{\rm exo}$ for a CO$_2$ mixing ratio of $\sim$0.96 must slightly deviate from the Martian and Venusian values, as conditions at these planets are rather different from this assumption. 
Equations~\ref{eq:T-exo} and \ref{eq:T-exoscaleing} illustrate these deviations mathematically as discussed in Sec.\,\ref{sec:composition_CO2}.

\begin{figure}
    \centering
    \includegraphics[width=\linewidth]{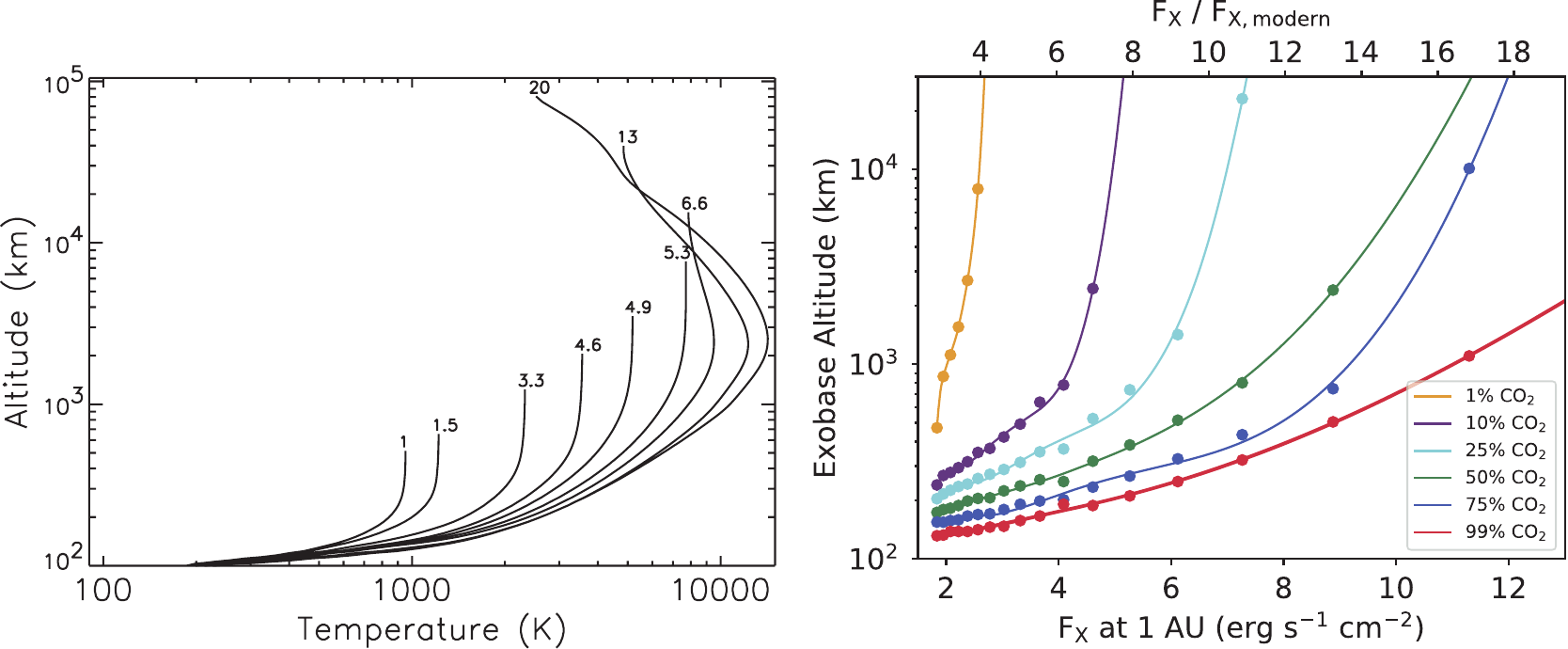}
    \caption{The left panel shows the thermospheric temperature profiles of an Earth-type N$_2$-O$_2$-dominated atmosphere with 400\,ppm CO$_2$ for different XUV surface fluxes up to the exobase level, according to the hydrodynamic 1-D model by \citet{tian2008}. The numbers on top of each profile give the XUV surface flux in present-day values. The right panel illustrates the exobase levels of N$_2$-CO$_2$-atmospheres as a function of the X-ray surface flux and the CO$_2$ mixing ratio simulated with a different upper atmosphere model by \citet{johnstone2018,Johnstone2021Atmosphere}. The various points are simulation results, the lines are fits to the simulations. Left panel adopted from \citet{tian2008}, right panel adopted from \citet{Johnstone2021Atmosphere}.}
    \label{fig:profilesXUV}
\end{figure}

If the XUV flux increases, modern Earth-type atmospheres react with an increase in exobase temperature and radius. \citet{tian2008} modelled the effect of high XUV fluxes on an Earth-like N$_2$-O$_2$-dominated atmosphere with 400\,ppm CO$_2$ and found that $T_{\rm exo}$ and $r_{\rm exo}$ increase from $\sim$900\,K to $\sim$\,8000 K  and 500\,km to $>$10\,000\,km, respectively, for a 6.6-fold increase in the XUV surface flux. The resulting thermospheric temperature profiles from \citet{tian2008} are shown in the left panel of Fig.~\ref{fig:profilesXUV}. 
Even if one neglected the atmospheric expansion and the resulting decrease of $g_{\rm exo}$ in Equation~\ref{eq:T-exoscaleing}, it would still predict a temperature increase to $T_{\rm exo}\sim5900$\,K for the aforementioned 6.6-fold increase in the XUV surface flux. This rough estimate is about 2000\,K lower than the modelled value by \citet{tian2008} as this also includes the increase of $r_{\rm exo}$ and the related decrease of $g_{\rm exo}$. For XUV fluxes above about 6 times the present-day value of $F_{\oplus}$, the Earth-type atmosphere adiabatically expands and escapes hydrodynamically into space. 
These results indicate that modern Earth-type atmospheres on planets in the HZ of low-mass stars may be strongly disfavoured (see also Fig.~\ref{fig:XUVEvo}). However, another recent upper atmosphere model by \citet{Nakayama2022} finds that modern Earth-type atmospheres are much more stable against XUV flux-induced escape due to strong cooling via radiative recombination and enhanced atomic line cooling. It will be an important future collaborative task to intercompare these models, their differences, and different results to understand better the thermal stability of Earth-type atmospheres.

The right panel of Fig.~\ref{fig:profilesXUV} further illustrates the variation of the exobase level of an N$_2$-CO$_2$-dominated atmosphere as a function of CO$_2$ partial pressure and the XUV surface flux as simulated by \citet{Johnstone2021Atmosphere}. One can see that atmospheres with higher CO$_2$ mixing ratios are more compact and hence more thermally stable. However, for $F_{\rm XUV}\gtrapprox\,15\,F_{\oplus}$, even CO$_2$-dominated atmospheres experience a strong expansion and significant loss rates as the CO$_2$ in the upper atmosphere gets increasingly dissociated. This behaviour also explains the decrease in the Jeans escape parameter, $\lambda_{\rm exo}$, in Fig.~\ref{fig:Fig_3b}; for an increase of the XUV surface flux from 1 to 14\,$F_{\oplus}$, $T_{\rm exo}$ and $r_{\rm exo}$ increase such that $\lambda_{\rm exo}$ decreases from 1000 to 0.5 for an atmospheric mixture of 1\% N$_2$ and 99\% CO$_2$, indicating hydrodynamic escape of the atmosphere.

Under present-day Earth-like conditions, Earth-type atmospheres are dominated toward the exobase in decreasing order by O, N$_2$, and N; the most dominant ions, albeit less abundant than the neutrals, are N$^+$, O$^+$, and H$^+$ \citep[e.g.,][]{Picone2002,johnstone2018}. Potential biosignature gases \citep[e.g.,][]{Seager2016} such as N$_2$O, NH$_3$, CH$_4$ and even O$_2$ and O$_3$ are less abundant in the upper atmosphere. Ozone, for instance, is predominantly formed in the stratosphere via the dissociation of O$_2$ by UV irradiation where it heats the atmosphere and produces the Earth's cold trap. However, there is some O$_3$ in the lower part of the upper atmosphere (mostly below 100\,km), which also contributes to atmospheric heating. Interestingly, O$_3$ heating increases over time whereas all other sources of heating in the upper atmosphere decrease with stellar age. This is due to O$_3$ absorption at wavelengths $>$200\,nm, a wavelength range that increases with increasing stellar age \citep{johnstone2018}. For early stellar ages and high XUV fluxes, the abundance of ions in the upper part of modern Earth-type atmospheres will strongly increase. In the model of \citet{Nakayama2022}, the ion fraction at the top of the upper atmosphere even reaches 100\% for highly irradiated atmospheres. In addition, these authors propose that highly-irradiated, but thermally stable O-rich atmospheres emit strongly in the NUV to optical wavelength range with an intensity of up to $2 \times 10^{20}$\,erg\,s$^{-1}$ This could be detectable during secondary transits in M dwarf systems such as Trappist-1. Current and future exoplanet observations are hence another crucial tool to verify or falsify certain upper atmosphere models and to enhance our knowledge of the stability, evolution, and galactic prevalence of Earth-type atmospheres.

\section{Role of Planetary Magnetic Fields}\label{sec:magnetic_field}
There have been many debates on the role of the planetary intrinsic magnetic fields on atmospheric escape \citep[e.g.][]{Lundin1989Natur.341..609L,Seki2001Sci...291.1939S,Ramstad2021SSRv..217...36R,Way2022}. The global magnetic field can deflect most of the stellar wind much further away from the planet and the upper atmosphere is protected by the magnetic barrier. A strong magnetic field results in the formation of a large magnetosphere, which enlarges the area that accumulates stellar wind energy inputs. As introduced in Sec.\,\ref{sec:intro_non-thermal}, the dominant atmospheric escape mechanisms can change between magnetized and unmagnetized planets. The effects of planetary magnetic fields include those that increase atmospheric escape and those that decrease it, while various factors need to be taken into account to determine whether the net escape rate will increase or decrease. In this Section, we review recent progress in our understanding of the role of the planetary intrinsic magnetic fields.

\subsection{Insight from Unmagnetized Planets (Mars/Venus) Observations}\label{sec:magnetic_field_unmag}
%
As mentioned above, there is a lack of consensus on the effects of the intrinsic magnetic field on atmospheric ion escape from terrestrial planets. Although Mars does not have a global intrinsic magnetic field, localized strong crustal magnetic fields exist mainly in its southern hemisphere. Thus, Mars has a built-in control experiment with a localized crustal magnetic field, having both magnetized and unmagnetized regions, and should host various escape processes that occur on both kinds of objects. Observations demonstrate that the induced magnetosphere's structure is influenced by the crustal magnetic fields, and its effect depends upon the local time of the strong crustal magnetic field region \citep[e.g.][]{Luhmann2015GeoRL..42.9087L,Masunaga2017JGRA..122.4089M}. Effects of crustal magnetic fields on ion escape from Mars have been investigated by comparing the northern and southern hemispheres.

\begin{figure}
    \centering
    \includegraphics[width=0.9\linewidth]{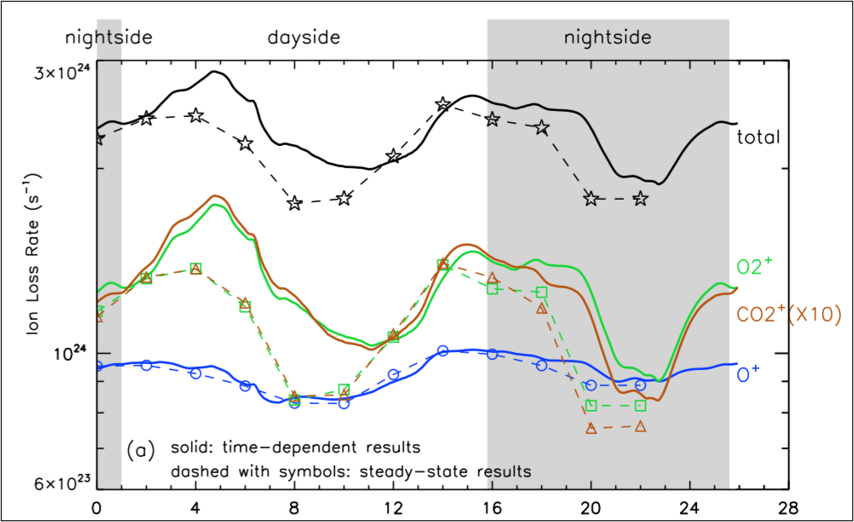}
    \caption{Variation of the ion loss rate from Mars with rotation of the strong crustal magnetic fields region [adopted from Fang et al., 2015].}
    \label{fig:7.1.1}
\end{figure}

Global MHD simulation with the rotating crustal magnetic field on Mars has shown that when simulations include crustal fields, ion loss changes by 0.1-30 times. When simulations rotate crustal fields through a day, ion loss changes by 15-50\%, as shown in Fig.\,\ref{fig:7.1.1} \citep[][]{Fang2015JGRA..12010926F}. The simulation results are consistent with observations of the polar plume. Observations also show that solar wind electric field effects are more dominant. Subsolar crustal magnetic fields tend to prevent the formation of the molecular ion plume under low dynamic pressure conditions \citep[][]{Sakakura2022JGRA..12729750S}. As for the tailward escape (see also Sec.\,\ref{sec:observations_mars}), the escape flux is larger in the -E hemisphere of the MSE coordinate than the +E, and indicates that the solar wind electric field effects are dominant. In the same +E or -E hemisphere, the existence of crustal magnetic fields reduces both the ion escape rate and O$^+$/O$_2^+$ ratio \citep[][]{Inui2019JGRA..124.5482I}.

\subsection{Insight from Magnetized Planet (Earth) Observations}
As introduced in Sec.\,\ref{sec:intro_non-thermal} and \ref{sec:observations_earth}, there are different non-thermal escape processes operating in Earth’s magnetosphere (i.e., the polar wind, cusp outflows, auroral outflows, plasma-spheric drainage plumes, and ENA production). The different paths of ion escape depend on the energy, mass, and species, as illustrated in Fig.\,\ref{fig:2.3.3_msphere-circulation}. Even though the Earth’s intrinsic magnetic field reduces the direct access of the solar wind to atmosphere and also traps part of the ion outflow inside the magnetosphere, observational evidences clearly show enhancements in escape for enhanced solar wind energy inputs \citep[][and reference therein]{Dandouras2021JGRA..12629753D}. These include the dependence of the escape rate of the heavy-ion outflow on solar wind dynamics pressure \citep{Schillings2019EP&S...71...70S} and geomagnetic activity \citet{Slapak2017AnGeo..35..721S} as shown in Fig.\,\ref{fig:2.3.5_Oplus}. Since about half of the incoming solar wind energy input is consumed as Joule Heating in the Earth’s magnetosphere \citep{Tenfjord2013JGR}, it is expected to provide favourable conditions for the escape processes. Magnetospheric dynamics such as magnetospheric convection and substorms/storms due to the enhanced solar wind energy input also increase plasma sheet ion escape from the magnetotail via plasmoid formation and enhance the loss from the dayside magnetopause as observed both for heavy ions \citep[][and reference therein]{Kronberg2014SSRv..184..173K} and the cold light ions \citep{Haaland2012JGRA..117.7311H}. 

During geomagnetic storms, a larger fraction of solar wind energy input is consumed in the ring current \citep[29\%, compared to average level of 15\%][]{Tenfjord2013JGR} and contributes to enhanced charge exchange, which is the major loss mechanism, in particular during the late recovery phase of the storm \citep{Keika2011JGRA..116.0J20K}. Plasmaspheric wind process, on the other hand, is a continuous ion outflow process in equatorial latitude driven by interchange instability of the corotating plasma in the dipole field \citep{Andre2006JASTP} and was detected under quiet magnetospheric condition \citep{Dandouras2013AnGeo..31.1143D}. Hence, the relative importance of the different escape processes are expected to vary depending on the state of the magnetosphere, as well as the relative importance of the internal (rotation) and solar wind-driven processes. In the latter context, observations at different solar system planets with different strengths and orientations of the dipole will help predicting possible effects of the magnetospheric processes on the escape at exoplanets. For example, importance of escape due to a loop-like plasmoid in the magnetotail was reported for Uranus \citep{DiBraccio2019GRL}, which has a slanted dipole tilt and rotational axis.  The observation suggests that both rotation and solar wind are driving the plasma sheet ion escape in the magnetotail, differently from the Earth.

\subsection{Insight from Early Earth Studies}
Due to the fact that Earth lacks almost any geological record in the Hadean ($\sim 4.6$ to 4 billion years ago), beyond the scarce samples of detrital zircons \citep[][]{Harrison2009annurev.earth.031208.100151,Lingam_Balbi_2024}, it has proven challenging to determine the strength of Earth’s surface magnetic field, or equivalently, its magnetic dipole moment. Hence, it is not surprising that researchers have arrived at distinct, and even opposing, conclusions. 

One school of thought holds that Earth has possessed a geomagnetic field strength comparable to that of modern Earth since {at least} $\sim 4.2$ billion years ago (Ga) based on the analysis of ferromagnetic inclusions in the zircons \citep[][]{Tarduno2015Sci...349..521T,Tarduno2020PNAS..117.2309T}, which is best summarized in Fig.\,3 of \citet{Tarduno2023nature}. Other groups have contended that, owing to the possibility of these inclusions being incorporated at a later date, it is impossible to establish the existence of the geodynamo prior to 3.5-3.7 Ga \citep[][]{Weiss2018Geo....46..427W,Borlina2020,Taylor2023}. Earth’s paleomagnetic field intensity is better resolved as we move into the Archean (4 to 2.5 Ga), as suggested by \citet{Nichols2024JGRB..12927706N}.

However, Earth’s magnetic field has not been constant throughout this period, due to phenomena such as magnetic field reversals. A striking example of this is the evidence for a very weak magnetic field in the Ediacaran Period (defined as 635 to 539 Ma), according to multiple paleomagnetic studies \citep[][]{Bono2019NatGe..12..143B,Thallner2021E&PSL.56817025T,Domeier2023ESRv..24204444D}. This period is characterised by a magnetic field strength $\sim 30$ times lower than today, and may have lasted for tens of Myr \citep[][]{Huang2024ComEE...5..207H}. 

There have been several attempts to try and link the state of Earth’s magnetic field in a particular period to our planet’s habitability by focusing on atmospheric escape rates \citep[e.g.][]{Bono2019NatGe..12..143B,Huang2024ComEE...5..207H,Nichols2024JGRB..12927706N}. While there may be potentially significant effects on Earth’s biosphere due to magnetic field changes \citep[cf.][]{Lingam2019ApJ...874L..28L,pan_li2023}, the latter's impact on atmospheric escape may be less important. The reason is that irrespective of the magnitude of Earth’s magnetic field (or dipole moment), the total atmospheric escape rate (in units of kg/s) is predicted to fluctuate only by a factor of $\sim 2$ at most \citep[][]{Gunell2018A&A...614L...3G}. Modelling indicates that these relatively modest variations in the escape rate may not have caused major atmospheric mass or composition fluctuations, thereby mitigating the effects of atmospheric escape on the biosphere in such scenarios \citep[][]{Lingam2019ApJ...874L..28L}.

\subsection{Predictions from Simulation Studies}

\begin{figure}
    \centering
    \includegraphics[width=\linewidth]{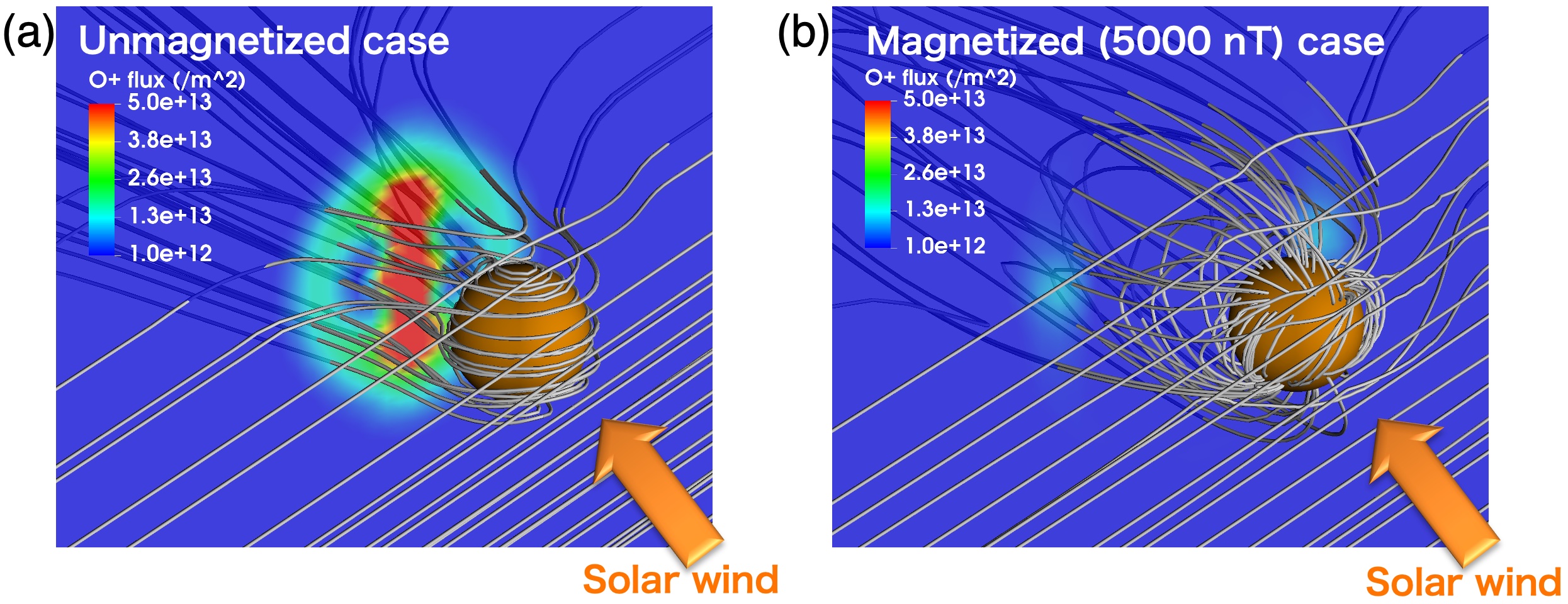}
    \caption{Results of global multi-fluid MHD simulations of the interaction between the solar wind and ancient Mars (a) without and (b) with the global (dipole) intrinsic magnetic field. Lines show the magnetic field lines and the colour contour indicates the tailward escape flux of planetary heavy ions.}
    \label{fig:7.4.1}
\end{figure}

There have been studies using global hybrid \citep[e.g.][]{Egan2019MNRAS.488.2108E}, multi-species MHD \citep[e.g.][]{Sakai2018GeoRL..45.9336S,Sakai2021JGRA..12628485S,Sakai2023JGRA..12830510S,Sakata2020JGRA..12526945S,Sakata2022JGRA..12730427S,Dong2017JGRA..122.4009D,Dong2020ApJ...896L..24D,Nishioka2023JGRA..12831405N}, and multi-fluid MHD \citep[][]{Sakata2024JGRA..12932320S} simulations of the interaction between the stellar wind and magnetized planets. Both hybrid and multi-species MHD simulations indicate that when a planet has a weak magnetic field, it can increase the ion escape rate and the molecular/atomic ion ratio in the ion escape. However, increasing the planetary magnetic field strength leads to decreases in the ion escape rate. This is demonstrated in Fig.\,\ref{fig:7.4.1}. Here, the ion escape rate significantly decreases in the strong magnetic field case (i.e., 5000 nT on the equatorial planetary surface, panel b), compared to the unmagnetized case (panel a). The transition of the magnetic field effects on the ion escape rate from increase to decrease depends on the relative strength between the stellar wind dynamic pressure and the magnetic pressure of the planetary dipole field. Namely, a weak intrinsic magnetic field increases ion loss rates when the solar wind dynamic pressure greatly exceeds the magnetic pressure (overpressure). The existence of an intrinsic magnetic field facilitates cusp outflows enabling more escape of molecular ions (O$_2^+$, CO$_2^+$) \citep[e.g.][]{Sakata2020JGRA..12526945S}. The orientation of the interplanetary magnetic field embedded in the stellar wind can also be an important controlling factor in ion escape \citep[e.g.][]{Sakai2021JGRA..12628485S,Sakai2023JGRA..12830510S}.

\begin{figure}
    \centering
    \includegraphics[trim=0 0 0 15,clip,width=0.5\linewidth]{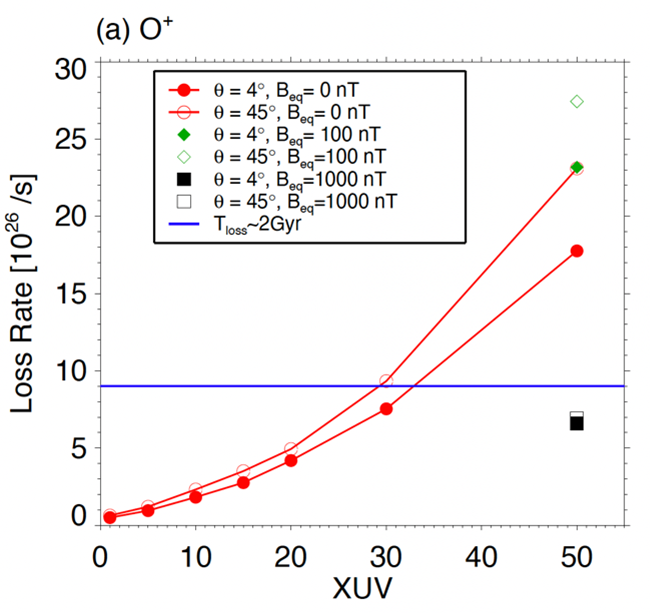}
    \caption{A summary of global MHD simulation results of ion (O$^+$) escape from a Venus-like exoplanet at TOI-700\,d location. Dependence on stellar XUV radiation is shown for two interplanetary magnetic field orientations (almost radial: $\theta\,=\,4^{\circ}$ and similar to Earth’s location: $\theta\,=\,45^{\circ}$). Green and black symbols show the results of magnetized cases with dipole strength of 100 and 1000 nT at the planetary equatorial surface, respectively \citep[][]{Nishioka2023JGRA..12831405N}.}
    \label{fig:7.4.3}
\end{figure}

There have been some studies that simulate ion escape from magnetized exoplanets \citep[e.g.][]{Dong2017JGRA..122.4009D,Dong2020ApJ...896L..24D,Nishioka2023JGRA..12831405N}. A strong intrinsic magnetic field suppresses ion escape, allowing the atmosphere to be retained longer than at unmagnetized planets. For example, as shown with red lines in Fig.\,\ref{fig:7.4.3}, stellar XUV flux must be within 30 times that of Earth to retain a modern Venus-like atmosphere for an unmagnetized exoplanet. However, a strong intrinsic magnetic field suppresses ion escape (black symbols in Fig.\,\ref{fig:7.4.3}), allowing the atmosphere to be retained even when XUV is 50 times that of Earth \citep[][]{Nishioka2023JGRA..12831405N}.

Similar results were obtained in the MHD simulations of hydrodynamically escaping atmospheres of close-in sub-Neptune-like and giant exoplanets interacting with stellar wind \citep[e.g.][]{Carolan2021MNRAS.508.6001C}. It was shown that the presence of an intrinsic magnetic field leads to a suppression of the outflow through the magnetotail and the day side while enhancing the outflow through the polar regions. The interplay of those effects leads to an increase in escape rates for weak planetary magnetic fields and the suppression of escape rates for strong ones. It was also shown that for a given planetary magnetic field, the escape increases with the increasing magnetic field strength of the stellar wind \citep[][]{Gupta2023ApJ...953...70G}. These results cannot be directly extrapolated to the escape from young terrestrial planets but suggest that the magnetic field effects during that period (at least, for XUV-driven escape) might be similar to the present day.

\section{Summary: General Implications for Exoplanetary Systems}\label{sec:discussion}
Upper atmosphere processes, their relative contribution to volatile losses, and possible consequences for atmospheric evolution depend strongly on planetary parameters such as mass, atmospheric composition, and the presence of a magnetic field, {as was at length discussed in Sec.\,\ref{sec:planetary_mass}, \ref{sec:composition}, and \ref{sec:magnetic_field}}. Local conditions set by the planet’s host star and orbit also play a role {(see Sec.\,\ref{sec:stellar_input}), in} particular, the incident XUV radiation {\citep[e.g.][]{kubyshkina2018A&A...619A.151K,Seki2015SSRv..192...27S}}, stellar wind properties {\citep[e.g.][]{Kislyakova2014A&A...562A.116K,Curry2015P&SS..115...35C}}, and plasma-related phenomena like Coronal Mass Ejections {\citep[CMEs; e.g.][]{Hazra2025MNRAS.536.1089H}} and Stellar Energetic Particles {\citep[in particular, for ENA production, e.g.][]{Yue2019JGRA..124.7786Y}}. Observations of the terrestrial planets in the Solar System show that all of these parameters should be considered as an ensemble {(see Sec.\,\ref{sec:observations})}. Effects connected to atmospheric composition, come out differently for planets of different masses and under different irradiation levels {(see Sec.\,\ref{sec:composition_CO2})}. Meanwhile, many relevant parameters remain poorly constrained for exoplanets and are expected to change drastically throughout their lifetimes {\citep[e.g.][]{Emsenhuber2021A&A...656A..69E}}. Nevertheless, we can make some {qualitative} predictions by employing available observations and theoretical modelling{, as described below. We note, that these predictions rely on the specific stellar irradiation and atmospheric escape models (i.e., affected by specific model assumptions), and should therefore be considered as an order-of magnitude estimates.}
\begin{figure}
    \centering
    \includegraphics[width=0.8\linewidth]{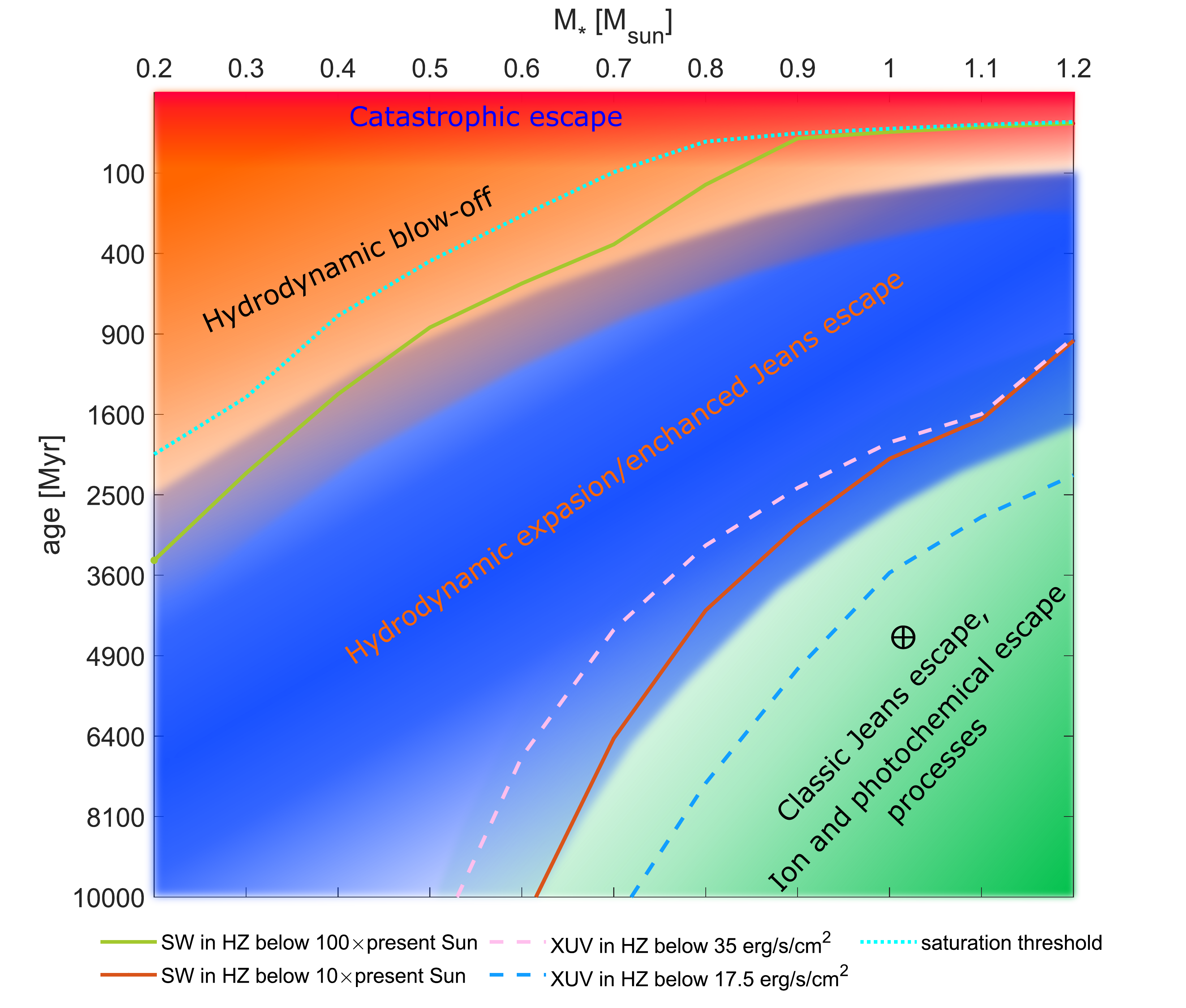}
    \caption{Illustration of thermal and non-thermal atmospheric escape processes as a function of stellar mass and age. Coloured areas indicate the most likely dominant escape processes for the given parameters. The thin cyan dotted line shows the time when the star of a given mass drops out of saturation, while pink and blue dashed lines show the time when XUV radiation in the HZ decreases below the stability threshold of N$_2$-dominated atmospheres with a 10\% and 1\% CO$_2$ mixing ratio, respectively {\citep[according to ][]{VanLooveren2024Trappist}}. Solid green and brown lines show where the stellar wind flow (density $\times$ velocity) decreases below 100 and 10 times that of the present Sun. The ``$\oplus$'' sign denotes the position of the Earth. All estimates are made assuming the stars to be moderate rotators using the Mors stellar evolution code \citep[][]{Johnstoneetal21} and empirical approximations for the stellar wind strength from \citet{Vidotto21}.}
    \label{fig:Escape_processes}
\end{figure}

Fig.~\ref{fig:Escape_processes} illustrates the expected contribution of the major atmospheric escape processes discussed in this work for planets within the classical liquid water habitable zone where mass versus age is plotted for a given planet's host star.
The highest bulk atmospheric escape rates (red area) are predicted during the stellar activity saturation phase (see \ref{sec:stellar_input:rotation}), when the gas disk has just dispersed, and a young (proto)planet is stripped of its surrounding nebula gas. During this phase, the hydrodynamic outflow can be powered by the high internal energy of a newborn planet with inflated atmosphere {\citep[e.g.][]{ginzburg2018}} and extreme stellar heating {\citep[e.g.][]{owen_wu2016boil-off}}, with large impacts contributing additional escape \citep[related to the planetary accretion stage that continues after the gas disk dispersal; e.g.][]{Schlichting2015Icar..247...81S,Emsenhuber2021A&A...656A..69E}. The escape rates and duration of this phase are mainly constrained by the planet's mass and size {\citep[e.g.][and references therein]{Kubyshkina2022AN....34310077K}}. This phase smoothly transitions into the hydrodynamic blow-off phase {(orange area)}. Here, the escape is mostly powered by stellar XUV radiation \citep[e.g.][]{king2021MNRAS.501L..28K}. The duration of the blow-off phase ($\sim$0.1-2.5\,Gyr) depends strongly on the duration of the host star's saturation phase, and hence on stellar mass and rotation {(see discussion in Sec.\,\ref{sec:stellar_input:rotation})}. It can spread significantly around the value predicted for a star evolving as moderate rotator shown in Fig.\,\ref{fig:Escape_processes}. The formation of a stable secondary atmosphere during both of these early phases is unlikely {\citep[see e.g. the estimates in][]{VanLooveren2024Trappist}}.

After hydrodynamic escape begins to cease, the atmosphere of an evolving planet undergoes an efficient thermal escape phase { (blue-shaded area in Fig.\,\ref{fig:Escape_processes})}. This is governed by an interplay between a strongly heated and expanded thermosphere-exobase regime, also called enhanced Jeans escape {\citep[e.g.][]{lammer2008,Zhu2014Icar..228..301Z}}. 
Whilst the thermosphere expands hydrodynamically upward, adiabatic cooling breaks the upflow before it becomes supersonic, and not all upward-flowing atmospheric atoms reach the escape velocity at the expanded exobase level. At this stage, a significant fractionation of light species is expected. Atmospheres are counterbalanced by outgassing and escape; mass and stellar input still dominate, but the composition of the outgassed atmospheres and magnetic fields become increasingly important {\citep[e.g.][]{Lammer2013oepa.book.....L,Lammer2020SSRv..216...74L}}.

The transition from enhanced to classical Jeans escape (green area) also depends on the planetary mass and atmospheric composition, controlling adiabatic and IR-cooling in the upper atmosphere {\citep[e.g.][]{johnstone2018}}. This transition is illustrated with the pink and blue dashed lines that correspond to the atmospheric stability criteria of atmospheres with N$_2$/CO$_2$ mixing ratios of 90\%/10\% and 99\%/1\% on planets with a mass of 1\,M$_{\oplus}$ {\citep[][]{VanLooveren2024Trappist}}. 

During this latter phase, non-thermal ion and photochemical escape processes can contribute to the total atmospheric mass loss similarly to or more than thermal Jeans escape {\citep[e.g.][]{lammer2008}}. During earlier epochs, only the polar winds on magnetised planets are expected to have escape comparable to the hydrodynamic outflow {\citep[e.g.][]{Kislyakova2020JGRA..12527837K}}. However, the cumulative effect of the stellar wind-driven non-thermal escape mechanisms (see Tab.\,\ref{tab:non-thermal}) at planets orbiting low-mass stars requires further consideration. The stellar wind flow in the vicinity of such planets may remain more than 10 times the average present-Earth level even after reaching the stability threshold of XUV-driven escape for Earth-like atmospheres {\citep[compare pink-dashed and brown solid lines in Fig.\,\ref{fig:Escape_processes};][]{Vidotto21}}.

Fig.~\ref{fig:Escape_processes} additionally highlights that different atmospheric escape phases generally last longer for stars with lower masses. In particular, HZ planets around stars lighter than $\sim$0.5\,$M_{\odot}$ may never leave the hydrodynamic/enhanced Jeans escape regime during the main sequence {\citep[e.g.][]{Johnstone2021Atmosphere}}. For this reason, atmospheric escape is expected to have a larger impact on the atmospheric evolution of such planets, both in terms of the element fractionation and the stability of the entire atmosphere.  This may result in the atmospheric composition diversity of M-dwarf planets being heavily shifted towards more stable atmospheres (e.g., CO$_2$ or water-rich) instead of those dominated by lighter elements {\citep[e.g.][]{Scherf2024AsBio..24..916S}}. {It can be crucial for planets' habitability as, for example, nitrogen and oxygen are key elements of amino acids, DNA and RNA. Life processes on Earth are influenced by the atmospheric composition and a change in the relative abundance of these two species can substantially modify biochemical reactions as for example the synthesis of amino acids \citep{Bada2013CSRev..42.2186B,Camprubi2019SSRv..215...56C}. As atmospheric escape shapes the atmospheric composition over time \citep[e.g.][]{Yamauchi2007AsBio...7..783Y,Airapetian_2017}, it is essential to evaluate how the composition of the particles that escape responds to changes to the external driving factors, and more specifically how the nitrogen/oxygen escape ratio evolves in response to the solar activity \citep{Yau1993JGR....9811205Y,Christon2002GeoRL..29.1058C,Lin2024FrASS..1162957L}. Nitrogen and oxygen, although having closely-spaced atomic and molecular masses, have very distinct dissociation and ionization energies, and exhibit different responses to varying conditions.}

In general, assessing the possible atmospheric compositions of extrasolar planets is strongly limited by our knowledge of the initial volatile budget in the planet's interior, which can only be loosely constrained using formation models and observations of protoplanetary discs and young systems {\citep[e.g.][]{Burn2024}}. Thus, even though we can predict atmospheric escape rates, the actual atmospheric lifetime will additionally depend on how long a specific atmosphere can be replenished through outgassing {\citep[e.g.][]{Godolt2019A&A...625A..12G}}. Furthermore, the initial composition affects the escape rates of specific elements and vice versa. This degeneracy might only be broken with direct observations of planetary atmospheres (e.g. in transmission spectra), which remains challenging even for large planets in close orbits.

Even more problematic is the effect of planetary magnetic fields. While some methods were suggested to measure the magnetic fields of giant exoplanets \citep[e.g.][]{Kavanagh2024A&A...692A..66K}, the prevalence of intrinsic magnetic fields on terrestrial-like exoplanets remains an area of speculation until observational confirmation. As outlined in Sec.\,\ref{sec:magnetic_field}, the influence of planetary magnetic fields on atmospheric escape is uncertain and depends both on its strength and configuration but also on planetary and stellar parameters. However, even if the effects of the magnetic field on the volatile escape can be negligible on long timescales, its presence is expected to affect potential observational signatures \citep[e.g.][]{Rumenskikh2025arXiv250101122R}. In particular, %
planetary magnetospheres' shape and dynamics can change drastically depending on current stellar wind conditions, stellar periodicities, short-term events, and planetary geomagnetic variability. This can result in highly inhomogeneous outflows.

In this study we do not tackle every factor responsible for atmospheric escape, as such a comprehensive discussion falls outside the scope of the paper. For instance, we do not address the role of the (magnetic) obliquity, which has been shown to exert modest effects on atmospheric escape rates \citep{dong2019}.
We also omit the discussion of planets on highly eccentric or inclined orbits, which can affect star-planet interaction and planetary migration history in various ways \citep[e.g.][]{Emsenhuber2021A&A...656A..69E}. Furthermore, the discussion ignores the presence of atmospheric hazes and dust, which can affect atmospheric opacities, thermal structure \citep[e.g.][]{lavvas2019ApJ...878..118L,Adams2019ApJ...874...61A} and atmospheric pressure gradients \citep[e.g.][]{Schlichting2022PSJ.....3..127S}. Our discussion also ignores the role of impacts in atmospheric escape and evolution, which can induce significant losses of primordial atmospheres \citep[e.g.,][]{Schlichting2015Icar..247...81S,Lammer2020Icar..33913551L,gillmann2022} and can be an important factor in the evolution of secondary atmospheres \citep[e.g.,][]{Pham2011,Shorttle2024}. We also omit to discuss the poorly understood effect of electron precipitation; as this heats the lower thermosphere. It could have been highly relevant during the transition period between enhanced and classic Jeans escape \citep{Tian2008b}.
Finally, we did not consider in detail the interactions between lower and upper atmospheres (and in particular, the diffusion limit on the volatile supply of upper atmospheres), which are discussed in more detail in \citet{chapter7}.%

\backmatter




\bmhead{Acknowledgements}
DK was supported by a Schr\"odinger Fellowship supported by the Austrian Science Fund (FWF) project number J4792 (FEPLowS).
MJW acknowledges support from NASA’s Interdisciplinary Consortium for Astrobiology Research (ICAR) Nexus for Exoplanet System Science (NExSS). MJW acknowledges support from the GSFC Sellers Exoplanet Environments Collaboration (SEEC), which is funded by the NASA Planetary Science Division’s Internal Scientist Funding Model.
The work by ID  was supported in part by CNES through order 4500081416.
AFL gratefully acknowledges support from the Italian Ministry of University and Research through the Italian National Institute for Astrophysics (INAF).
MS thanks the Austrian Science Fund (FWF) for the support of the VeReDo research project, grant I6857-N.
MP is grateful for the support from the Swedish National Space Agency (Dnr 2023-00183).
KS was supported by Grant-in-Aid for Scientific Research (A) of JSPS (Japan Society for the Promotion of Science) KAKENHI Grant Number JP22H00164 and JP25H00684.


\section*{Declarations}

\begin{itemize}
\item Funding: Austrian Science Fund (FWF) project number J4792, CNES order 4500081416, Austrian Science Fund (FWF) grant I6857-N, Swedish National Space Agency (Dnr 2023-00183), Grant-in-Aid for Scientific Research (A) of JSPS (Japan Society for the Promotion of Science) KAKENHI Grant Number JP22H00164 and JP25H00684
\item Conflict of interest: There is no conflict of interests to the best of our knowledge 
\item Ethics approval and consent to participate: not applicable
\item Consent for publication: not applicable
\item Data availability: not applicable
\item Materials availability: not applicable
\item Code availability: not applicable
\item Author contribution: not detailed
\end{itemize}
%

\bigskip

\begin{appendices}
\section{Terms and abbreviations used in the present study}\label{apx:notations}
In Tab.\,\ref{tab:notations} we summarise the important terms and notations commonly used in the chapter. For the convenience of the reader, the terms are sorted alphabetically.

\begin{table}
    \caption{List of terms and notations}
    \label{tab:notations}
    \centering    
    \begin{tabularx}{\textwidth}{X|c|X}
            \toprule
        Term & Notation & Comment \\
        \midrule
        \endhead
        {Astrosphere} & {} & {The magnetic obstacle formed by the stellar magnetic field's interaction with the interstellar medium.} \\ \midrule
        {Corotation Interaction Region} & {CIR} & {The stellar wind plasma structure formed by interaction of the faster wind flow (e.g., originating from coronal holes) with the slower one. Characterised by the forward pressure wave at the leading edge and a reverse pressure wave at the trailing edge \citep[][]{Gosling1999SSRv...89...21G}. It can produce shock waves at large distances from the star.} \\ \midrule
        {Energetic neutral atom} & {ENA} & {Neutral atmospheric particle with energy sufficient to escape planetary gravitational bound. This energy can be acquired through photochemical reactions or collisions with energetic ions.} \\ \midrule
        {Escape velocity} & {$v_{\rm esc}$} & {The minimum speed for an object (atmospheric particle) to escape from a planetary body.} \\ \midrule
        {Exobase} & {$_{\rm exo}$} & {The atmosphere level where barometric conditions are no longer valid (particle's mean free path becomes larger than atmospheric scale height).} \\ \midrule
        {Exosphere} & {} & {The uppermost atmospheric layer where no collisions between particles take place.}\\ \midrule
        {Extreme Ultraviolet} & {EUV} & {Stellar high-energy radiation in 10-91.2\,nm wavelength range (most relevant for thermal escape). The upper boundary is set by the atomic hydrogen photoionisation threshold.}\\ \midrule
        {Geomagnetic (sub)storm} & { } & {Short- (order of hours or less) or long-term (up to a few days) perturbations in geomagnetic activity level (defined through Kp, quantifying disturbances in horizontal MF component, AE, auroral electrojet index, and Dst, disturbance storm time), indicating disturuptions in the magnetospheric convection.}\\ \midrule
        {Habitable zone} & {HZ} & {Here we only consider the definition of the habitable zone as the orbital range where liquid water can be sustained on a planetary surface. For more discussion see, e.g. \citet{Scherf2024AsBio..24..916S}.} \\ \midrule
        {Heterosphere} & {} & {The atmospheric layer where molecular diffusion dominates, fractionating lighter species (with a larger scale height, $H_i$ from heavier species (with a smaller $H_i$). Roughly equivalent to upper atmosphere.} \\ 
       \bottomrule
       \end{tabularx}       
    \end{table}
%
\begin{table}
    \centering    
    \begin{tabularx}{\textwidth}{X|c|X}
            \toprule
        Term & Notation & Comment \\
        \midrule
        \endhead
        {Homosphere} &   & {The atmospheric layer where turbulent eddy diffusion dominates, leading to a well mixed bulk gas with a single scale height, $H$. Roughly equivalent to lower atmosphere.} \\ \midrule
        {(Magneto)hydrodynamic} & {(M)HD} & {The approach describing a given medium (e.g., planetary atmospheres or stellar wind plasma) using fluid dynamic equations. It is applicable as long as the medium remains collisional (mean free path shorter than the local scale height/collisional timescale is much shorter than other timescales in the system).} \\ \midrule
        {(Interplanetary) Coronal Mass Ejection} & {(I)CME} & {The significant ejection of stellar plasma into the astrosphere, most commonly accelerated through the reconnection event in stellar coronae. CMEs propagate along stellar wind flow, manifesting as closed magnetic field regions (magnetic clouds), a bulge of plasmas with higher density and magnetic field and lower proton temperature.} \\ \midrule
        {Induced magnetosphere} & {} & {Same as magnetosphere but the planetary magnetic field emerges from the solar wind interaction with a planetary ionosphere instead of the intrinsic dynamo.} \\ \midrule 
        {Ionosphere} & {} & {The ionized layer produced within the thermosphere.} \\ \midrule
        {Jeans parameter} & {$\lambda_{\rm exo}$} & {Ratio of the planet's escape velocity to the thermal speed of atmospheric particles at the exobase, squared.} \\ \midrule
        {Jeans parameter (generalized)} & {$\Lambda$} & {Same as Jeans parameter, but calculated at the planetary radius (photosphere).} \\ \midrule
       {Lower atmosphere} & {} & {The Mesosphere, stratosphere, and troposphere} \\ \midrule        
       {Planetary radius/Planetary photosphere} & {$R_{\rm pl}$} & {Radius defined at the altitude where the atmosphere becomes optically thin to visible light. For secondary atmospheres, it is $\simeq$ radius of the solid part of a planet whereas for H/He atmospheres it lies at atmospheric pressures $\sim$100\,mbar.} \\ \midrule
       {(Electro)magnetic field} & {MF} & {--} \\ \midrule
       {Magnetosphere} & {} & {An obstacle formed by the interaction of the intrinsic/induced planetary MF with the stellar wind} \\ 
       \bottomrule
       \end{tabularx}
       \label{tab:notations_2}
\end{table}

\begin{table}
    \centering    
    \begin{tabularx}{\textwidth}{X|c|X}
            \toprule
        Term & Notation & Comment \\
        \midrule
        \endhead
        {Mesosphere} & {} & {The third layer from the bottom of an atmosphere, above the stratosphere and below the thermosphere, where the temperature decreases with increasing altitude.} \\ \midrule
       {Solar System} & {SS} & {--} \\ \midrule
       {Stellar Wind (Solar Wind in SS)} & {SW} & {A transonic outflow of the fully ionized stellar material (plasma) with a frozen-in MF } \\ \midrule
       {Stratosphere} & {} & {The second-lowest atmospheric layer where on Earth the temperature increases with altitude as a result of the UV absorption by the ozone layer. Not necessarily present at all planets.} \\ \midrule
       {Thermosphere} & {} & {The atmospheric layer where photoionization, photodissociation, and electron dissociative recombination occur resulting in strong atmospheric heating.} \\ \midrule
       {Troposphere} & {} & {The lowest atmospheric layer where most weather phenomena take place.} \\ \midrule 
        {Plasma sheet} & {} & {The region in the planetary magnetosphere's tail (stretched in antistellar direction) where ions from some of the ionospheric regions (typically transported through the open-line lobes) get trapped within closed magnetic field lines. It typically embeds a much thinner current sheet in the middle, where the neutral sheet separates magnetic fields of opposite polarities.} \\ \midrule
       {Upper atmosphere} & {} & {The thermosphere and exosphere.} \\ \midrule 
       {X-ray and Extreme ultraviolet} & {XUV} & {Stellar high-energy radiation with photon energies above 13.6\,eV (wavelengths shorter than 91.2 nm). In the context of non-thermal escape, this definition commonly also includes UV wavelengths ($\leq380$\,nm) that ionize elements heavier than helium (e.g., O$^+$ production).} \\ 
        \bottomrule
       \end{tabularx}
       \label{tab:notations_3}
\end{table}

\end{appendices}

\bibliography{upper_atmospheres}

\end{document}